\documentclass[aps,prb,twocolumn,superscriptaddress]{revtex4-1}
\usepackage{graphicx,psfrag,color}
\usepackage{mathbbol,amsmath,amsfonts,amssymb,bm,dsfont}
\usepackage{wasysym}
\usepackage{hyperref}
\hypersetup{
    colorlinks,
    citecolor=blue,
    filecolor=blue,
    linkcolor=blue,
    urlcolor=blue
}

\def\d{{\mathbf{d}}}
\def\m{{\mathbf{m}}}
\def\s{{\mathbf{s}}}

\def\j{{\mathbf{j}}}
\def\n{{\mathbf{n}}}
\def\k{{\mathbf{k}}}
\def\q{{\mathbf{q}}}
\def\r{{\mathbf{r}}}
\def\R{{\mathbf{r}}}

\begin{document}

\title{Generalization of Bloch's theorem for arbitrary boundary conditions: \\ 
Interfaces and topological surface band structure} 

\author{Emilio Cobanera}
\affiliation{SUNY Polytechnic Institute, 100 Seymour Rd, Utica, NY 13502, USA }
\affiliation{\mbox{Department of Physics and Astronomy, Dartmouth
College, 6127 Wilder Laboratory, Hanover, NH 03755, USA}}

\author{Abhijeet Alase}
\affiliation{\mbox{Department of Physics and Astronomy, Dartmouth
College, 6127 Wilder Laboratory, Hanover, NH 03755, USA}}

\author{Gerardo Ortiz}
\affiliation{\mbox{Department of Physics, Indiana University,
Bloomington, Indiana 47405, USA}}
\affiliation{\mbox{Department of Physics, University of Illinois, 
1110 W Green Street, Urbana, Illinois 61801, USA}}

\author{Lorenza Viola}
\affiliation{\mbox{Department of Physics and Astronomy, Dartmouth
College, 6127 Wilder Laboratory, Hanover, NH 03755, USA}}


\begin{abstract}
We describe a method for exactly diagonalizing clean $D$-dimensional lattice
systems of independent fermions subject to arbitrary boundary conditions in one 
direction, as well as systems composed of two bulks meeting at a planar interface. 
The specification of boundary conditions and interfaces can be easily adjusted to describe 
relaxation, reconstruction, or disorder away from the clean bulk regions of the system. Our 
diagonalization method builds on the {\it generalized Bloch theorem} 
[A. Alase {\em et al.}, Phys. Rev. B {\bf 96}, 195133 (2017)]
and the fact that the bulk-boundary separation of the Schr\"odinger equation is compatible 
with a partial Fourier transform operation. Bulk equations may display unusual features 
because they are relative eigenvalue problems for non-Hermitian, bulk-projected Hamiltonians.
Nonetheless, they admit a rich symmetry analysis that can simplify considerably the structure 
of energy eigenstates, often allowing a solution in fully analytical form.
We illustrate our extension of the generalized Bloch theorem to multicomponent systems 
by determining the exact Andreev bound states for a simple SNS junction. We then analyze 
the Creutz ladder model, by way of a conceptual bridge from one to higher dimensions. 
Upon introducing a new Gaussian duality transformation that maps the Creutz ladder to a system of 
two Majorana chains, we show how the model provides a first example of a short-range chiral topological 
insulators hosting topological zero modes with a power-law profile. Additional applications include the 
complete analytical diagonalization of graphene ribbons with both zigzag-bearded and armchair boundary 
conditions, and the analytical determination of the edge modes in a chiral \(p+ip\) two-dimensional 
topological superconductor. Lastly, we revisit the phenomenon of Majorana flat bands and anomalous 
bulk-boundary correspondence in a two-band gapless $s$-wave topological superconductor.
Beside obtaining sharp indicators for the presence of Majorana modes through the use of the 
boundary matrix, we analyze the equilibrium Josephson response of the system, showing how the 
presence of Majorana flat bands implies a substantial enhancement in the \(4\pi\)-periodic supercurrent.
\end{abstract}

\date{\today}
\maketitle


\section{Introduction}

This paper is the logical continuation of Ref.\,[\onlinecite{PRB1}],
referred to as Part I henceforth. In Part I, we described a method for the exact 
diagonalization of clean systems of independent fermions subject to arbitrary 
boundary conditions (BCs), and illustrated its application in several prototypical 
one-dimensional ($D=1$) tight-binding models \cite{PRB1,JPA,PRL}. 
Our broad motivation was, and remains, to develop an analytic approach 
for exploring and quantitatively characterizing the interplay between {\em bulk} and 
{\em boundary} physics, in a minimal setting where translation symmetry is  
broken {\em only} by BCs. On a fundamental level, such an understanding is 
a prerequisite toward building a complete physical picture of the 
bulk-boundary correspondence for mean-field topological electronic matter. 
For systems classified as topologically non-trivial \cite{chiu16}, there exist at least one 
bulk invariant and one boundary invariant whose values must coincide \cite{prodanBook}.
Bulk invariants are insensitive to BCs by construction, but what is the impact of BCs 
on boundary invariants?  Likewise, with an eye toward applications, what are design principles 
and ultimate limitations in engineering boundary modes in topological materials?    
   
Our method of exact diagonalization provides an insightful
first step towards answering these questions, because it can 
be casted neatly as a generalization of Bloch's theorem to arbitrary BCs. As 
we showed, in the generic case the exact energy eigenstates of a 
single-particle Hamiltonian are linear combinations of {\em generalized Bloch 
states}. The latter are uniquely determined by the analytic continuation 
of the Bloch Hamiltonian (or some closely-related matrix function)
off the Brillouin zone to {\it complex} values of the crystal momentum. 
In essence, the problem of diagonalizing the single-particle Hamiltonian boils 
down to finding all linear combinations of generalized Bloch states which satisfy 
the BCs. As long as the bulk is disorder-free and couplings have finite range, 
BCs can be encoded in a {\it boundary matrix}, whose shape is generally 
independent of the number of lattice sites. Any change in the energy levels and eigenstates 
induced by a change in BCs is thus directly and efficiently computable from the boundary matrix in principle. 

The generalized Bloch theorem properly accounts for two types of energy 
eigenstates that do not exist once translation invariance is imposed via 
Born-von-Karman (periodic) BCs: perfectly localized modes and localized modes 
whose exponential decay exhibits a power-law prefactor.  While such ``power-law
modes'' have been well documented in numerical investigations of long-ranged tight-binding 
models \cite{longrange}, it was a surprise to find them in short-range models\cite{PRB1,JPA} -- notably, 
the topological zero-modes of the Majorana chain display power-law behavior in a 
parameter regime known as the ``circle of oscillations". As shown in 
Part I, both types of exotic modes appear precisely when the transfer matrix of 
the model fails to be invertible.  The generalized Bloch theorem may be thought of 
as bestowing {\it exact solvability} in the same sense as the algebraic 
Bethe ansatz does: the linear-algebraic task of diagonalizing the single-particle
Hamiltonian is mapped to one of solving a small (independent of 
the number of sites) system of {\em polynomial} equations. While 
in general, if the polynomial degree is higher than four, the roots 
must be found numerically, whenever this polynomial system can be solved 
analytically, one has managed to solve the original linear-algebraic 
problem analytically as well. In fact, fully analytical solutions
are less rare than one might think, and either emerge in special parameter regimes, 
or by suitably adjusting BCs.  

In this paper, Part II, we extend the scope of our generalized Bloch 
theorem even further, with a twofold goal in mind. First, while in Part I 
we presented the basic framework for calculating energy eigenstates 
of fermionic $D$-dimensional lattice systems with surfaces, for simplicity 
we restricted to a setting where the total system Hamiltonian retains 
translation invariance along $D-1$ directions parallel to the surfaces. 
In more realistic situations in surface physics, however, this assumption is 
invalidated by various factors, including surface reconstruction and 
surface disorder. Establishing procedures for exact 
diagonalization of \(D\)-dimensional clean systems subject to 
arbitrary BCs (surface disorder included) on two parallel hyperplanes 
is thus an important necessary step. We accomplish this in 
Sec.\,\ref{theoryrecap}, by allowing for BCs to be adjusted in order to 
conveniently describe surface relaxation, reconstruction, or disorder in 
terms of an appropriate boundary matrix. 

As a second main theoretical extension, we proceed to show in Sec.\,\ref{interfaces} 
how to diagonalize ``multi-component'' systems that host 
hyperplanar interfaces separating clean bulks, that is, ``junctions". 
Surface and interface problems are conceptually related: BCs are but effective models of 
the interface between the system of interest and its ``complement'' or 
environment. While it is well appreciated that exotic many-body phenomena can take 
place at interfaces, there are essentially no known principles to guide 
interface engineering (see e.g. Ref.\,[\onlinecite{diez15}] for an
instructive case study). It is our hope that our characterization of interfaces 
in terms of {\it interface matrices} will shed some light on the problem of finding 
such guiding principles, at least within the mean-field approximation.  
As a concrete illustration, we include an exact 
calculation of the Andreev bound states in a simple model of a clean 
superconducting-normal-superconducting (SNS) junction, complementing 
the detailed numerical investigations reported in Ref. [\onlinecite{bena12}]. 

In addition to the SNS junction, we provide in Sec.\,\ref{high_dim} several 
explicit applications of our diagonalization procedures to computing surface 
band structures in systems ranging from insulating ladders to $p$- and 
and $s$-wave topological superconductors (TSCs) in $D=2$ lattices. 
The ladder model of domain-wall fermions introduced by Creutz 
\cite{Creutz,CreutzPRD,CreutzRMP} 
serves as a bridge between one to higher dimensions. 
For some values of the magnetic flux, the Creutz ladder can be classified as a 
topological insulator in class A and we find that it displays {\em topological power-law modes}. 
To the best of our knowledge, this is the first example of such power-law modes 
in a short-range insulator. In addition, we uncover a Gaussian duality mapping the Creutz 
ladder to a dual system consisting of two Majorana chains (see Ref.\,[\onlinecite{equivalence}]
for other examples of dualities bridging distinct classes in the mean-field topological
classification of electronic matter, and Ref. [\onlinecite{dualitygeneral}]  for the general approach 
to dualities). 

Moving to $D=2$ systems, we first consider graphene 
ribbons with two types of edges, ``zigzag-bearded'' and ``armchair'' (in the terminology of 
Ref.\,[\onlinecite{kohmoto07}]), in order to also provide an 
opportunity for direct comparison within our method and other analytical
calculations in the literature. As a more advanced application, we 
compute in closed form the surface band structure of the chiral 
\(p+ip\) TSC \cite{read00}. This problem is well under
control within the continuum approximation\cite{bernevig}, 
but not on the lattice. This distinction is important because the phase 
diagram of lattice models is richer than one would infer from the 
continuum approximation. As a final, technically harder example of a surface 
band-structure calculation, we investigate a two-band, gapless $s$-wave TSC that 
can host symmetry-protected Majorana flat bands and is distinguished by a 
non-unique, anomalous bulk-boundary correspondence 
\cite{swavePRL,swavePRB}.

We conclude in Sec.\,\ref{outlook} by iterating our key points and highlighting some 
key open questions. To ease the presentation, most technical details of our calculations 
are deferred to appendixes, including the analytic diagonalization 
of several paradigmatic $D=1$ models with boundaries. For reference, a summary of 
all the model systems we explicitly analyzed so far using the generalized Bloch 
theorem approach
is presented in Table \ref{MainTable}.  

\begin{table*}
\centering
\begin{tabular}{|l|c|c|c|c|} \hline
 {\bf $D=1$ and quasi-($D$=1) systems} & {\bf PC} & {\bf Boundary Conditions} & {\bf Some Key Results} & {\bf See} \\ \hline 
 \hline
 {\sf Single-band chain}  & yes & open/edge impurities & full diagonalization & Part I, Sec.\,V.A\\ \hline
 {\sf Anderson model}  & yes & open & full diagonalization & Part I, Sec.\,V.B\\ \hline
 {\sf Majorana Kitaev chain}  & no & open & full diagonalization & Refs.\,[\onlinecite{PRL,JPA}]; \\  
                                              &       &         & power-law Majorana modes & Part I, Sec.\,V.C \\ \hline                                     
 {\sf Two-band $s$-wave TSC}  & no & open/twisted &  \(4\pi\)-periodic supercurrent & Part I, Sec\,VI.B\\ 
  & & &  without parity switch & \\ \hline 
 {\sf BCS chain} & no & open & full diagonalization & App.\,\ref{appBCS}\\ \hline
 {\sf Su-Schrieffer-Heeger model} &  yes & reconstructed & full diagonalization & App.\,\ref{basic_examples} \\ \hline
 {\sf Rice-Mele model} &  yes &reconstructed & full diagonalization & App.\,\ref{basic_examples} \\ \hline 
 {\sf Aubry-Andr\'e-Harper model} &  yes & reconstructed & full diagonalization & App.\,\ref{basic_examples} \\ 
       (period-two) & & &  & \\ \hline 
 {\sf Creutz ladder}  & yes & open &  power-law topological modes& Sec.\,\ref{creutzladder}, App.\,\ref{creupendix}\\ \hline
 {\sf Majorana ladder} &  no & open & SC dual of Creutz ladder & Sec.\,\ref{majoranaladder}\\ \hline  
 {\sf SNS junction}      & no  & junction &    Andreev bound states         & Sec.\,\ref{interfaces}\\ \hline  \hline     
 {\bf $D=2$ systems} & &  & &  \\ \hline \hline 
 {\sf Graphene} (including   & yes & zigzag-bearded (ribbon) & full diagonalization & Sec.\,\ref{zbsec} \\ 
       modulated on-site potential)&  & armchair (ribbon)& full diagonalization & Sec.\,\ref{armpitsec}\\ \hline
 {\sf Harper-Hofstadter model}  & yes & open (ribbon) &  closed-form edge bands and states   & Ref.\,[\onlinecite{Qiaoru}] \\ \hline
 {\sf Chiral p+ip TSC} & no & open (ribbon) & closed-form edge bands and states & Sec.\,\ref{pwavetoponductor} \\
    				 &     &                        &  power-law surface modes & \\  \hline
 {\sf Two-band $s$-wave TSC}  & no & open/twisted & \(k_\parallel\)-resolved DOS  
 & Sec.\,\ref{2Dswavetoponductor} \\ 
 &  & &localization length  at zero energy &\\ 
 & & & enhanced $4\pi$-periodic supercurrent & \\
 \hline
\end{tabular}
\caption{
Summary of representative models analyzed in this work along with Part I (Ref. [\onlinecite{PRB1}]) and 
Ref. [\onlinecite{Qiaoru}], by using the generalized Bloch theorem approach.  
Some emerging key results are highlighted in the fourth column. 
PC: particle-conserving, DOS: density of states, SC: superconductor (or superconducting, 
depending on context).  Additional models that are amenable to solution by our approach include Majorana 
chains with twisted BCs \cite{Katsura17} or longer-range (e.g., next-nearest-neighbor) couplings 
\cite{Liu}, dimerized Kitaev chains \cite{Zhou}, period-three hopping models \cite{Kevin}, 
as well as time-reversal-invariant TSC
wires with spin-orbit coupling \cite{Aligia}, to name a few.
}
\label{MainTable}
\end{table*}

\section{Tailoring the generalized Bloch theorem to surface physics problems}
\label{theoryrecap}

As mentioned, the main aim of this section is to describe how the generalized 
Bloch theorem may be tailored to encompass BCs encountered in realistic 
surface-physics scenarios, which need not respect translation invariance along
directions parallel to the interface, as we assumed in Part I. Notwithstanding, 
the key point to note is that the bulk-boundary separation introduced
in Part I goes through {\em regardless} of the nature of the BCs. As a result, 
the bulk equation describing a clean system can always be decoupled by a partial 
Fourier transform into a family of ``virtual" chains parametrized
by the conserved component of crystal momentum \(\k_\parallel\).
{\em If} the BCs conserve \(\k_\parallel\), then they also reduce to BCs for 
each virtual chain. If they do {\em not}, then the BCs hybridize the generalized 
Bloch states associated to the individual virtual chains. In general, the boundary matrix
will then depend on all crystal momenta in the surface Brillouin zone.

\subsection{Open boundary conditions}

We consider a clean system of independent fermions embedded on a $D$-dimensional lattice 
with associated Bravais lattice $\Lambda_D$. Let $d_{\rm int}$ denote the number of fermionic 
degrees (e.g., the relevant orbital and spin degrees) enclosed by a primitive cell attached to each 
point of $\Lambda_D$. Now let us terminate this system along two parallel lattice hyperplanes, 
or {\it hypersurfaces} henceforth -- resulting in open (or ``hard-wall'') BCs. 

\begin{figure}[t]
\begin{center}
\hspace*{3mm}\includegraphics[width=8cm]{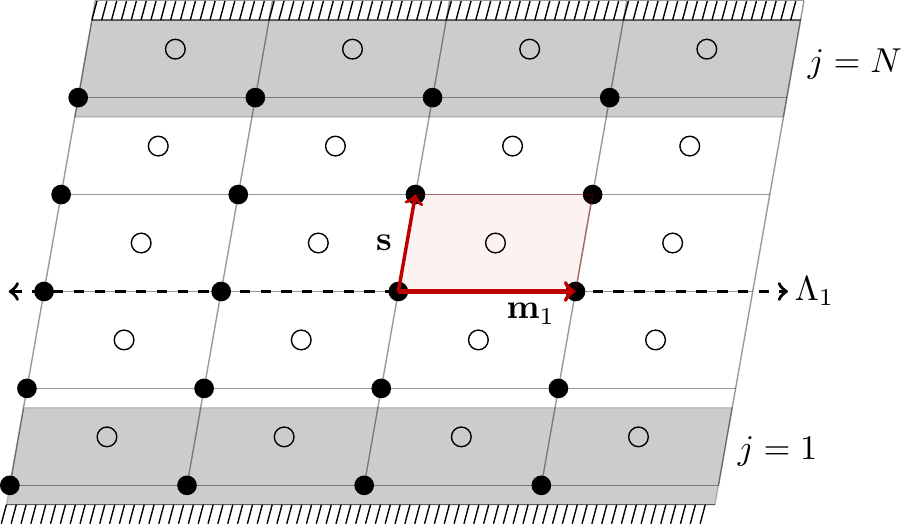}
\end{center}
\vspace*{-2mm}
\caption{
(Color online) Sketch of a $D=2$ lattice system with nearest-neighbor (NN)  hopping
subject to arbitrary BCs.
The filled and hollow circles together form a Bravais lattice $\Lambda_2$.
The two $D=1$ edges of the system are shown by horizontal lines decorated by pattern.
The surface mesh $\Lambda_1$ is generated by $\m_1$, and consists of all points
connected by dashed black lines. $\m_1$ and $\s$ generate the Bravais lattice $\bar{\Lambda}_2$,
formed only by the filled circles. A primitive cell of $\bar{\Lambda}_2$ (shaded brown)
encloses two points of $\Lambda_2$. In this case, assuming there are $d_{\rm int}$ 
internal degrees associated to each point of $\Lambda_2$, we get
$\bar{d}_{\rm int} = 2d_{\rm int}$. Since, for NN hopping, $R=1$, the operator $W$ that 
implements the BCs has its support on single-particle states in the boundary region 
(shaded gray). 
\label{fig_surface}}
\end{figure}

The terminated system is translation-invariant along $D-1$ lattice vectors parallel to the 
hypersurfaces, so that we can associate with it 
a Bravais lattice $\Lambda_{D-1}$ of spatial dimension $D-1$,
known as the {\it surface mesh} \cite{bechstedt}. If $\m_1,\dots,\m_{D-1}$ 
denote the primitive vectors of $\Lambda_{D-1}$,
then any point $\j_\parallel \in \Lambda_{D-1}$ can be expressed 
as $\j_\parallel = \sum_{\mu=1}^{D-1}j_\mu \m_\mu$, where $j_\mu$ are integers.
Let us choose a lattice vector $\s$ of $\Lambda_D$ that is not in the surface mesh 
(and therefore, not parallel to the two hypersurfaces). 
We will call $\s$ the {\it stacking vector}. Since $\{\m_1,\dots,\m_{D-1},\s\}$ are not 
the primitive vectors of $\Lambda_D$ in general, the Bravais lattice $\bar{\Lambda}_D$
generated by them may cover only a subset of points in $\Lambda_D$. 
Therefore, in general, each primitive cell of $\bar{\Lambda}_D$
may enclose a number $I>1$ of points of $\Lambda_D$. 
As a result, there are a total of $\bar{d}_{\rm int} =Id_{\rm int}$ fermionic degrees of freedom
attached to each point $\j_\parallel + j\s$ of $\bar{\Lambda}_D$ with $j$ an integer (see Fig. \ref{fig_surface}).
Let us denote the corresponding creation (annihilation) 
operators by $c^\dagger_{\j_\parallel j 1}, \dots, c^\dagger_{\j_\parallel j \bar{d}_{\rm int}}$ 
($c_{\j_\parallel j 1},\dots,c_{\j_\parallel j \bar{d}_{\rm int}}$). For each $\j_\parallel$ in the surface 
mesh, we define the array of the basis of fermionic operators by
\[
\hat{\Phi}^\dagger_{\j_\parallel} \equiv 
\begin{bmatrix}\hat{\Phi}^\dagger_{\j_\parallel,1} & \cdots & 
\hat{\Phi}^\dagger_{\j_\parallel,N}\end{bmatrix},
\;\;
 \hat{\Phi}^\dagger_{\j_\parallel,j} \equiv 
\begin{bmatrix}c^\dagger_{\j_\parallel j1} & \cdots & 
c^\dagger_{\j_\parallel j\bar{d}_{\rm int}}\end{bmatrix},
\]
where the integer $N$ is proportional to the separation between the two hypersurfaces. 
For  arrays, such as $\hat{\Phi}^{\dagger}_{\j_\parallel}$ and 
$\hat{\Phi}_{\j_\parallel}^{\;}$, 
we shall follow the convention that the arrays appearing on the left (right) 
of a matrix are row (column) arrays. 
In the above basis, the many-body Hamiltonian of the system, subject to
open BCs on the hypersurfaces, can be expressed as \cite{PRB1} 
\begin{eqnarray*}
\widehat{H}_N = \hspace{-3mm}
\sum_{\j_\parallel,\r_\parallel \in \Lambda_{D-1}} \hspace{-3mm}
\Big[\hat{\Phi}^\dagger_{\j_\parallel}K_{\r_\parallel}\hat{\Phi}^{\;}_{\j_\parallel+\r_\parallel}
+\frac{1}{2}(\hat{\Phi}^\dagger_{\j_\parallel}\Delta_{\r_\parallel}
\hat{\Phi}_{\j_\parallel+\r_\parallel}^\dagger +\text{H.c.})\Big], 
\label{Hamtransinv} 
\end{eqnarray*}
where $\j_\parallel, \r_\parallel$ are vectors in the surface mesh, and
$K_{\r_\parallel}$, $\Delta_{\r_\parallel}$ are $N\bar{d}_{\rm int}\times N\bar{d}_{\rm int}$ 
hopping and pairing matrices that satisfy
\( K_{-\r_\parallel}=K_{\r_\parallel}^\dagger$, 
$\Delta_{-\r_\parallel}=-\Delta_{\r_\parallel}^{\rm T} \)
by virtue of fermionic statistics, with the superscript ${\rm T}$ denoting 
the transpose operation.  Thanks to the assumptions of clean, finite-range system, 
these are {\em banded block-Toeplitz matrices} \cite{JPA}: explicitly, if $R \geq 1$ is 
the range of hopping and pairing, we may write 
$[S_{\r_\parallel}]_{jj'} \equiv S_{\r_\parallel,j'-j} \equiv S_{\r_\parallel,r}$, 
with 
\[ S_{\r_\parallel,r} =0 \quad \text{if} \quad |r|>R, \;\; \forall \r_\parallel,\quad \mbox{where } S=K,\Delta . \] 


Next, we enforce periodic BCs along the directions $\m_1,\dots,\m_{D-1}$ in which translation 
invariance is retained, by restricting to 
those lattice points $\j_\parallel = \sum_{\mu=1}^{D-1} j_\mu \m_\mu$ where  
for each $\mu$, $j_\mu$ takes values from $\{1,\dots,N_\mu\}$,
$N_\mu$ being a positive integer.
Let $\n_1,\dots,\n_{D-1}$ denote the primitive vectors of the surface reciprocal 
lattice, which is the $(D-1)$-dimensional lattice reciprocal to the surface mesh $\Lambda_{D-1}$,
satisfying $\m_\mu \cdot \n_\nu = 2\pi \delta_{\mu\nu}$ for $\mu,\nu=1,\dots,D-1$.
The Wigner-Seitz cell of the surface reciprocal lattice is the {\it surface Brillouin zone},
denoted by SBZ. In the Fourier-transformed basis defined by
\begin{equation}
\label{phikperp}
\hat{\Phi}_{\k_\parallel}^\dagger \equiv  
\sum_{\j_\parallel}^{\Lambda_{D-1}} \frac{e^{i\k_\parallel\cdot \j_\parallel}}{\sqrt{N_S}}\hat{\Phi}_{\j_\parallel}^\dagger,
\quad N_S = N_1\dots N_{D-1},
\end{equation}
where $\k_\parallel = \sum_{\mu=1}^{D-1} \frac{k_\mu}{N_\mu}\n_\mu$ and the integers
$k_\mu$ are crystal momenta in the SBZ, we can then express the relevant many-body Hamiltonian 
in terms of ``virtual wires'' labeled by $\k_\parallel$. 
That is, 
\begin{eqnarray}
\widehat{H}_N &\equiv&  \sum_{\k_\parallel \in \text{SBZ}}
\widehat{H}_{\k_\parallel,N} ,\quad \text{where}  \label{HamOBC} \\
\widehat{H}_{\k_\parallel,N} &=& \frac{1}{2}(\hat{\Phi}_{\k_\parallel}^\dagger K_{\k_\parallel}\hat{\Phi}^{\;}_{\k_\parallel}
- \hat{\Phi}^{\;}_{-\k_\parallel} K_{-\k_\parallel}^{*} \hat{\Phi}_{-\k_\parallel}^\dagger \notag \\
&+& \hat{\Phi}_{\k_\parallel}^\dagger \Delta_{\k_\parallel}\hat{\Phi}_{-\k_\parallel}^\dagger
-  \hat{\Phi}_{-\k_\parallel}\Delta_{-\k_\parallel}^{*} \hat{\Phi}_{\k_\parallel}) + \frac{1}{2}\text{Tr } K_{\k_\parallel}. 
\notag
\end{eqnarray}
Here, Tr denotes trace and the $N\bar{d}_{\rm int} \times N\bar{d}_{\rm int}$ matrices $S_{\k_\parallel}$,  
for $S = K,\Delta$, have entries
\[[S_{\k_\parallel}]_{jj'} \equiv S_{\k_\parallel,j'-j} \equiv S_{\k_\parallel,r}\equiv
\sum_{\r_\parallel} e^{i\k_\parallel\cdot \r_\parallel}S_{\r_\parallel,r},   \]
and the finite-range assumption requires that 
\begin{equation}
\label{Range}
S_{\k_\parallel,r} =0 \ \,  \text{if} \  |r|>R,  \;\; \forall \k_\parallel \in \text{SBZ},\ \ \mbox{where } S=K,\Delta .
\end{equation}

\subsection{Arbitrary boundary conditions}
\label{sub:abc}

\begin{figure}
\includegraphics[width=8cm]{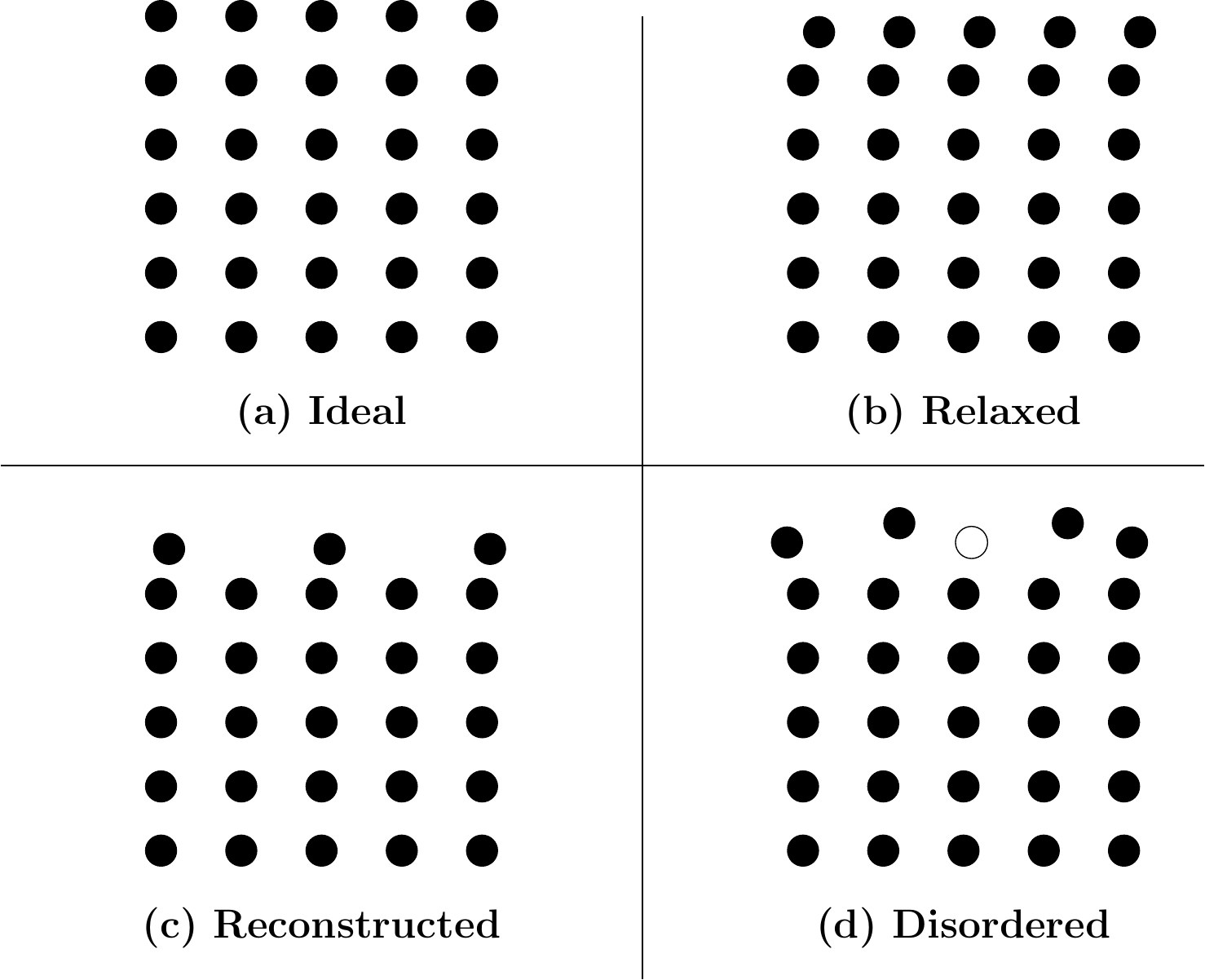}
\caption{(a) Sketch of a $D=2$ crystal with ideal surface. The remaining panels 
show the same crystal with (b) relaxed, (c) reconstructed, and (d) disordered surface. 
The unfilled circle in panel (d) shows a surface impurity atom.
\label{fig_reconst}}
\end{figure}

Physically, non-ideal surfaces may result from processes such as 
surface relaxation or reconstruction, as well as from the presence of 
surface disorder (see Fig. \ref{fig_reconst}). In our setting, these may be 
described as effective BCs, modeled by a Hermitian operator of the form 
\[
\widehat{W} \equiv \sum_{\j_\parallel,\j'_\parallel} \Big[
\hat{\Phi}^\dagger_{\j_\parallel}W^{(K)}_{\j_\parallel,\j'_\parallel}\hat{\Phi}^{\;}_{\j_\parallel}
+\frac{1}{2}(
\hat{\Phi}^\dagger_{\j_\parallel}W^{(\Delta)}_{\j_\parallel,\j'_\parallel}\hat{\Phi}^{\dagger}_{\j_\parallel'}
+\text{H.c.})\Big], \]
subject to the constraints from fermionic statistics, 
\begin{eqnarray*}
W^{(K)}_{\j'_\parallel,\j_\parallel}=\big[W^{(K)}_{\j_\parallel,\j'_\parallel}\big]^\dagger,\quad 
W^{(\Delta)}_{\j'_\parallel,\j_\parallel}=-\big[W^{(\Delta)}_{\j_\parallel,\j'_\parallel}\big]^{\rm T}. 
\end{eqnarray*}
Since such non-idealities at the surface are known to influence 
only the first few atomic layers near the surfaces, we assume that $\widehat{W}$ 
affects only the first $R$ boundary slabs of the lattice, so that (see also Fig. \ref{fig_surface})
\[ \big[W^{(S)}_{\j_\parallel,\j'_\parallel}\big]_{jj'} = 0 \quad \forall \j_\parallel,\j'_\parallel,\quad  S=K,\Delta, \]
if $j$ or $j'$ take values in $\{R+1,\dots,N-R\}$. 

The total Hamiltonian subject  to arbitrary BCs is 
\[ \widehat{H} \equiv \widehat{H}_N+\widehat{W}. \]
Let $j\equiv b =1, \ldots, R; N-R+1, \ldots, N$ label boundary lattice sites. 
While in Part I we also assumed $\widehat{W}$ to be periodic along $\m_1,\dots,\m_{D-1}$ 
[case (a) in Fig. \ref{fig_reconst}], in general only $\widehat{H}_N$ will be able to be 
decoupled by Fourier-transform, whereas $\widehat{W}$ will retains cross-terms of the form 
\begin{eqnarray*}
&&[W_{\q_\parallel,\k_\parallel}^{(S)}]_{bb'} = \sum_{\j_\parallel,\j'_\parallel}
e^{i(\k_\parallel\cdot \j'_\parallel-\q_\parallel\cdot\j_\parallel)}\big[W_{\j_\parallel,\j'_\parallel}^{(S)}\big]_{bb'},
\quad S=K,\Delta.
\end{eqnarray*}
If the system is not particle-conserving, let us reorder the 
fermionic operator basis according to \cite{PRB1}
\begin{eqnarray*}
\hat{\Psi}_{\k_\parallel}^\dagger \equiv \begin{bmatrix}
\hat{\Psi}_{\k_\parallel,1}^\dagger  & \cdots & \hat{\Psi}_{\k_\parallel,N}^\dagger
\end{bmatrix},\quad
\hat{\Psi}_{\k_\parallel,j}^\dagger \equiv  \begin{bmatrix}\hat{\Phi}_{\k_\parallel,j}^\dagger &
\hat{\Phi}^{\;}_{-\k_\parallel,j} 
\end{bmatrix} .
\end{eqnarray*} 
The single-particle Hamiltonian can then be expressed as
\begin{align}
& H=H_N+W= \label{spHam} \\
&=\sum_{\k_\parallel}|\k_\parallel\rangle\langle \k_\parallel|\otimes H_{\k_\parallel,N}
+\sum_{\q_\parallel,\k_\parallel}|\q_\parallel\rangle\langle \k_\parallel|\otimes W_{\q_\parallel,\k_\parallel},
\notag 
\end{align}
where $H_{\k_\parallel,N}$ is the single-particle (BdG) Hamiltonian corresponding to 
Eq. \eqref{HamOBC}. In terms of the shift matrix
$T\equiv \sum_{j=1}^{N-1}|j\rangle\langle j+1|$
implementing a shift along the direction $\s$, and letting $r=j'-j$ as before, we have 
\begin{eqnarray}
&&H_{\k_\parallel,N}=\mathds{1}_N\otimes h_{\k_\parallel,0}+
\sum_{r=1}^R \, [T^r\otimes h_{\k_\parallel,r}+\text{H.c.}],
\label{spHkN} \\
&&h_{\k_\parallel,r} = \sum_{\r_\parallel}e^{i \k_\parallel\cdot\r_\parallel }h_{\r_\parallel,r},\quad
h_{\r_\parallel,r} =
\begin{bmatrix} 
K_{\r_\parallel,r} & \Delta_{\r_\parallel,r} \\ -\Delta_{\r_\parallel,r}^* & -K_{\r_\parallel,r}^*
\end{bmatrix}, \notag
\end{eqnarray}
whereas the single-particle boundary modification $W_{\q_\parallel,\k_\parallel}$ in Eq. \eqref{spHam} is given by 
\begin{eqnarray*}
W_{\q_\parallel,\k_\parallel} &=& \begin{bmatrix} 
W^{(K)}_{\q_\parallel,\k_\parallel} & W^{(\Delta)}_{\q_\parallel,\k_\parallel} \\
-[{W^{(\Delta)}_{-\q_\parallel,-\k_\parallel}}]^* & -[{W^{(K)}_{-\q_\parallel,-\k_\parallel}}]^*
\end{bmatrix} .
\end{eqnarray*}
In the simpler case where the system is particle-conserving, then
$h_{\r_\parallel,r} = K_{\r_\parallel,r}$ and $W_{\q_\parallel,\k_\parallel} = 
W^{(K)}_{\q_\parallel,\k_\parallel}$.

Reflecting the different ways in which a surface may deviate from its ideal structure (Fig. \ref{fig_reconst}), 
we may consider BCs as belonging to three different categories of increasing complexity:

\begin{itemize}
\item {\it Relaxed BCs---}
In the process of surface relaxation, the atoms in the surface slab displace from their ideal position
in such a way that the surface (and the bulk) layers remain translation invariant
along $\m_1,\dots,\m_{D-1}$, as assumed in Part I. Therefore, $\k_\parallel$ remains a good 
quantum number, and
$W_{\q_\parallel,\k_\parallel}=\delta_{\q_\parallel,\k_\parallel}W_{\k_\parallel,\k_\parallel}$. 
In particular, \(W_{\q_\parallel,\k_\parallel}=0\) for each $\q_\parallel,\k_\parallel$ for open BCs, 
which falls in this category.
 
\item {\it Reconstructed BCs---}
If the surfaces undergo reconstruction, then the total system can have lower periodicity 
than the one with ideal surfaces. This scenario is also referred to 
as {\it commensurate} surface reconstruction \cite{bechstedt}. In this case, $W$ may retain some 
cross-terms of the form $W_{\q_\parallel,\k_\parallel}$.
However, not all values $\k_\parallel$ are expected to have cross-terms in this way, and the system can still 
be block-diagonalized. For example, for $2\times 1$ reconstruction of the (111) surface of Silicon crystals, 
each block of the Hamiltonian will consist of only $2\times 1=2$ values of $\k_\parallel$, whereas
for its $7\times7$ reconstruction, each block includes $49$ values of $\k_\parallel$ \cite{bechstedt}.

\item {\it Disordered BCs---}
If the surface reconstruction is {\it non-commensurate}, or if the 
surface suffers from disorder, then the Hamiltonian cannot be block-diagonalized 
any further in general. Non-commensurate reconstruction of a surface is likely to 
happen in the case of adsorption. 
\end{itemize}

Our setting is general enough to model adsorption as well as thin layer deposition
up to a few atomic layers. In the following, unless otherwise stated, we will assume that the 
system is subject to the most general type of disordered BCs.

\subsection{Generalized Bloch theorem}
\label{sub:gbt}

The first needed ingredient toward formulating the generalized Bloch theorem is a 
description of the eigenstates of the single-particle Hamiltonian $H_{\k_\parallel, N}$ 
of the virtual wire labeled by $\k_\parallel$, given in Eq. \eqref{spHkN}. Let 
\[ d\equiv \left\{ \begin{array}{lcl}
\bar{d}_{\rm int} & \text{if} & \Delta=0 =W^{(\Delta)} ,\\
2\bar{d}_{\rm int} & \text{if} & \Delta \ne 0 \; \text{or} \; W^{(\Delta)} \ne 0 .
\end{array}\right. \] 
Then, the projector
\begin{eqnarray*}
P_B =\bm{1} \otimes \sum_{j=R+1}^{N-R}|j\rangle\langle j|\otimes \mathds{1}_d ,
\end{eqnarray*}
determined by the range \(R\) of the virtual chains is the 
{\it bulk projector}, where we have used the completeness relation 
$\bm{1}=\sum_{\k_\parallel \in \text{SBZ}}|\k_\parallel\rangle\langle \k_\parallel|$. 
By definition, the matrix $W$ describing BCs  satisfies  \( P_B W=0 \), whereby it 
follows that \(P_{B}H=P_B(H_N+W)=P_B H_N\). Accordingly, building on the exact 
bulk-boundary separation also used in Part I, the {\em bulk equation} to be solved reads
\begin{eqnarray}
\label{bulkeq}
P_{B}H_{N}|\psi\rangle=\epsilon P_{B}|\psi\rangle , \quad \epsilon \in {\mathbb R}. 
\end{eqnarray}

To proceed, we need to introduce some auxiliary matrices and states. First and foremost there is the 
\(d\times d\) analytic continuation of the Bloch Hamiltonian \cite{PRL}, which now takes the form 
\begin{equation} 
H_{\k_\parallel}(z)  \equiv h_0+\sum_{r=1}^R \, (z^rh_{\k_\parallel,r}+z^{-r}h_{\k_\parallel,r}^\dagger),
\quad z\in {\mathbb C}, 
\label{HBloch}
\end{equation} 
acting on a $d$-dimensional internal space spanned by 
states $\{|m\rangle,\ m=1,\dots,d\}$. 
If the matrix \(h_{\k_\parallel,R}\) is {\em invertible}, 
then \(H_{\k_\parallel}(z)\) is essentially everything one needs to proceed. 
Otherwise, the related matrix polynomial
\begin{eqnarray}
K_{\k_\parallel}^-(\epsilon,z) \equiv z^R(H_{\k_\parallel}(z)-\epsilon\mathds{1}_d)
\end{eqnarray}
is of considerable importance. We will also 
need the \(dv\times dv\) generalized Bloch Hamiltonians with block entries
\begin{align}
\label{genhb}
&[H_{\k_\parallel,v}(z)]_{xx'} \equiv \\
&\frac{\partial_z^{x'-x}H_{\k_\parallel}(z)}{(x'-x)!}= 
\frac{H_{\k_\parallel}^{(x'-x)}(z)}{(x'-x)!},\quad 1\le x\le x'\le v,\nonumber
\end{align}
with \(H_{\k_\parallel}^{(0)}(z)=H_{\k_\parallel}(z)\) given in Eq. \eqref{HBloch}.
In array form,
\[
H_{\k_\parallel,v}(z)=
\begin{bmatrix}
\ \ H^{(0)} & H^{(1)}     & \frac{1}{2} H^{(2)}  & \cdots       & \frac{1}{(v-1)!}H^{(v-1)}          \\
 0          &\! \ddots    &\! \! \ddots          &\! \! \ddots  & \vdots             \\
\vdots      &\! \ddots    &\! \ddots             &\! \! \ddots  & \frac{1}{2}H^{(2)} \\
\vdots      &             &\! \ddots             &\! \ddots     & H^{(1)}            \\
0           &  \cdots     & \cdots              &0             & H^{(0)}            
\end{bmatrix},
\]
where the label $(z)$ and the subscript $\k_\parallel$ were dropped for brevity.
The \(dv\times dv\) block matrix \(K_{\k_\parallel,v}^-(\epsilon,z)\) is 
defined by the same formula. The important difference between these 
two matrices is that \(K_{\k_\parallel,v}^-(\epsilon,z)\) is well defined 
at \(z=0\), whereas \(H_{\k_\parallel,v}(z)\) is not. These block matrices 
act on column arrays of \(v\) internal states, which can be expressed 
in the form
\( |u\rangle=
\begin{bmatrix}
|u_1\rangle& \dots& |u_{v}\rangle
\end{bmatrix}^{\rm T},  \)
where each of the entries is an internal state. 

For fixed but arbitrary \(\epsilon\), the expression 
\begin{eqnarray}
\label{thepoly}
P_{\k_\parallel}(\epsilon,z)\equiv \det K_{\k_\parallel}^-(\epsilon,z)
\end{eqnarray}
defines a family of polynomials in \(z\). We call a given 
value of \(\epsilon\) {\it singular} \cite{JPA} if \( P_{\k_\parallel}(\epsilon,z)\)
vanishes identically for all $z$ for some value of \(\k_\parallel\). 
Otherwise, \(\epsilon\) is {\it regular}. At a singular value
of the energy, \(z\) becomes independent of \(\epsilon\)
for some \(\k_\parallel\). Physically, singular energies correspond 
to {\em flat bands}, at fixed \(\k_\parallel\). As explained in Part I, 
flat bands are not covered by the generalized Bloch theorem and 
require separate treatment \cite{FBRemark}. In the following, we will 
concentrate on the {\em generic case where \(\epsilon\) is regular}.  

For regular energies, \(P_{\k_\parallel}(\epsilon,z)\) can be factorized 
in terms of its {\em distinct} roots as
\[ P_{\k_\parallel}(\epsilon,z)=c\prod_{\ell=0}^n(z-z_\ell)^{s_\ell}, 
\quad c\in {\mathbb C},  \]
with \(c\) a non-vanishing constant and \(z_0=0\) by convention. 
If zero is not a root, then \(s_0=0\). The \(z_\ell,\ \ell=1,\dots,s_\ell\), 
are the distinct non-zero roots of multiplicity \(s_\ell\geq 1\). 
It was shown in Ref.\,[\onlinecite{JPA}] that the number of solutions 
of the kernel equation
\begin{eqnarray}
\label{amps}
(H_{\k_\parallel,s_\ell}(z_\ell)-\epsilon\mathds{1}_{ds_\ell})|u\rangle=0
\end{eqnarray}
coincides with the multiplicity \(s_\ell\) of  \(z_\ell\).
We will denote a complete set of independent solutions of 
Eq.\,\eqref{amps} by $|u_{\ell s}\rangle,$ $s=1,\dots,s_\ell$,
where each $|u_{\ell s}\rangle$ has $d \times 1$ block-entries
\[
|u_{\ell s}\rangle = \begin{bmatrix}|u_{\ell s1}\rangle & \dots & |u_{\ell s s_\ell}\rangle\end{bmatrix}^{\rm T}.
\] 
Moreover, if we define
\[ K_{\k_\parallel}^-(\epsilon)\equiv 
K_{\k_\parallel,s_0}^-(\epsilon,z_0=0)\equiv 
K_{\k_\parallel}^+(\epsilon)^\dagger, \]
then it is also the case that the kernel equations 
\begin{eqnarray*}
K_{\k_\parallel}^-(\epsilon)|u\rangle=0,\quad K_{\k_\parallel}^+(\epsilon)|u\rangle=0
\end{eqnarray*}
have each \(s_0\) solutions. We will denote a basis of solutions 
of these kernel equations by \(|u^\pm_s\rangle,\) \(s=1,\dots,s_0\),
each with block entries
\[ |u^\pm_{s}\rangle = \begin{bmatrix}|u^\pm_{s1}\rangle & \dots & |u^\pm_{s s_0}\rangle\end{bmatrix}^{\rm T}.
\] 

In order to make the connection to the lattice degrees of freedom, 
let us introduce the lattice states 
\begin{eqnarray}
\label{zstate}
\!\!|z,v\rangle\equiv\!
\sum_{j=1}^N \frac{j^{(v-1)}}{(v-1)!}z^{j-v+1}|j\rangle\!=\!
\frac{1}{(v-1)!}\partial_z^{v-1}|z,1\rangle, \quad 
\end{eqnarray}
with \(j^{(0)}=1\) and \(j^{(v)}=(j-v+1)(j-v+2)\dots j\) for \(v\) a positive integer. The states
\begin{align}
|\k_\parallel\rangle|\psi_{\k_\parallel \ell s}\rangle
\equiv & \sum_{v=1}^{s_\ell}|\k_\parallel\rangle|z_\ell,v\rangle|u_{\ell s v}\rangle, 
\quad s=1,\dots, s_\ell,\nonumber\\
|\k_\parallel\rangle|\psi^-_{\k_\parallel s}\rangle\equiv &
\sum_{j=1}^{s_0}|\k_\parallel\rangle|j\rangle|u_{sj}^-\rangle,\quad s=1,\dots,s_0,\nonumber\\
|\k_\parallel\rangle|\psi^+_{\k_\parallel s}\rangle\equiv 
&\sum_{j=1}^{s_0}|\k_\parallel\rangle|N-j+s_0\rangle|u_{sj}^+\rangle,
\quad s=1,\dots,s_0, 
\label{asin}
\end{align}
form a complete set of independent solutions of the {bulk equation}, Eq. \eqref{bulkeq}. 
Intuitively speaking, these states are eigenstates of the Hamiltonian
``up to BCs''. For regular energies as we assumed, there are 
exactly \(2Rd=2s_0+\sum_{\ell=1}^ns_\ell\) solutions of the bulk 
equation for each value of \(\k_\parallel\) \cite{JPA,PRB1}.
The solutions associated to the non-zero roots are {\it extended 
bulk solutions}, and the ones associated to \(z_0=0\) are
{\it emergent}. Emergent bulk solutions are perfectly localized around 
the edges of the system in the direction perpendicular to the hypersurfaces.

It is convenient to obtain a more uniform description of solutions 
of the bulk equation by letting 
\begin{eqnarray}
\label{asinto}
\!\! |\psi_{\k_\parallel \ell s}\rangle=
\left\{
\begin{array}{lcl}
|\psi_{\k_\parallel s}^-\rangle &\mbox{if} & \ell=0;\ s=1,\dots,s_0,\\
|\psi_{\k_\parallel \ell s}\rangle &\mbox{if}& \ell=1,\dots,n;\ s=1,\dots,s_\ell,\\
|\psi_{\k_\parallel s}^+\rangle &\mbox{if} & \ell=n+1;\ s=1,\dots,s_0,
\end{array}\right. \quad
\end{eqnarray}
Also, let \(s_{n+1}\equiv s_0\). Then, the {\it ansatz} 
\[ |\epsilon,\bm{\alpha}\rangle 
\equiv \sum_{\k_\parallel \in \text{SBZ}}\sum_{\ell=0}^{n+1}\sum_{s=1}^{s_\ell}
\alpha_{\k_\parallel\ell s}|\k_\parallel\rangle|\psi_{\k_\parallel\ell s}\rangle
\] 
describes the most general solution of the bulk equation 
in terms of \(2Rd\) amplitudes \(\bm{\alpha}\) for each 
value of \(\k_\parallel\). We call it an ansatz because the states 
\(|\epsilon,\bm{\alpha}\rangle\) provide the appropriate search space for 
determining the energy eigenstate of the full Hamiltonian \(H=H_N+W\). 

As a direct by-product of the above analysis, it is interesting to 
note that a {\em necessary} condition for $H$ to admit an eigenstate of 
exponential behavior localized on the left (right) edge is that some of the roots $\{z_\ell\}$ 
of the equation $\det K^-_{\k_\parallel}(\epsilon,z)=0$ be inside (outside)
the unit circle. Therefore, one simply needs to compute all roots of  
$\det K^-_{\k_\parallel}(\epsilon,z)$ to know whether localized edge states 
may exist in principle. 

We are finally in a position to impose arbitrary BCs. As before, let \(b=1,\dots,R; N-R+1,\dots,N\)
be a variable for the boundary sites. Then the {\it boundary matrix} \cite{PRL,JPA,PRB1} 
is the block matrix 
\begin{align*}
&[B(\epsilon)]_{\q_\parallel b,\k_\parallel\ell s}=\\
&=\delta_{\q_\parallel,\k_\parallel}\langle b|(H_{\k_\parallel,N}-\epsilon\mathds{1}_{dN}|\psi_{\k_\parallel\ell s}\rangle
+\langle b|W(\q_\parallel,\k_\parallel)|\psi_{\ell s}\rangle ,
\end{align*}
with non-square \(d\times 1\) blocks (one block per boundary site
\(b\) and crystal momentum \(\k_\parallel\)). By construction,
\begin{align*}
(H-\epsilon\mathds{1})|\epsilon,\bm{\alpha}\rangle=
\sum_{\q_\parallel,b}\sum_{\k_\parallel,\ell,s}
|\q_\parallel\rangle|b\rangle [B(\epsilon)]_{\q_\parallel b,\k_\parallel \ell s}\alpha_{\k_\parallel\ell s}, 
\end{align*}
for {\it any} regular value of \(\epsilon\in \mathds{C}\). Hence, an 
ansatz state represents an energy eigenstate if and only if
\[
\sum_{\k_\parallel,\ell,s}[B(\epsilon)]_{\q_\parallel b,\k_\parallel \ell s}\alpha_{\k_\parallel \ell s}=0
\quad \forall \,\q_\parallel, b ,
\]
for all boundary sites \(b\) and crystal momenta \(\q_\parallel\), or, 
more compactly, \(B(\epsilon)\bm{\alpha}=0\). We are finally in a 
position to state our generalized Bloch theorem for clean systems
subject to arbitrary BCs on two parallel hyperplanes, and extending 
Theorem 3 in Part I: 

\medskip

{\it Theorem (Generalized Bloch theorem).} 
Let $H=H_N+W$ denote a single-particle Hamiltonian 
as specified above [Eq. \eqref{spHam}], 
for a slab of thickness \(N>2Rd\). Let \(B(\epsilon)\) be
the associated boundary matrix. If \(\epsilon\) is an eigenvalue 
of \(H\) and a regular energy of \(H(z)\), the corresponding 
eigenstates of \(H\) are of the form
\[ |\epsilon,\bm{\alpha}_\kappa\rangle 
= \sum_{\k_\parallel}\sum_{\ell=0}^{n+1}\sum_{s=1}^{s_\ell}
\alpha_{\k_\parallel \ell s}^{(\kappa)} \, |\k_\parallel\rangle|\psi_{\k_\parallel\ell s}\rangle,\quad
\kappa=1,\dots,\mathcal{K}, \]
where the amplitudes 
\( \bm{\alpha}_\kappa\) are determined as a complete set of 
independent solutions of the kernel equation 
\( B(\epsilon)\bm{\alpha}_{\kappa}=0\), 
and 
the degeneracy \(\mathcal{K}\) of the energy level \(\epsilon\)
coincides with the dimension of the kernel of the boundary matrix,
\(\mathcal{K}=\dim {\rm Ker\,}B(\epsilon)\). 

\medskip  

In the above statement, the lower bound \(N>2dR\) on the 
thickness of the lattice is imposed in order to ensure that the emergent 
solutions on opposite edges of the system have
zero overlap and are thus necessarily independent. It can be 
weakened to \(N>2R\) in the generic case where \(\det h_{\k_\parallel,R}\neq 0\), because 
in this case \(s_0=0\) and there are no emergent solutions. 

Based on the generalized Bloch theorem, an algorithm for numerical
computation of the electronic structure was given in Part I, directly applicable to the 
case of relaxed BCs. In particular, it was shown that the complexity of the algorithm 
is independent of the size $N$ of each virtual wire. In the most general case of disordered 
BCs we consider here, however, since the boundary matrix can have 
cross-terms between the virtual wires, we correspondingly have to deal with a single 
(non-decoupled) boundary matrix of size $2RdN^{D-1} \times 2RdN^{D-1}$. Finding 
the kernel of this boundary matrix has time complexity $\mathcal{O}(N^{3D-3})$, which will
be reflected in the performance of the overall algorithm.

The generalized Bloch theorem relies on the complete solution of the bulk equation, given 
in Eq. (\ref{bulkeq}). 
Since the latter describes an unconventional {\em relative} eigenvalue problem for 
the (generally) {\em non-Hermitian} operator \(P_B H_N\), the standard symmetry analysis 
of quantum mechanics does not immediately apply. 
It is nonetheless possible to decompose the solution spaces of the bulk equation 
into symmetry sectors, if the Hamiltonian
obeys unitary symmetries that also commute with the bulk projector $P_B$. 
Assume that a unitary operator $\mathcal{S}$ commutes with {\em both} $H=H_N+W$ and 
$P_B$. Then any vector in the bulk solution space satisfies
\[
P_B(H_N+W-\epsilon\mathds{1})|\psi\rangle = 0 \Rightarrow 
\mathcal{S}^\dagger P_B(H_N+W-\epsilon\mathds{1})\mathcal{S}|\psi\rangle = 0 .
\]
This implies that the bulk solution space is invariant under the action of $\mathcal{S}$.
Therefore, there exists a basis of the bulk solution space in which the action of $\mathcal{S}$
is block-diagonal. This leads to multiple eigenstate ans\"{a}tze, each labeled by an eigenvalue
of $\mathcal{S}$. Further, $\mathcal{S}^\dagger P_B \mathcal{S}=0$ implies
that the boundary subspace (i.e., the kernel of $P_B$) is also invariant under $\mathcal{S}$.
After finding a basis of the boundary subspace in which $\mathcal{S}$ is block-diagonal,
the boundary matrix itself splits into several matrices, each labeled by an eigenvalue 
of $\mathcal{S}$. We will use this strategy in some of the applications in 
Sec.\,\ref{interfaces} and Sec.\,\ref{high_dim}. We also discuss in Appendix 
\ref{app:condition} how symmetry conditions can help identifying a criterion 
for the absence of  localized edge modes, which may be of independent
interest.

\section{Interface physics problems}
\label{interfaces}

\subsection{Multi-component generalized Bloch theorem}

As mentioned, a second extension of our theoretical framework 
addresses the exact diagonalization of systems with internal boundaries, 
namely, interfaces between distinct bulks. In the spirit of keeping technicalities 
to a minimum, we focus on the simplest setting whereby two bulks with identical
reduced Brillouin zones are separated by one interface.
The extension to multi-component systems is straightforward, and can be 
pursued as needed by mimicking the procedure to be developed next.
 
Since the lattice vectors for the two bulks forming the interface are the same, the primitive 
vectors of the surface mesh $\{\m_\mu,\ \mu=1,\dots,D-1\}$, the stacking vector $\s$,
and the basis $\{\d_{\bar \nu},\ {\bar \nu}=1,\dots,I-1\}$ are shared by both bulks.
Let us further assume that the latter are described by systems that are half-infinite in the 
directions $-\s$ and $\s$, respectively.
The bulk of system number one (left, $i=1$) occupies sites \(\{\j =  \j_\parallel +j\s +\d_{\bar \nu},\ 
j =0,-1,\dots,-\infty\}\), whereas the bulk of system number two (right, $i=2$) occupies the 
remaining sites, corresponding to \(j=1,\dots,\infty\) in the direction $\s$. In analogy to the case 
of a single bulk treated in Sec. \ref{theoryrecap}, we may 
write single-particle Hamiltonians for the left and right bulks in terms of appropriate shift operators,  
namely, 
\[
T_{1}\equiv \sum_{j=-\infty}^{-1} |j\rangle\langle j+1|,\quad 
T_{2}\equiv \sum_{j=1}^{\infty}|j\rangle\langle j+1|.
\]
Then 
$H_i = \sum_{\k_\parallel}|\k_\parallel\rangle
\langle \k_\parallel|\otimes H_{i,\k_\parallel}$, where 
\begin{eqnarray*}\label{hamsigma}
H_{i\k_\parallel}=\mathds{1}\otimes h_{i\k_\parallel0}+\sum_{r=1}^{R_i}
\big[T_{i}^r\otimes h_{i\k_\parallel r}+\text{H.c.}\big],
\end{eqnarray*}
with the corresponding bulk projectors given by 
\begin{eqnarray*}
P_{B_1}\! \equiv \!\!\sum_{j=-\infty}^{-R_1}\!\!\bm{1}\otimes|j\rangle\langle j|\otimes \mathds{1}_d,\;\;
P_{B_2}\!\equiv \!\!\sum_{j=R_2+1}^{\infty}\!\!\bm{1}\otimes|j\rangle\langle j|\otimes \mathds{1}_d.
\end{eqnarray*}
The projector onto the interface is 
\(P_\partial=\mathds{1}-P_{B_1}-P_{B_2}\).

The Hamiltonian for the total system is of the form
\[ H=H_1+W+H_2,  \]
with \(P_{B_i}W=0,\ i=1,2\). In this context, \(W\) describes an {\em internal BC}, that is,
physically, it accounts for the various possible ways of joining the two bulks. For simplicity,
let us assume that $W$ is translation-invariant in all directions parallel to the interface, so that 
we may write $W = \sum_{\k_\parallel}|\k_\parallel\rangle
\langle \k_\parallel|\otimes W_{\k_\parallel}$. The next step is to split the Schr\"odinger equation 
\((H-\epsilon\mathds{1})|\epsilon\rangle=0\) into a bulk-boundary system of equations \cite{PRL,PRB1}.  
This is possible by observing that an arbitrary state of the total system may be decomposed 
as \(|\Psi\rangle=P_1|\Psi\rangle+P_2|\Psi\rangle\) in terms of the left and right projectors
\[ P_1 \equiv \sum_{j=-\infty}^{0}\bm{1}\otimes|j\rangle\langle j|\otimes \mathds{1}_d,\quad 
P_2 \equiv \sum_{j=1}^\infty\bm{1}\otimes|j\rangle\langle j|\otimes \mathds{1}_d, \]
and that the following identities hold:
\[P_{B_1}(H_1-\epsilon\mathds{1})P_2=0=P_{B_2}(H_2-\epsilon\mathds{1})P_1.\]
Hence, the bulk-boundary system of equations for the interface (or junction) takes the form 
\begin{eqnarray*}
\begin{array}{r}
P_{B_1}(H_1-\epsilon\mathds{1})P_1|\epsilon\rangle=0,\\
P_\partial(H_1+W+H_2-\epsilon\mathds{1})|\epsilon\rangle=0,\\
P_{B_2}(H_2-\epsilon\mathds{1})P_2|\epsilon\rangle=0.
\end{array}
\end{eqnarray*}

We may now solve for fixed but arbitrary \(\epsilon\) the bottom and top bulk equations just as 
in the previous section. The resulting simultaneous solutions of the two bulk equations are expressible as
\begin{eqnarray}
\label{ansatzmultbulk}
|\epsilon,\bm{\alpha}_{\k_\parallel}\rangle&=&
|\epsilon,\bm{\alpha}_{1\k_\parallel}\rangle + |\epsilon,\bm{\alpha}_{2\k_\parallel}\rangle\nonumber\\
&=&\sum_{i=1,2}\sum_{\k_\parallel}|\k_\parallel\rangle \otimes
\big(\sum_{\ell=0}^{n_i}\sum_{v=1}^{s_{i \ell}} 
\alpha_{i \ell s}|\psi_{i \k_\parallel\ell s}\rangle , \quad 
\end{eqnarray}
where $\{|\psi_{i \ell s}\rangle = \sum_{v=1}^{s_{i \ell}}P_i |z_\ell,v\rangle|u_{i \ell s v}\rangle\}$ 
are solutions of the bulk equation for the $i$th bulk. 
In such situations, we extend the definition of the lattice state 
$|z,v\rangle$ to a bi-infinite lattice by allowing the index $j$ in 
Eq.\,\eqref{zstate} to take all integer values. We refer to 
$|\epsilon,\bm{\alpha}_{i\k_\parallel}\rangle,\ i=1,2$, as the eigenstate 
ansatz for the $i$th bulk. For $|\epsilon,\bm{\alpha}_{\k_\parallel}\rangle$ to be an 
eigenstate of the full system, the column array of complex amplitudes
\(
\bm{\alpha}_{\k_\parallel}=
\begin{bmatrix}
\bm{\alpha}_{1\k_\parallel}& \bm{\alpha}_{2\k_\parallel}
\end{bmatrix}^{\rm T}
\)
must satisfy the boundary equation \(B(\epsilon)\bm{\alpha}_{\k_\parallel}=0\),
in terms of the interface boundary matrix, 
\[
[B_{\k_\parallel}(\epsilon)]_{b, i \ell s}=
\langle b|(H_{1\k_\parallel}+W+H_{2\k_\parallel}-\epsilon\mathds{1})|\psi_{i \k_\parallel\ell s}\rangle, 
\]
where the boundary index \(b\equiv -R_1+1,\dots,0; 1,\dots,R_2\).

\begin{figure*}
\includegraphics[width=1\textwidth]{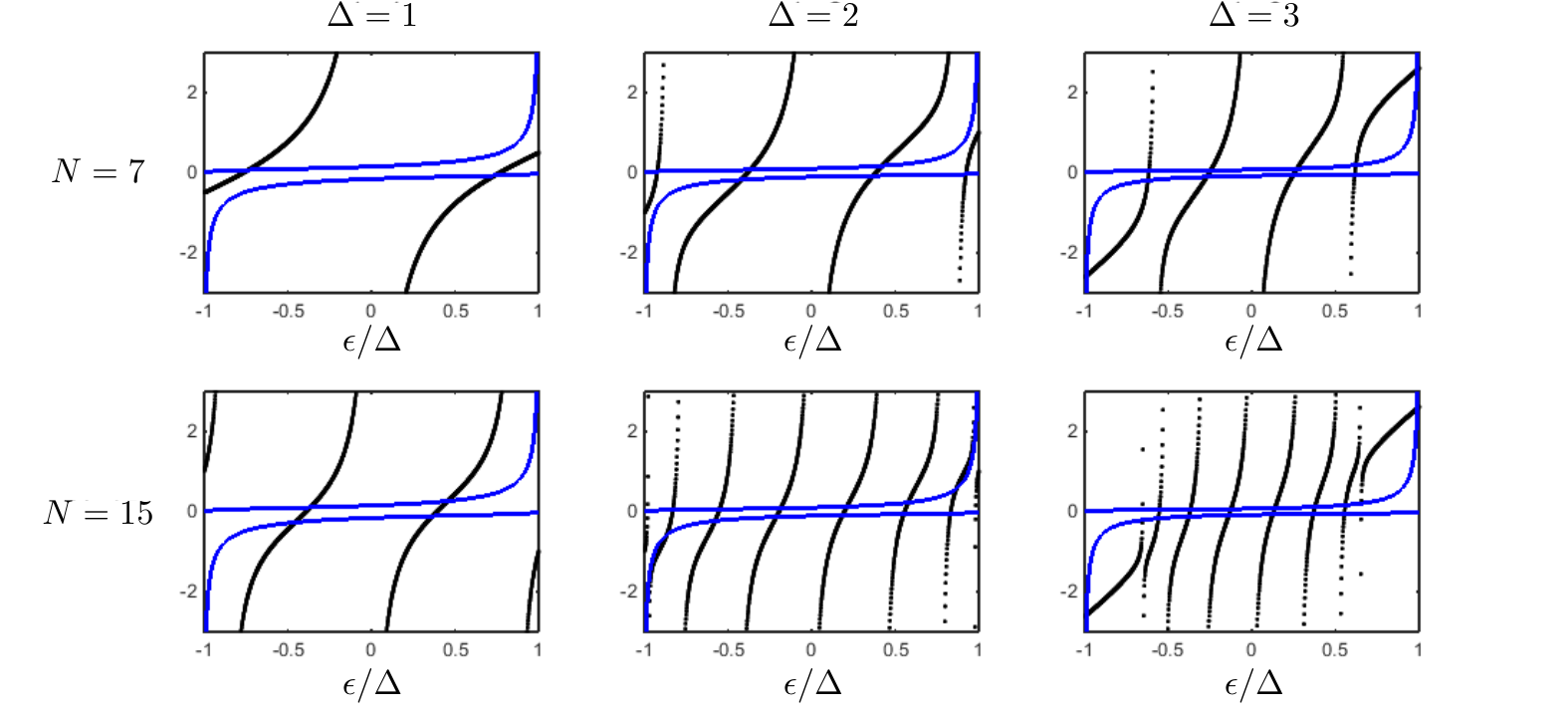}
\vspace*{-5mm}
\caption{(Color online) Bound modes of an SNS junction. The figure shows 
plots of two independent constraints derived from the boundary matrix 
against the ratio $\epsilon/\Delta$, as energy is swept from $-\Delta$ to $\Delta$ 
(with reference to Appendix\,\ref{snsApp}, these constraints are the functions on the left hand-side (blue solid lines) and 
right hand-side (black dotted lines) of Eq.\,\eqref{snsboundaryeq1} 
and Eq.\,\eqref{snsboundaryeq2}. Each intersection
of the two distinct sets of lines indicates the emergence of a bound 
state. The plots in the top (bottom) panels correspond to $N=7$ 
($N=15$), respectively.
Parameters $t=1,t'=0.5$ are fixed for all the plots. The number 
of intersections shows an expected increment as we increase the 
length of the normal region ({\sf N}) from $N=7$ to $15$, without changing other 
parameters. For a fixed value of $N$, the number of intersections increases 
as we change $\Delta=1$ to $\Delta=2$, but it stays constant for $\Delta=2$ vs.  
$\Delta=3$, except for additional states pinned to the energies near $\epsilon=\pm\Delta$.}
\label{snsplot}
\end{figure*}

\subsection{Application to SNS junctions}

We illustrate the generalized Bloch theorem for interfaces 
by outlining an analytical calculation of the Andreev 
bound states for an idealized SNS junction. 
The equilibrium Josephson effect, namely, the phenomenon 
of supercurrent  flowing through a junction of two superconducting 
leads connected via a normal link, is of great importance for theoretical 
understanding of superconductivity, as well as for its applications 
in SC circuits. One of the questions this phenomenon poses is to understand 
how exactly a weak link with induced band-gap due to superconducting proximity 
effect can carry a supercurrent. An answer to this question invokes the formation
of bound states in the band gap of the weak link, known as the ``Andreev
bound states'', that allow transport of Cooper pairs \cite{lesovik11}.

We model a basic $D=1$ SNS junction as a system formed by 
attaching a finite metallic chain (a ``normal dot", denoted by {\sf N}) to two 
semi-infinite SC chains (``superconducting leads", denoted by {\sf S1} and {\sf S2}).
Following Ref.\,[\onlinecite{bena12}], we describe the SC leads in terms of a $D=1$ BCS 
pairing Hamiltonian, 
\begin{eqnarray}
\widehat{H}_{\sf S}  = - \sum_{j,\sigma}t c^\dagger_{j\sigma}c_{j+1\sigma}
 -  \sum_j \Delta c_{j\uparrow}^\dagger c_{j\downarrow}^\dagger+\text{H.c.}, 
\label{BCS}
\end{eqnarray}
where we have assumed zero chemical potential.  
This Hamiltonian can be diagonalized analytically for open BCs, 
see Appendix\,\ref{appBCS} (see also Refs.\,[\onlinecite{arutyunov08,ortiz14,ortiz16}] for 
a critical discussion of $D=1$ models of superconductivity).
The normal dot is modeled by NN hopping of strength $t$.
The links connecting the SC regions to the metallic one 
have a weaker hopping strength,  $t' <t$. 
The Hamiltonian of the full system is thus 
\( \widehat{H}_{\sf SNS}=\widehat{H}_{\sf S1}+
\widehat{H}_{\sf S2}+\widehat{H}_{\sf T}+
\widehat{H}_{\sf N},\)
where $\widehat{H}_{\sf S1}$ and $\widehat{H}_{\sf S2}$ denote the SC Hamiltonians
for the leads, $\widehat{H}_{\sf N}$ describes the normal metal, and $\widehat{H}_{\sf T}$ is the 
tunneling Hamiltonian, of the form 
\begin{equation}
\widehat{H}_{\sf T}=- \!\!
\sum_{\sigma=\pm1} [t' (c_{-2{\sf L} \sigma}^{\dagger}c_{-2{\sf L}+1 \sigma}+
c_{2{\sf L}-1\sigma}^{\dagger}c_{2{\sf L} \sigma})+\text{H.c.}].
\label{Htunnel}
\end{equation}
The region {\sf S1} extends from $j=-\infty$ on the left to $j=-2{\sf L}$, whereas 
{\sf S2} extends from $j=2{\sf L}$ to $j=\infty$, so that the length the of the metallic 
chain is $N\equiv 4{\sf L}-1$.  

The technical implementation of our diagonalization procedure for 
junctions is described in full detail in Appendix \ref{snsApp}. Let us summarize 
the key results here (see also Fig.\,\ref{snsplot} for illustration). 
The structure of the boundary 
equations makes clear the dependence of the number of bound states with the 
length $N$ of the normal dot and the pairing amplitude $\Delta$. 
When the metal strip is completely disconnected from the SC,
that is, when $t'=0$, the stationary states of the normal dot (standing 
waves) are labelled by the quantum numbers    
\(k=\frac{\pi q}{2{\sf L}} +\frac{\pi}{4{\sf L}},\) \(q=0,1,\dots,2{\sf L}-1\),
typical of the lattice-regularized infinite square well. 
Each of these states at energy less than $\Delta$ turns into a 
bound state with a slightly different value of energy for 
weak tunneling. For a fixed value of $\Delta$, increasing $N$ 
allows for more solutions of the boundary equations, and so for 
more Andreev bound states. 
Conversely, for fixed \(N\) the number of bound modes does not increase 
with the value of $\Delta$ once $|\Delta|>|t|$. Instead, we find pinning 
of bound states near energy values $\epsilon=\pm\Delta$ as \(N\) increases. 
These pinned states, that appear only if $|\Delta|>|t|$, are characterized physically 
by a large penetration depth in the superconducting regions {\sf S1} and {\sf S2}.

\section{Surface bands in higher-dimensional systems}
\label{high_dim}

In this section we illustrate the application of the generalized Bloch 
theorem to computing surface bands. Our goal is to gain 
as much insight as possible on the interplay between bulk
properties -- topological or otherwise -- and BCs toward 
establishing the structure of surface bands. We consider first 
a prototypical ladder system, the Creutz ladder \cite{Creutz}, 
as a stepping stone going from one dimension to two.  We next examine 
a graphene ribbon, partly because there has been a considerable 
amount of analytical work on the surface band structure of
this system. Thus, this permits benchmarking 
our generalized Bloch theorem against other approaches. In this regard, we 
emphasize that our method yields analytically {\it all} of the eigenstates 
and eigenvalues of a graphene strip, not just the surface ones. 

Our two final illustrative systems are $D=2$ TSCs. Specifically, we 
first compute the surface band structure of the chiral \(p+ip\) 
TSC analytically, with emphasis on the interplay between the phase 
diagram of the {\it lattice} model and its surface physics. A key 
point here is to gain physical insight into the emergence of {\em chiral surface 
bands} from the point of view of the boundary matrix. We conclude 
by providing an exact, albeit non analytical, solution for the {\em Majorana 
surface flat bands} of a time-reversal invariant gapless $s$-wave TSC model.
Here, we both revisit the anomalous bulk-boundary correspondence 
that this model is known to exhibit \cite{Deng14} through the eyes of the 
boundary matrix, and leverage access to the system's eigenstates to 
characterize physical equilibrium properties.  Notably, we predict that the 
presence of a Majorana surface flat band implies a substantial enhancement in the 
equilibrium $4\pi$-periodic Josephson supercurrent as compared to a gapped $D=2$ TSC
that hosts only a finite number of Majorana modes.

\subsection{The Creutz ladder}
\label{creutzladder}

\begin{figure*}[t]
\includegraphics[width = 18cm]{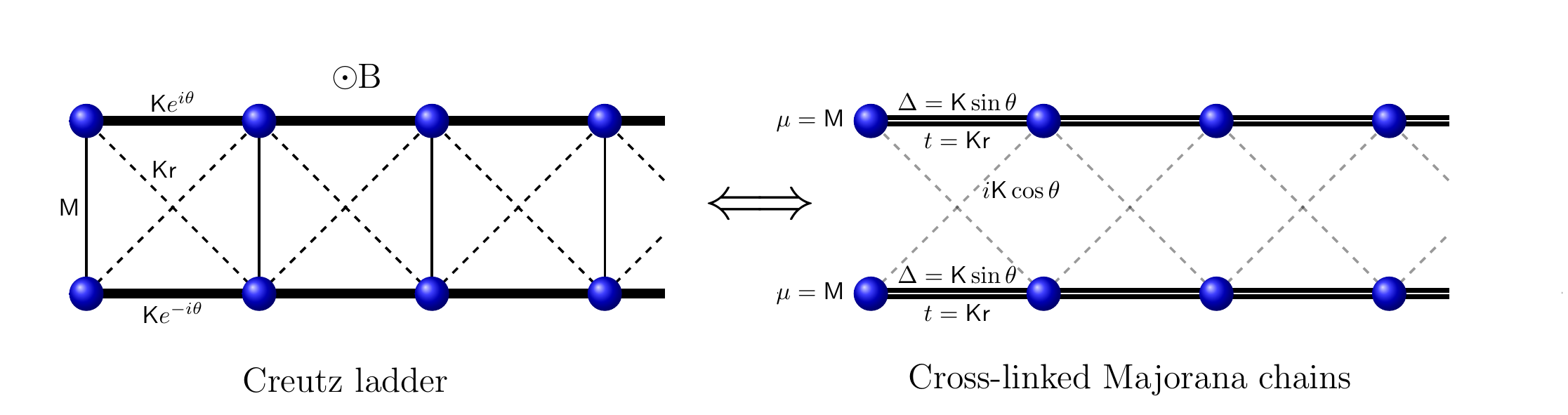}
\caption{(Color online) 
Schematic of the duality between the Creutz ladder (left) and cross-linked Majorana chains (right). 
\label{fig:duality}}
\end{figure*}

The ladder model described by Hamiltonian 
\begin{align}
\label{CreutzHam}
\widehat{H}=&
-\sum_j\big[{\sf K}(e^{i\theta}a_{j}^\dagger a_{j+1}+
e^{-i\theta}b_{j}^\dagger b_{j+1}+\text{H.c.})+\nonumber\\
&+{\sf r} {\sf K}(a_j^\dagger b_{j+1}+b_j^\dagger a_{j+1}+\text{H.c.})+
{\sf M} (a_j^\dagger b_j+b_j^\dagger a_j)\big].
\end{align}
is typically referred to as the Creutz ladder after its proponent 
\cite{Creutz,CreutzRMP, CreutzPRD}, and is schematically depicted in 
Fig. \ref{fig:duality}(left).
Here, $a_j$ and $b_{j}$ denote 
fermionic annihilation operators for fermions at site \(j\) 
of two parallel chains visualizable as the sides of a ladder. 
Fermions  on each side of the ladder are characterized by an 
inverse effective mass ${\sf K}$. There is a homogeneous magnetic 
field perpendicular to the plane of the ladder, responsible 
for the phase $e^{i\theta}$($e^{-i\theta}$) for hopping along the 
upper (lower) side of the ladder. Hopping along rungs of the ladder occur 
with amplitude \({\sf M}\), whereas diagonal hoppings occur with amplitude 
${\sf K}{\sf r}$. 

The Creutz ladder is known to host mid-gap bound states when $|{\sf M}|<|2{\sf K}{\sf r}|$ and 
$\theta\ne0,\pi$. Such states are called {\em domain-wall fermions}
in lattice quantum field theory. The domain-wall fermions of 
the Creutz ladder are, for the most part, not topologically 
protected or mandated by the bulk-boundary correspondence. If
\(\theta\neq \pm\pi/2\), the Creutz ladder may be classified
as a $D=1$ model in class \(A\), thus the domain-wall
fermions are not protected. However,
if \(\theta=\pm \pi/2\), then the Creutz ladder enjoys a chiral
symmetry, and with a canonical transformation of the fermionic basis,
the single-particle Hamiltonian can be made real (see Appendix \,\ref{creupendix}). 
In this parameter regime, the model belongs to class BDI,
which is topologically non-trivial in $D=1$. Interestingly, this was the parameter
regime analyzed in depth in the original work \cite{Creutz}.
We reveal some of these features analytically 
for ${\sf r}=\pm1$ in Appendix \ref{creupendix}. 
Ladder systems are not quite $D=1$, but are not
$D=2$ either. Ultimately, it is more convenient 
to investigate ladders in terms of the basic generalized
Bloch theorem of Part I. For this reason, we have chosen to relegate 
a detailed discussion of the diagonalization of the Creutz ladder to 
Appendix \ref{creupendix}. 
In the following, we highlight two related new results:
a many-body duality transformation that maps the Creutz ladder 
to a pair of Majorana chains, and the existence of 
edge modes with a power-law prefactor. 

\subsubsection{The dual Majorana ladder} 
\label{majoranaladder}

The Gaussian duality transformation \cite{equivalence} 
\begin{align*}
&a_j\mapsto 
\mathcal{U}_{\sf d} a_j\mathcal{U}_{\sf d}^\dagger
={\sf c} \, a_j+i{\sf s}\, a_j^\dagger-i{\sf c} \, b_j
+{\sf s} \ b_j^\dagger ,\\
&b_j\mapsto 
\mathcal{U}_{\sf d} b_j\mathcal{U}_{\sf d}^\dagger
={\sf s} \, a_j-i{\sf c}\, a_j^\dagger-i{\sf s} \, b_j
-{\sf c} \ b_j^\dagger,
\end{align*}
with \(\mathcal{U}_{\sf d}\) a unitary transformation in Fock space
and (${\sf c}=\frac{\cos \varphi}{\sqrt{2}}$, ${\sf s}=\frac{\sin \varphi}{\sqrt{2}}$),
transforms the Creutz ladder model to a dual SC.  
Specialized to $\varphi=\pi/4$, the dual SC Hamiltonian 
is \(
\mathcal{U}_{\sf d}\widehat{H}\mathcal{U}_{\sf d}^\dagger=
\widehat{H}_a+\widehat{H}_b+\widehat{H}_{ab}\),
with
\begin{align*}
&\widehat{H}_a=\!-\sum_{j}\big[t a_{j}^\dagger a_{j+1}+\frac{\mu}{2}a_j^\dagger a_j
+\Delta \,  a_ja_{j+1} +\text{H.c.} ] , \\
&\widehat{H}_b=\!-\sum_{j}\big[t b_{j}^\dagger b_{j+1}+\frac{\mu}{2}b_j^\dagger b_j
+\Delta \,  b_jb_{j+1} +\text{H.c.} ] , \\
t&\equiv {\sf r}{\sf K},\quad \Delta\equiv {\sf K}\sin\theta,\quad \mu\equiv {\sf M},
\end{align*}
and, finally, 
\begin{align*}
\widehat{H}_{ab}=\!-\sum_j\big[i {\sf K} \cos \theta (b^\dagger_ja_{j+1}+b^\dagger_{j+1}a_j-\text{H.c.})-{\sf M}\big]. 
\end{align*}
We conclude that the dual system may be described as a ladder 
consisting of Majorana chains on each side, connected by electron 
tunneling and with no pairing term associated to the rungs of the 
ladder [see Fig. \ref{fig:duality}(right)]. Moreover, the Majorana chains (the sides of the ladder) 
decouple if \(\theta=\pm \pi/2\), in which case the Creutz ladder 
displays chiral symmetry. Since these two decoupled Majorana chains 
have real parameter values, the dual system also belongs to the 
topologically non-trivial class D.

The fermion number operator \(\hat{N}_F\equiv \sum_j(a_j^\dagger a_j
+b_j^\dagger b_j)\), regarded as the broken particle conservation 
symmetry of the Majorana ladder, maps by the inverse of the duality transformation
to a broken symmetry 
\(\hat{N}_{C}\equiv \mathcal{U}^\dagger_{\sf d} \hat{N}_F\mathcal{U}_{\sf d}\)
of the Creutz ladder. In other words, we expect the insulating spectral 
gap of the Creutz ladder to close whenever the symmetry \(\hat{N}_{C}\) is 
restored,
unless there is a stronger factor at play. This symmetry is restored 
for \({\sf K}\sin(\theta)=0\), which is indeed a gapless regime unless \({\sf K}=0\),
because then the Creutz ladder reaches the atomic limit. A similar
explanation of the insulating gap for the Peierls chain in terms of 
a hidden broken symmetry was given in Ref.\,[\onlinecite{equivalence}], where 
fermionic Gaussian dualities were investigated in higher dimensions as well.

\subsubsection{Topological power-law modes}

The generalized Bloch theorem identifies regimes
in which the domain-wall fermions of the Creutz ladder may
display power-law behavior. From the analysis in Appendix \ref{creupendix},
power-law modes are forbidden only if ${\sf M}=0,$ $\theta=\pm\pi/2$ 
and ${\sf K}\ne0,$ ${\sf r}\ne \pm1$. For arbitrary values of ${\sf K},{\sf r},\theta,{\sf M}$, 
one can expect in general a finite number of values of $\epsilon$ 
for which the full solution of the bulk equation includes power-law modes, 
potentially compatible with the BCs. Let us point out for illustration the 
power-law modes of the Creutz ladder in the parameter regime
$\theta=\pi/2,\ {\sf M}=2{\sf K}\sqrt{{\sf r}^2-1}$, with ${\sf r}>1$. In this 
regime the Creutz ladder is dual to two decoupled Kitaev chains, 
each individually on its ``circle of oscillations'' in its phase 
diagram \cite{hegde16}. The topological power-law modes of 
the Kitaev chain have been explicitly described in Part I (see Sec. V C).
Therefore, the power-law topological edge modes of the Creutz ladder may 
be found by way of our duality transformation. Alternatively, there is a 
shortcut at the single-particle level.

Let us rewrite the Creutz ladder in terms of a new set of fermionic degrees
of freedom
\begin{align}
\label{tildeferms}
\widetilde{a}_j=\frac{1}{\sqrt{2}}(a_j+b_j), \quad \widetilde{b}_j=\frac{i}{\sqrt{2}}(a_j-b_j).
\end{align}
Unlike for our previous duality transformation, the result is another particle-conserving Hamiltonian.
The associated single-particle Hamiltonian is 
\begin{eqnarray}
&&\widetilde{H}_N=\mathds{1}_{N}\otimes \tilde{h}_0+
T\otimes \tilde{h}_1+T^\dagger\otimes \tilde{h}_1^\dagger,  \label{tildeCreutz} \\
&& \tilde{h}_{0}=-\begin{bmatrix}{\sf M} & 0\\
0 & -{\sf M}
\end{bmatrix}, \nonumber\\
&& \tilde{h}_1=-\begin{bmatrix}{\sf K}({\sf r}+\cos\theta) & {\sf K}\sin\theta\\
-{\sf K}\sin\theta & {\sf K}(-{\sf r}+\cos\theta)
\end{bmatrix}. \nonumber
\end{eqnarray}
For \(\theta=\pi/2\), and with the identifications \(t={\sf K}{\sf r}, \Delta={\sf K}\sin \theta, \mu={\sf M}\) already
introduced, the above \(\widetilde{H}_N\) becomes identical to the single-particle Hamiltonian for the Majorana chain 
of Kitaev. Moreover, if \({\sf M}=\mu=2{\sf K}\sqrt{{\sf r}^2-1}\), it follows that \((\mu/2t)^2+(\Delta/t)^2=1\).
This is the aforementioned coupling regime known as the ``circle of oscillations". Hence, by 
simply translating the calculations of Part I, Sec. V C, we obtain the topological power-law mode
\begin{align*}
|\epsilon=0\rangle =
\sum_{j=1}^{\infty} j \, w^{j-1}|j\rangle
\begin{bmatrix}
1 \\ -1
\end{bmatrix},\quad 
w \equiv -\Big(\frac{{\sf r}-1}{{\sf r}+1}\Big)^{1/2},
\end{align*}
of the Creutz ladder (in the particle-conserving representation of Eq.\,\eqref{tildeferms}).
To our knowledge, this provides the first example of a topological power-law zero mode in a particle-conserving 
Hamiltonian in class AIII.

\subsection{Graphene ribbons}
\label{secgraphene}

In this section we investigate NN tight-binding models on the honeycomb (hexagonal) lattice, with graphene as 
the prime motivation \cite{castroneto09}. The surface band structure 
of graphene sheets or ribbons is well understood, even analytically in limiting cases 
\cite{mao10,kohmoto07,delplace11,Iachello}. 
As emphasized in Ref. [\onlinecite{yao09}], a perturbation that breaks inversion symmetry can have interesting 
effects on these surface bands. With this in mind, in our analysis below we include a sublattice potential and show that the 
Hamiltonian for a ribbon subject to zigzag-bearded BCs 
can be fully diagonalized in closed form. 

\begin{figure*}
\hspace*{-5mm}\includegraphics[angle=0, width=.9\columnwidth]{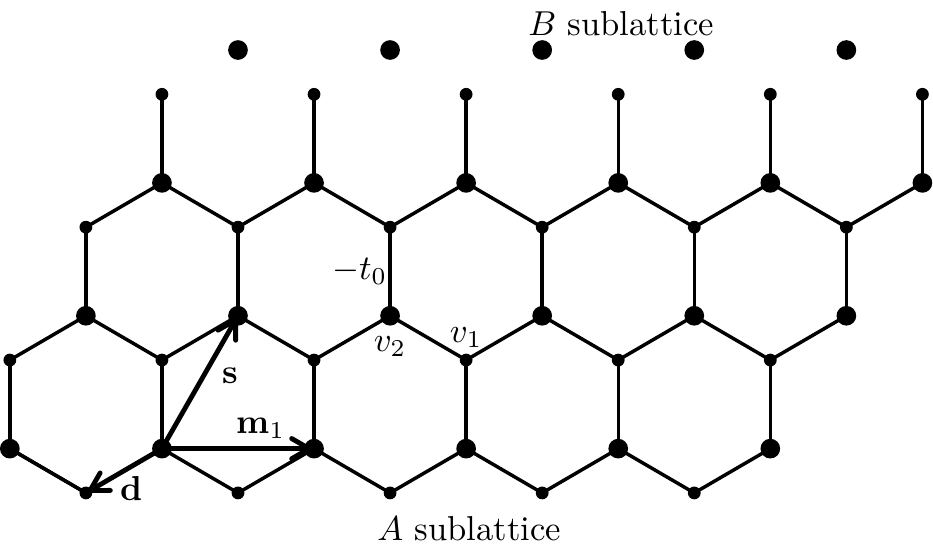}
\hspace*{1.5cm}\includegraphics[angle=0, width=6.5cm]{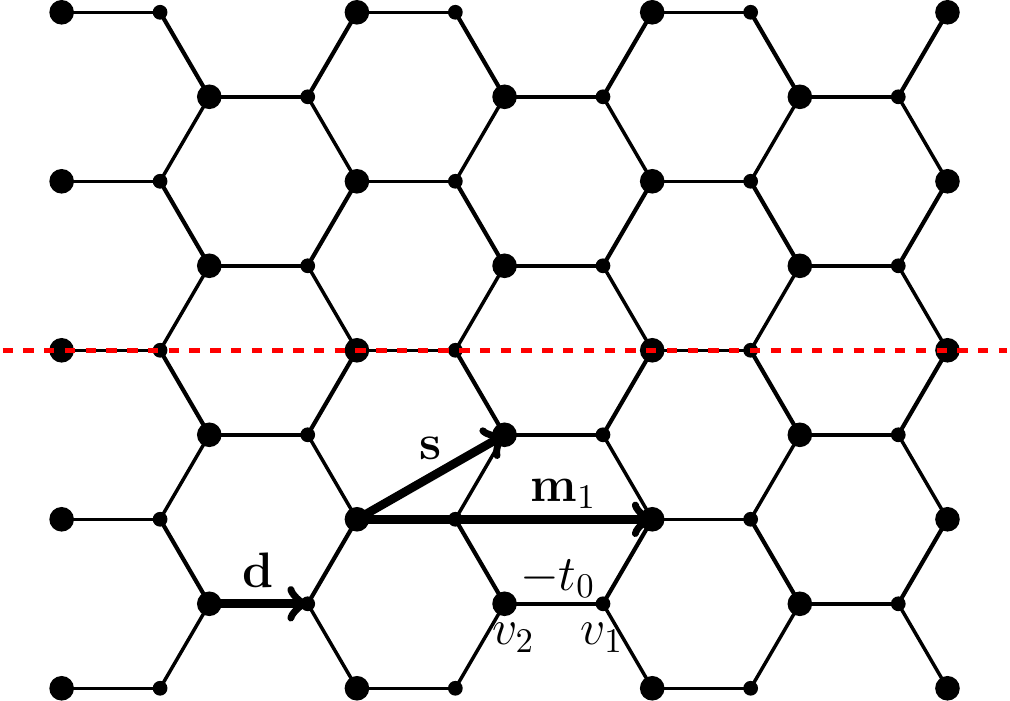}
\caption{(Color online) Graphene ribbon, periodic or infinite 
in the horizontal 
\(\m_1\) direction. Left: The ribbon is terminated in the vertical direction by a zigzag edge 
on the bottom and a ``bearded" edge on top. The decoupled \(B\) sites at the  
top are auxiliary degrees of freedom. 
Right: The ribbon is terminated by armchair edges. The system has mirror 
symmetry about the dashed (red) line. In both cases, 
on-site potentials \(v_1\) and \(v_2\) are associated with the $A$ and $B$
sublattice, respectively.
}
\label{zigcomb}
\end{figure*}

\subsubsection{Zigzag-bearded boundary conditions}
\label{zbsec}

The honeycomb lattice is bipartite, with triangular sublattices \(A\) and \(B\) 
displaced by \(\d\) relative to each other, see Fig.\,\ref{zigcomb}(left). 
We parametrize the lattice sites \(\mathbf{R}\) as 
\begin{align*}
\mathbf{R}(j_1,j,m)=
\left\{
\begin{array}{lcl}
j_1\m_1+j\s +\d & \mbox{if} & m=1\\
j_1\m_1+j\s & \mbox{if} & m=2
\end{array}\right., \quad \\
\m_1\equiv a
\begin{bmatrix}1\\0\end{bmatrix},\ 
\s=\frac{a}{2}\begin{bmatrix}1\\\sqrt{3}\end{bmatrix},\ 
\d=-\frac{a}{2\sqrt{3}}\begin{bmatrix}\sqrt{3}\\1\end{bmatrix},
\end{align*}
with \(j_1,j\in\mathds{Z}\), \(a=1\) being the lattice parameter and 
$m=1$ ($m=2$) denoting the $A$ ($B$) sublattice. 
The localized (basis) states are
\(|\j\rangle|m=1\rangle\) and \(|\j\rangle|m=2\rangle\), and so the
sublattice label plays the role of a pseudospin-$1/2$ degree of freedom. 
The ribbon we consider is translation-invariant in the \(\m_1\) direction
and terminated along $\s$, with single-particle Hamiltonian 
$H_N = \sum_{\k_\parallel \in \text{SBZ}}|\k_\parallel\rangle\langle\k_\parallel |
\otimes H_{\k_\parallel,N}$, where
\begin{eqnarray*}
H_{\k_\parallel,N}
& =& \mathds{1}_N\otimes 
\begin{bmatrix}
v_1& -t_0(1+e^{-ik_\parallel})\\
-t_0(1+e^{ik_\parallel})& v_2
\end{bmatrix}\\ 
& + &
\Big(T\otimes
\begin{bmatrix}
0& 0\\
-t_0& 0
\end{bmatrix} + \text{H.c.}\Big) ,
\end{eqnarray*}
and the \(2\times 2\) 
matrices act on the sublattice degree of freedom. Notice that \(H_N\) 
is chirally symmetric if the on-site potentials
\(v_1=0=v_2\), and the edges of the ribbon 
are of the zigzag type, see Fig.\,\ref{zigcomb}(left). 
While in the following we shall set \(v_1=0\) for simplicity, it 
is easy to restore \(v_1\) anywhere along the way if desired. 
In particular, \(v_1=-v_2\) is an important special case \cite{yao09}.

The analytic continuation of the Bloch Hamiltonian is 
\begin{align*}
H_{k_\parallel}(z)
=\begin{bmatrix}
0   & -t_1(k_\parallel)e^{-i\phi_{k_\parallel}}-t_0z^{-1}\\
-t_1(k_\parallel)e^{i\phi_{k_\parallel}}-t_0z & v_2 
\end{bmatrix},
\end{align*}
\[ t_1(k_\parallel) \equiv t_0\sqrt{2(1+\cos(k_\parallel))},\ \ 
e^{i\phi_{k_\parallel}} \equiv t_0(1+e^{ik_\parallel})/t_1(k_\parallel). \]
This analysis reveals the formal connection between graphene and 
the  Su-Schrieffer-Heeger (SSH) model: just compare the above $H_{k_\parallel}(z)$ 
with $H(z)$ in Eq.\,\eqref{precisely0}.

We impose BCs in terms of an operator \(W\) such that 
\begin{align*}
\langle k_\parallel|W|k_\parallel'\rangle
\!=\!\delta_{k_\parallel,k_\parallel'}
|N\rangle\langle N|
\otimes 
\begin{bmatrix}
0& t_1(k_\parallel)e^{-i\phi_{k_\parallel}}\\
t_1(k_\parallel)e^{i\phi_{k_\parallel}}& 0
\end{bmatrix} .
\end{align*}
In real space, this corresponds to 
\begin{align*}
W=\,&\mathbf{1}\otimes|N\rangle\langle N|\otimes
\begin{bmatrix}
0& -t_0\\
-t_0& 0
\end{bmatrix}+\\
&\quad \quad \quad +
\Big(
\mathbf{T}\otimes |N\rangle\langle N|\otimes 
\begin{bmatrix}
0& 0\\
-t_0& 0
\end{bmatrix}+\text{H.c.}
\Big),
\end{align*}
The meaning of these BCs is as follows:
for the modified ribbon Hamiltonian described by \(H=H_N+W\), the sites
\(|j_1\rangle|j=N\rangle|B\rangle\) are decoupled from the rest of the
system and each other, see Fig.\,\ref{zigcomb}(left). The termination
of the actual ribbon, consisting of the sites connected to each 
other, is of the zigzag type on the lower edge, and ``bearded"
on the upper edge. From a geometric perspective, this ribbon
is special because every \(B\) site is connected to exactly three
\(A\) sites, but not the other way around.   

At this point we may borrow results from dimerized chains that we
include in Appendix \,\ref{basic_examples}, to which we refer for full detail. 
The energy eigenstates that
are perfectly localized on the upper edge (consisting
of decoupled sites) constitute a flat surface band at energy \(v_2\).
For $|k_\parallel|>2\pi/3$, the energy eigenstates 
localized on the lower edge constitute a flat 
surface band at \(v_1=0\) energy. Explicitly, these zero modes are
\begin{align*}
|\epsilon=0,k_\parallel\rangle=&
|k_\parallel\rangle|z_1(k_\parallel)\rangle
\begin{bmatrix}
(t_1(k_\parallel)^2-t_0^2)e^{-i\phi_{k_\parallel}}/t_1(k_\parallel)\\
0
\end{bmatrix},\\
\quad z_1(k_\parallel)\equiv&-e^{i\phi_{k_\parallel}}\frac{t_1(k_\parallel)}{t_0}=-(1+e^{ik_\parallel}).
\end{align*}
While their energy is insensitive to \(k_\parallel\), their characteristic localization 
length is not; specifically, 
\begin{eqnarray}
\ell_{\rm loc}(k_\parallel)= -\frac{1}{\ln(|z_1(k_\parallel)|)}=- \frac{2}{\ln(2+2\cos(k_\parallel))}.\quad
\label{locl}
\end{eqnarray}
For \(k_\parallel\neq\pm\frac{2\pi}{3}\), the bulk states are
\begin{align*}
|\epsilon_n(k_\parallel,q)\rangle
\!=\! |k_\parallel\rangle|\chi_1(q)\rangle \!
\begin{bmatrix} t_1(k_\parallel)e^{-i\phi_{k_\parallel}}\\ -\epsilon_n(k_\parallel,q)\end{bmatrix}
\! + \! |k_\parallel\rangle|\chi_2(q)\rangle \! \begin{bmatrix}t_0\\ 0\end{bmatrix},
\end{align*}
with
\begin{align}
\label{chiwf1}
&|\chi_1(q)\rangle
\equiv 2i\sum_{j=1}^N\sin\!\big(\pi qj/N\big)e^{-i\phi j}|j\rangle,\\
\label{chiwf2}
&|\chi_2(q)\rangle 
\equiv 2i\sum_{j=1}^N\sin\!\big(\pi q(j-1)/N\big)e^{-i\phi( j-1)}|j\rangle,
\\
&\epsilon_n(k_\parallel,q)=\\
&=\frac{v_2}{2}+
(-1)^n\sqrt{\frac{v_2^2}{4}+t_1(k_\parallel)^2+t_0^2+2t_1(k_\parallel)t_2\cos\!\big(\frac{\pi}{N}q\big)}
\nonumber,
\end{align}
for $n=1,2$. Since \(t_1(k_\parallel=\pm\frac{2\pi}{3})=t_0\), 
the virtual chains \(H_{k_\parallel,N}\)
are gapless if \(v_2=0\), reflecting the fact that 
graphene is a semimetal. 
The energy eigenstates 
are similar but simpler than the ones just described. 

\subsubsection{Armchair terminations}
\label{armpitsec}

The graphene ribbon with zigzag terminations can be 
described in terms of smooth terminations of the triangular
Bravais lattice with two atoms per unit cell. In contrast, 
armchair terminations require a fairly different description 
of the underlying atomic array. Figure\,\ref{zigcomb}(right) 
shows how to describe this system in terms of a {\it centered rectangular} 
Bravais lattice \cite{bechstedt} with two atoms per unit cell and smooth parallel
terminations. In this case, we parametrize the lattice sites 
\(\mathbf{R}\) as 
\begin{align*}
\mathbf{R}(j_1,j,m)=
\left\{
\begin{array}{lcl}
j_1\m_1+j\s +\d & \mbox{if} & m=1\\
j_1\m_1+j\s & \mbox{if} & m=2
\end{array}\right.,\\
\m_1\equiv a
\begin{bmatrix}\sqrt{3}\\0\end{bmatrix},\ 
\s=\frac{a}{2}\begin{bmatrix}\sqrt{3}\\1\end{bmatrix},\ 
\d=\frac{a}{\sqrt{3}}\begin{bmatrix}1\\0\end{bmatrix}, 
\end{align*}
where as before \(j_1,j\in\mathds{Z}\), $a=1$, and $m\in \{1,2\}$ labels the 
sublattice. The total single-particle Hamiltonian can now be taken to read $H=H_N+W$, with $W=0$ and   
$H_N = \sum_{\k_\parallel \in \text{SBZ}}|\k_\parallel\rangle\langle\k_\parallel |
\otimes H_{\k_\parallel,N}$, where
\[
H_{\k_\parallel,N}
=\mathds{1}_N\otimes 
\begin{bmatrix}
v_1& -t_0\\
-t_0 & v_2
\end{bmatrix}+
\Big(T\otimes
\begin{bmatrix}
0& -t_0e^{-ik_\parallel}\\
-t_0& 0
\end{bmatrix} + \text{H.c.}\Big), \]
and the analytic continuation of the Bloch Hamiltonian for
each $k_\parallel$ is
\[ H_{k_\parallel}(z) \!=\!
\begin{bmatrix} v_1 & -t_0(1+ze^{-ik_\parallel} + z^{-1})\\
-t_0(1+z^{-1}e^{ik_\parallel} + z) & v_2
\end{bmatrix}. \]

The diagonalization of the Hamiltonian proceeds from here on 
as before. There is, however, a shortcut based on Appendix \ref{app:condition},
which explains in addition the absence of edge modes in this system. Let 
\(
T_{k_\parallel}\equiv e^{-ik_\parallel/2}T
\). In terms of this \(k_\parallel\)-dependent matrix, 
\[
H_{\k_\parallel,N}\!=\!
\mathds{1}_N\otimes 
\begin{bmatrix}
v_1& \!\!-t_0\\
-t_0 &\!\! v_2
\end{bmatrix}
-t_0
\Big(T_{k_\parallel}+T_{k_\parallel}^\dagger\Big)\otimes
\begin{bmatrix}
0                     & e^{-ik_\parallel/2}\\
e^{ik_\parallel/2}&  0
\end{bmatrix} \!.\]
\begin{widetext}
It follows that the (unnormalized) energy eigenstates of the graphene 
ribbon with armchair terminations are
\begin{eqnarray*}
\label{armpit}
|\epsilon_{q,\pm}\rangle = 
|k_\parallel\rangle 
\sum_{j=1}^{N}|j\rangle e^{ik_\parallel j/2}\sin [\pi qj/(N+1)]
\begin{bmatrix}-2t_0(1+e^{-ik_\parallel/2}\cos[\pi q/(N+1)])\\ \epsilon_{q,\pm} \end{bmatrix}
,\qquad q=1,\dots,N,
\end{eqnarray*}
where $\epsilon_{q,+}$ and $\epsilon_{q,-}$ are the two roots (in $\epsilon$) 
of the quadratic equation
\[ \epsilon^2-v_2\epsilon-
t_0^2 - 4t_0^2\cos (k_\parallel/2)\cos[\pi q/(N+1)] 
- 4t_0^2\cos^2[\pi q/(N+1)] =0. \]
These are the $2N$ energy eigenstates of the system for each value
of $k_\parallel$. 
\end{widetext}

\subsection{A chiral \(p+ip\) superconductor}
\label{pwavetoponductor}

The spinless \(p+ip\) SC of Ref.\,[\onlinecite{read00}] is the prototype of spinless
superconductivity in $D=2$. The model may be regarded as the mean-field approximation 
to an exactly-solvable (by the algebraic Bethe ansatz) 
pairing Hamiltonian \cite{rombouts10}. It belongs to class D in the Altland-Zirnbauer 
classification, and thus, according to the ten-fold way, it admits an integer ($\mathbb{Z}$) 
topological  invariant.  There has been hope for some time that the related phenomenon
of triplet superconductivity is realized in layered perovskite strontium 
ruthenate \(\rm Sr_2RuO_4\), but the matter remains controversial \cite{Sr2RuO4}. 
The many-body model Hamiltonian can be taken to be
\begin{align*}
&\hat{H}=
-t\sum_{\R}(c_{\R+\s}^\dagger c_{\R}+
c_{\R+\m}^\dagger c_{\R}+\text{H.c.})\\
&-\Delta\sum_{\R}(c_{\R}c_{\R+\s}-ic_{\R}c_{\R+\m}+\text{H.c.})-
(\mu-4t)\sum_\R c^\dagger_\R c_\R,
\end{align*}
on the square lattice of unit lattice spacing and with standard 
unit vectors \(\s,\m\) pointing in the \(x\) and \(y\) 
directions, respectively. The parameters \(t,\Delta\) are real numbers. 
The corresponding single-particle Hamiltonian is
\begin{align*}
H&=- [(\mu-4t)\mathds{1}+t (T_\s+T_\s^\dagger)+t(T_\m+T_\m^\dagger)]\otimes \tau_z+\\
&+i\Delta(T_\s-T_\s^\dagger)\otimes \tau_y+i\Delta(T_\m-T_\m^\dagger)\otimes\tau_x , 
\end{align*}
in terms of shift operators \(T_\s \equiv \sum_\r |\r\rangle\langle \r+\s|,$ $T_\m\equiv \sum_\r |\r\rangle\langle \r+\m|\) 
which
can be adjusted to describe relevant BCs 
(open-open, open-periodic, periodic-open, and periodic-periodic). 

\subsubsection{Closed-form chiral edge states}

If energy is measured in units of \(t\), then the parameter
space of the model can be taken to be two-dimensional after
a gauge transformation that renders \(\Delta>0\). We shall 
focus on the line \(\Delta=1=t \), in which \(\mu\) is the
only variable parameter. The Bloch Hamiltonian is
\begin{align*}
H(\k)&=
\begin{bmatrix}
e(\k)& \Delta(\k)\\
\Delta(\k)^*& -e(\k)
\end{bmatrix},\\
\Delta(\k)&\equiv 2i\sin k_1 -2\sin k_2,\\
e(\k)& \equiv -2 \cos k_1 - 2\cos k_2-\mu +4,
\end{align*}
for \( \k=(k_1,k_2)\in [-\pi,\pi)\times [-\pi,\pi)\).
The resulting single-particle bulk dispersion then reads   
\begin{eqnarray*}
\epsilon(k_1,k_2)^2 &=& \mu^2 -8\mu +24  + 4(\mu-4)(\cos k_1+\cos k_2)\\
&+&8\cos k_1 \cos k_2,
\end{eqnarray*}
and it is fully gapped unless \(\mu=0,4,8\). The gap closes at \(\k=0\) if \(\mu=0\),
\(\k=(-\pi,0)\) and \(\k=(0,-\pi)\) if \(\mu=4\), and at 
\(\k=(-\pi,-\pi)\) if \(\mu=8\). For $0<\mu <8$, the system is in the 
weak-pairing topologically non-trivial phase with odd fermion number parity
in the ground state. The phase transition to the trivial strong-pairing phase happens 
at $\mu=0$ \cite{read00}.

We now impose open BCs in the \(x\) direction while keeping 
the \(y\) direction translation invariant, that is, \(k_2=k_\parallel\). Accordingly, we need the analytic 
continuation of the Bloch Hamiltonian in \(k_1\). Let us introduce the compact notation
\[
\omega \equiv -2\cos k_\parallel-\mu+4,\quad 
\xi  \equiv -2\sin k_\parallel , \]
so that \( H_{k_\parallel}(z)=h_{k_\parallel,0}+zh_1+z^{-1}h_1^\dagger
\), with 
\begin{align}
\label{h1here}
h_{k_\parallel,0}= 
\begin{bmatrix}
\omega &  \xi \\
\xi & -\omega
\end{bmatrix},\quad
h_1=
\begin{bmatrix}
-1& 1\\
-1& 1
\end{bmatrix}.
\end{align}
The condition 
\(\det(H_{k_\parallel}(z)-\epsilon\mathds{1}_2)=0\) is then equivalent to 
the equation 
\begin{align}
\label{thisfirst}
\epsilon^2&=\omega^2+\xi^2+4-2\omega\, (z+z^{-1}).
\end{align}
Note that the replacement \(z+z^{-1}\mapsto 2\cos k_1\) recovers
the bulk dispersion relation. Moreover, if \(2<\mu<6\), there are values of \(k_\parallel\)
for which \(\omega=0\) and the dispersion relation becomes flat. From \(H_{k_\parallel}(z)\) it is 
immediate to reconstruct the family of virtual chain Hamiltonians 
\begin{align*}
H_{k_\parallel,N}&=\mathds{1}_N\otimes h_{k_\parallel,0}+T\otimes h_1+ T^\dagger\otimes h_1^\dagger.
\end{align*}
From the point of view of any one of these chains, mirror symmetry is 
broken by the NN pairing terms.
This fact is important, because then the boundary matrix
is {\em not} mirror-symmetric either, which will ultimately lead to 
surface states of opposite chirality on the left and right edges. 

The number of edge degrees of freedom is \(2Rd=4\) for each 
value of \(k_\parallel\). Since \(h_1\) [Eq. \eqref{h1here}] is not invertible, 
and Eq.\,\eqref{thisfirst} is a polynomial 
of degree \(2\) in \(z\), the complete eigenstate 
ansatz is formed out of four independent states (one ansatz state 
for each \(k_\parallel\)): two extended states associated to the roots 
\(z_\ell=z_\ell(\epsilon,k_\parallel),\) \(\ell=1,2,\)
of Eq.\,\eqref{thisfirst}, 
and two emergent states of finite
support localized on the edges of the virtual chains \(H_{k_\parallel,N}\). 
With hindsight, we will ignore the emergent states and focus on
the reduced ansatz, namely,  
\[
|\epsilon\rangle=\alpha_1|z_1,1\rangle|u_1\rangle+\alpha_2z_2^{-N+1}|z_2,1\rangle|u_2\rangle.
\]
The state \(|z_1,1\rangle|u_1\rangle\) should represent a
surface state for the left edge, \(z_2^{-N+1}|z_2,1\rangle|u_2\rangle\) 
one for the right edge, with 
\begin{eqnarray}\label{upip}
|u_\ell\rangle=
\begin{bmatrix}
\xi +z_\ell-z_\ell^{-1}\\
-\omega +\epsilon+z_\ell+z_\ell^{-1}
\end{bmatrix}
\end{eqnarray}
satisfying the equation 
\(H_{k_\parallel}(z_\ell)|u_\ell\rangle=\epsilon|u_\ell\rangle\). The boundary equations 
\(P_\partial(H_{k_\parallel,N}-\epsilon\mathds{1}_{2N})|\epsilon\rangle=0\)
are encoded in the boundary matrices 
\begin{eqnarray*}
B_{k_\parallel}(\epsilon)=-
\begin{bmatrix}
h_1^\dagger|u_1\rangle & z_2^{-N-1}h_1^\dagger |u_2\rangle\\
z_1^{N+1}h_1|u_1\rangle& h_1|u_2\rangle
\end{bmatrix},
\end{eqnarray*}
which are, however, non-square \(4\times 2\) matrices as  we have 
ignored the two emergent states that in principle appear in 
the ansatz. Nonetheless, since \(h_1\) is a matrix of rank one, 
we can extract a square boundary matrix, namely, 
\begin{align*}
\tilde{B}_{k_\parallel}(\epsilon) \!=\!\begin{bmatrix}
z_1(\xi -\omega+\epsilon+2z_1)& \! z_2^{-N}(\xi-\omega+\epsilon+2z_2)\\
z_1^N(\xi+\omega-\epsilon-2z_1^{-1})& \! z_2(\xi+\omega-\epsilon-2z_2^{-1})
\end{bmatrix} ,
\end{align*}
that properly captures the BCs for our reduced trial states.
Surface states are characterized 
by the condition \(|z_1|=|z_2^{-1}|<1\). Hence, in the large-\(N\) limit, 
one may set \(z_1^N=z_2^{-N}=0\). Within this approximation, 
the left and right edges are effectively decoupled by virtue of their large 
spatial separation. 

\begin{figure}[t]
\label{niceplots}
\includegraphics[width=\columnwidth]{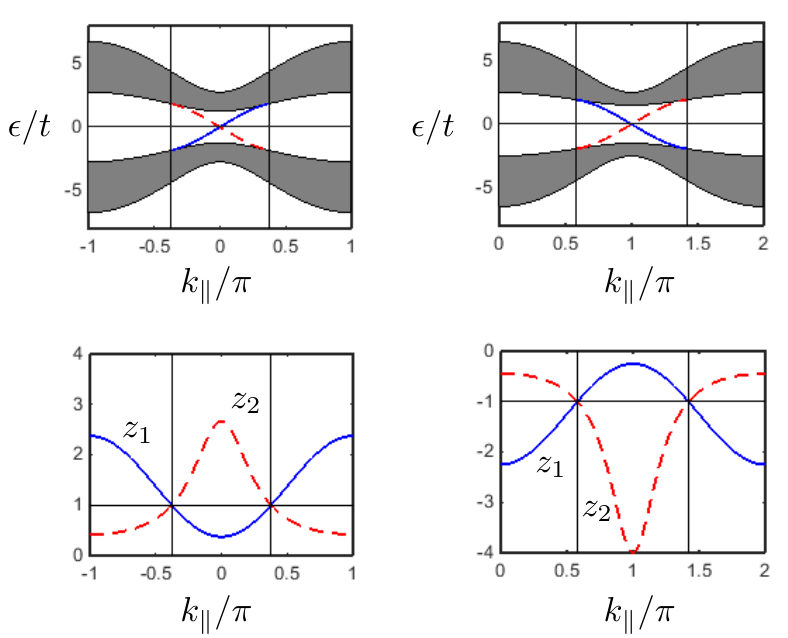}
\caption{(Color online) Surface bands for \(\mu=1.5\), centered at \(k_\parallel=0\) (top left panel),
and \(\mu=6.5\), centered at \(k_\parallel=-\pi\) (top right panel). The shaded (gray) region 
shows the bulk bands. The electrons on the right edge (dashed red curve) propagate
to the right only, and those on the left edge (solid blue curve) to the left only, that
is, the surface bands are chiral. The lower panels show the 
behavior of $z_1$ (solid blue curve) and $z_2$ (dashed red curve) with $k_\parallel$.
Notice how \(z_1\) (\(z_2\)) enters (exits) the unit circle precisely when the 
surface bands touch the bulk bands, as marked by vertical solid black lines.
\label{fig:pip}}
\end{figure}

In summary, the left surface band is determined by the polynomial system 
\begin{align}
\label{rightedge}
\left\{
\begin{array}{l}
0=\xi -\omega +\epsilon+2z_1\\
0=\epsilon^2+2\omega \, (z_1+z_1^{-1})-(\omega^2+\xi^2+4)
\end{array}\right..
\end{align}
In the following, we will focus on the cases \(0<\mu<2\) or \(6<\mu<8\) for simplicity 
(these parameter regimes are in the weak pairing phase and satisfy \(\omega\neq 0\)
for all values of \(k_\parallel\)). Notice that 
\begin{eqnarray}
\label{upipeasy}
|u_1\rangle=(\xi+z_1-z_1^{-1})\begin{bmatrix} 1\\ -1\end{bmatrix}
\end{eqnarray}
due to the (top) boundary equation in Eq.\,\eqref{rightedge} (recall
also Eq.\,\eqref{upip}).
The physical solutions\cite{footpip} are surprisingly simple. They are
\begin{align*}
\epsilon&\equiv \epsilon_{\rm left}(k_\parallel)=-\xi=2\sin k_\parallel,\\
z_1&=z_1(k_\parallel)=\frac{\omega}{2}=2-\frac{\mu}{2}-\cos k_\parallel.
\end{align*}
These functions of \(k_\parallel\) represent the dispersion relation
and ``complex momentum" of surface excitations on the left 
edge
for those values of \(k_\parallel\) (and {\em only} those values) such that
\(|z_1(k_\parallel)|<1\) (see Fig. \ref{fig:pip}). Notice that {\em the edge band is chiral}.
The surface band touches the bulk band at the two values of 
\(k_\parallel\) such that \(|z_1(k_\parallel)|=1\). The (unnormalized) surface states 
are, for large \(N\),
\[
| \epsilon_{\rm left}(k_\parallel)\rangle=\sum_{j=1}^N\Big(2-\frac{\mu}{2}-\cos k_\parallel \Big)^j|k_\parallel\rangle 
|j\rangle\begin{bmatrix} 1\\ -1\end{bmatrix}.
\]

Similarly, the right surface band is determined by the polynomial system
\begin{align}
\label{leftedge}
\left\{
\begin{array}{l}
0=\xi+\omega-\epsilon-2z_2^{-1}\\
0=\epsilon^2+2\omega\,(z_2+z_2^{-1})-(\omega^2+\xi^2+4)
\end{array}\right..
\end{align}
Due to the boundary equation, 
\begin{eqnarray}
\label{upipeasy}
|u_2\rangle=(\xi+z_2-z_2^{-1})\begin{bmatrix} 1\\ 1\end{bmatrix} ,
\end{eqnarray}
the physical solutions are
\begin{align*}
\epsilon&\equiv \epsilon_{\rm right}(k_\parallel)=\xi=-2\sin k_\parallel,\\
z_2&=z_2(k_\parallel)=\frac{2}{\omega}= \Big(2-\frac{\mu}{2}-\cos k_\parallel \Big)^{-1}.
\end{align*}
This surface band is also chiral, but with the {\it opposite}
chirality to that of the left edge. The right surface band 
touches the bulk band at the pair of values of \(k_\parallel\) such 
that \(|z_2(k_\parallel)|=1\). These values of \(k_\parallel\), are the same 
as those computed for the surface band on the 
left edge, due 
to the fact that \(z_1(k_\parallel)=z_2(k_\parallel)^{-1}\). It is not obvious 
from comparing  Eqs.\,\eqref{rightedge} and \eqref{leftedge} 
that this basic relationship should hold, but the actual 
solutions do satisfy it. The (unnormalized) surface states 
are, for large \(N\),
\[
| \epsilon_{\rm right}(k_\parallel)\rangle=\sum_{j=1}^N
\Big(2-\frac{\mu}{2}-\cos k_\parallel \Big)^{-(j-N+1)}|k_\parallel\rangle |j\rangle\begin{bmatrix} 1\\ 1\end{bmatrix}.
\]

The root \(z_1(k_\parallel)\) (\(z_2(k_\parallel)\)) is entirely outside
(inside) the unit circle if \(\mu<0\) or \(\mu>8\). This is a direct
indication that the system does not host surface bands in these parameter regimes.
In Fig.\,\ref{fig:pip}, we show the surface bands for two
values of the chemical potential, one for each topologically 
non-trivial phase. The location of the surface bands in the 
Brillouin zone is not determined by the dispersion relation, 
which is itself independent of \(\mu\), but by the behavior of 
the wavefunctions as witnessed by \(z_1(k_\parallel)=z_2(k_\parallel)^{-1}\).

\subsubsection{Power-law zero modes}

Here we return to the basic model Hamiltonian with three parameters \(t,\Delta, \mu\). 
We consider a sheet of material rolled into a cylinder along the \(y\)-direction and half-infinite
in the \(x\)-direction. The virtual wires are 
\begin{align*}
H_{k_\parallel}=&\, 1\otimes h_{k_\parallel,0}+T\otimes h_{k_\parallel,1}
+T^\dagger \otimes h_{k_\parallel,1}^\dagger, \\
h_{k_\parallel,0}=&\begin{bmatrix}
-(\mu-4t)-2t\cos k_\parallel & -2\Delta \sin k_\parallel\\
-2\Delta \sin k_\parallel & (\mu-4t)+2t\cos k_\parallel
\end{bmatrix}, \\
h_{k_\parallel,1}=& 
\begin{bmatrix}
-t& \Delta\\
-\Delta & t
\end{bmatrix}.
\end{align*}
The crystal momenta \(k_\parallel=-\pi,0\) have special significance. Since
the off-diagonal entries of \(h_0\) vanish at these momenta, the virtual $D=1$ systems
can be interpreted as one-dimensional SCs. In particular, 
\begin{align*}
h_{0,0}=
\begin{bmatrix}
-(\mu-2t)& 0\\
0        & \mu-2t
\end{bmatrix}, \ 
h_{-\pi,0}=
\begin{bmatrix}
-(\mu-6t)  & 0\\
0        & \mu-6t
\end{bmatrix}
\end{align*}
and so the virtual chains \(H_{-\pi}\) 
and \(H_{0}\) are precisely the Majorana chain of Kitaev,
at two distinct values of an effective chemical potential \(\mu'=-(\mu-4t)\mp 2t\)
{\it for the chain}. We have investigated this paradigmatic system 
by analytic continuation in Refs.\,[\onlinecite{PRL,JPA,PRB1}].
If \(\mu<0\) or \(\mu>8t\), both chains are in their topologically
trivial regime. If \(0<\mu<4t\), then \(H_{0}\) is in the
non-trivial regime, but not \(H_{-\pi}\). The opposite is
true if \(4t<\mu<8t\). This analysis explains why is it that the fermionic parity of 
the ground state of the \(p+ip\) SC is odd in the weak pairing phase \cite{read00}, 
and suggests that one should expect surface bands crossing zero energy at 
\(k_\parallel=0\) (\(k_\parallel=-\pi\)) for \( 0<\mu<4\) (\(4<\mu<8\)). We already saw some
some of these bands in the previous section.   

Let us focus here on the virtual Kitaev chain at \(k_\parallel=0\). Its effective
chemical potential is \(\mu'=\mu-2t\). Suppose we are in a parameter regime
\[
4\Delta^2=\mu(4t-\mu),\quad 0<\mu<4t,
\]
of the full two-dimensional model. Then the \(H_{k_\parallel=0}\) virtual Kitaev chain
is in the topologically nontrivial parameter regime
\[\left (\frac{\mu'}{2t}\right)^2+\left(\frac{\Delta}{t}\right)^2=1, \quad -2t<\mu'<2t.\]
It is shown in Part I that the Majorana zero modes display an exotic power-law
profile in this regime. For the \(p+ip\) TSC these remarks imply the 
following power-law zero-energy surface mode:
\[
|\epsilon=0, k_\parallel=0\rangle=\sum_{j=1}^\infty\sum_{j_1=1}^{N_1} 
j\left(\frac{-2(t-\Delta)}{\mu-2t}\right)^{j}|j_1\rangle|j\rangle. 
\]

\subsection{Majorana flat bands in a gapless $s$-wave topological superconductor}
\label{2Dswavetoponductor}

A {\em gapless} SC is characterized by a vanishing single-particle excitation gap  at particular 
$\k$-points (or regions) of the Brillouin zone, whereas the SC order parameter 
remains non-vanishing. An example in $D=2$ was 
analyzed in Ref.\, [\onlinecite{Deng14}], where the nodeless 
character of the $s$-wave pairing in a two-band system was tuned 
to a gapless SC phase by introducing a suitable spin-orbit coupling. A remarkable 
feature of this system is the presence of zero-energy Majorana 
modes whose number grows with system size -- a {\em continuum} in the thermodynamic 
limit, namely, a Majorana flat band (MFB) -- 
as long as the system is subject to open BCs along one of the two spatial 
directions, but {\em not} the other.  This anomalous bulk-boundary correspondence was 
attributed to an asymmetric (quadratic vs. linear) closing of the bulk excitation gap near 
the critical momenta. In this section, we revisit this phenomenon and show that the 
indicator of bulk-boundary correspondence we introduced in Ref.\,[\onlinecite{PRL}] 
captures it precisely. Furthermore, in the phase hosting a MFB, we demonstrate 
by combining our Bloch ansatz with numerical root evaluation, that the characteristic 
length of the MFB wavefunctions diverges as we approach the critical values of 
momentum, similarly to what was observed in graphene [Eq. (\ref{locl})].
Finally, by comparing   
the equilibrium Josephson current in the gapless TSC to the one of a corresponding 
gapped model, we show how, similar to the case of the local DOS at the surface \cite{Deng14}, 
the presence of a MFB translates in principle into a substantial enhancement of the 
$4\pi$-periodic supercurrent. 
 
\begin{figure*}
\includegraphics[width = 17.5cm]{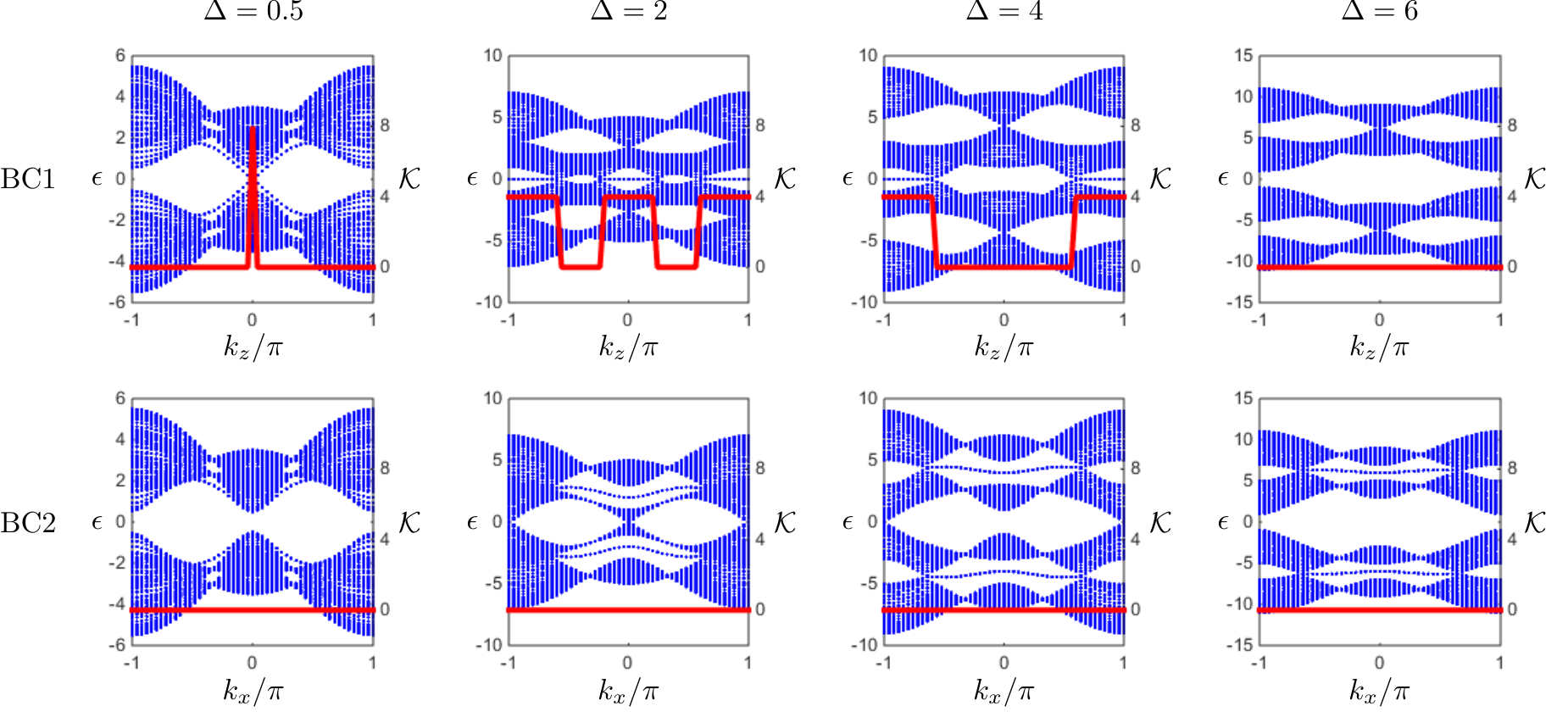}
\vspace*{-3mm}
\caption{(Color online) 
Energy spectrum (blue scatter plot) and degeneracy indicator $\mathcal{K}_{k_z}(0)$ for the zero 
energy level (red solid line) in the large-$N$ limit for BC1 (top panel) vs.  
BC2 (bottom panel) for various values of the SC pairing $\Delta$. The other 
parameters are $\mu=0$, $t=\lambda=u_{cd}=1$, 
$N_x = 120,$ $N_y=30$. 
\label{s-wavefig}}
\end{figure*} 

\subsubsection{Analysis of anomalous bulk-boundary correspondence\\ via boundary matrix}

The relevant model Hamiltonian in real space is
\[ \widehat{H}\! = \!\frac{1}{2}\sum_{\j}\Big(\hat{\Psi}^\dagger_{\j} h_{\bm{0}}\hat{\Psi}_{\j}-4\mu\Big) + 
\frac{1}{2}\sum_{\r=\hat{x},\hat{z}}\!\Big(\sum_{\j}\hat{\Psi}^\dagger_{\j} h_{\r}\hat{\Psi}_{\j+\r}+ \,\text{H.c.}\Big),
\]
with respect to a local basis of fermionic operators given by
$\hat{\Psi}^\dagger_\j \equiv \begin{bmatrix}c^\dagger_{\j,\uparrow} & c^\dagger_{\j,\downarrow}&
d^\dagger_{\j,\uparrow} & d^\dagger_{\j,\downarrow}&
c_{\j,\uparrow} & c_{\j,\downarrow}&
d_{\j,\uparrow} & d_{\j,\downarrow}\end{bmatrix}$. Here,  
\begin{eqnarray*}
h_{\bm{0}} &=& -\mu\tau_z + u_{cd}\tau_z\nu_z -\Delta \tau_x \nu_y \sigma_x,\\
h_{\hat{x}(\hat{z})} &=& -t\tau_z\nu_z +i\lambda \nu_x\sigma_{x(z)} ,
\end{eqnarray*}
with Pauli matrices $\tau_v, \nu_v, \sigma_v,\  v= x,y,z$ 
for the Nambu, orbital, and spin space, respectively. This Hamiltonian 
can be verified to obey time-reversal and particle-hole symmetry, as well as 
a chiral symmetry $U_K \equiv \tau_x\nu_z$. 
The topological response of the system was studied in Ref.\,[\onlinecite{Deng14}] using a 
$\mathds{Z}_2 \times \mathds{Z}_2$ indicator 
$(Q_{k_\parallel=0}, Q_{k_\parallel=\pi})$, where $Q_{k_\parallel}$ stands for the parity 
of the partial Berry phase sum for the value of transverse momentum \cite{Notenotation} $k_\parallel$.
The bulk-boundary correspondence of the system was studied subject to two 
different configurations: BC1, in which the system is periodic along $\hat{z}$ and
open along $\hat{x}$, and BC2, in which the system is periodic along $\hat{x}$
and open along $\hat{z}$. A MFB emerges along the open edges for BC1 in the 
phase characterized by $(Q_{k_z=0}, Q_{k_z=\pi})=(1,1)$. No MFB exists in 
the configuration BC2.

To shed light into this anomalous bulk-boundary correspondence using our
generalized Bloch theorem framework, consider first the configuration BC1. 
Then, if $N_x$ denotes the size of the lattice along the $\hat{x}$ direction, 
$\widehat{H}$ decouples into $N_x$ virtual wires, parametrized by the transverse 
momentum $k_z$. These virtual $D=1$ Hamiltonians have the form 
\begin{eqnarray*}
H_{k_z,N_x} & =&  \frac{1}{2}\sum_{j=1}^{N_x}\Big(\hat{\Psi}^\dagger_{j,k_z} 
h_{k_z,0}\hat{\Psi}_{j,k_z}-4\mu\Big)  \\
&+& \frac{1}{2} \sum_{j=1}^{N_x-1} \! \Big(\hat{\Psi}^\dagger_{j,k_z} h_{k_z,1}\hat{\Psi}_{j+1,k_z}+\text{H.c.}\Big),
\end{eqnarray*}
where \( h_{k_z,0} \equiv h_{\bm{0}} + (e^{ik_z}h_{\hat{z}} + \text{H.c.})$ and $h_{k_z,1} \equiv h_{\hat{x}}. \)
The {\em total} number of Majorana modes hosted by each such chain (on its two ends) 
is given by the degeneracy indicator introduced in Part I [Sec. VI], namely, 
\( \mathcal{K} (0) \equiv  \text{dim}\ \text{ker} [{B}_\infty(0)], \)
where ${B}_\infty(0)$ is the boundary matrix in the large-$N$ 
limit that we obtain after appropriately rescaling the extended bulk solutions
corresponding to $|z_\ell|>1$, and removing the un-normalizable extended
solutions corresponding to $|z_\ell|=1$. 
We calculate the above degeneracy indicator $\mathcal{K}(0) \equiv \mathcal{K}_{k_z}(0)$ for each wire 
parametrized by $k_z$, by evaluating the boundary matrix numerically. 
Representative results are shown in the top panel of Fig.\,\ref{s-wavefig}.
When the system is in a phase characterized by $(Q_{k_z=0}, Q_{k_z=\pi})=(1,-1)$ ($\Delta=2$) 
and $(Q_{k_z=0}, Q_{k_z=\pi})=(-1,-1)$ ($\Delta=4$) there are $\mathcal{O}(N)$ chains, each of 
them hosting four Majoranas (two pairs per edge). This is reflected in the four-fold degeneracy 
for a continuum of values of $k_z$. The values of $k_z$ at which the excitation gap closes are also the points 
at which the indicator changes its nature. 

The same analysis may be repeated for BC2, in which case periodic BCs are imposed 
along $\hat{x}$ instead. 
The resulting virtual $D=1$ systems are now parametrized by $k_x$,
with explicit expressions for the internal matrices given by
\(h_{k_x,0} = h_{\bm{0}} + (e^{ik_x}h_{\hat{x}} + \text{H.c.})\) and \(h_{k_x,1} = h_{\hat{z}}.\)
In the BC2 configuration, the degeneracy indicator remains zero, showcasing
the absence of MFBs, see bottom panel of Fig.\,\ref{s-wavefig}. 

\subsubsection{Penetration depth of flat-band Majorana modes}

\begin{figure}
\includegraphics[width=0.85\columnwidth]{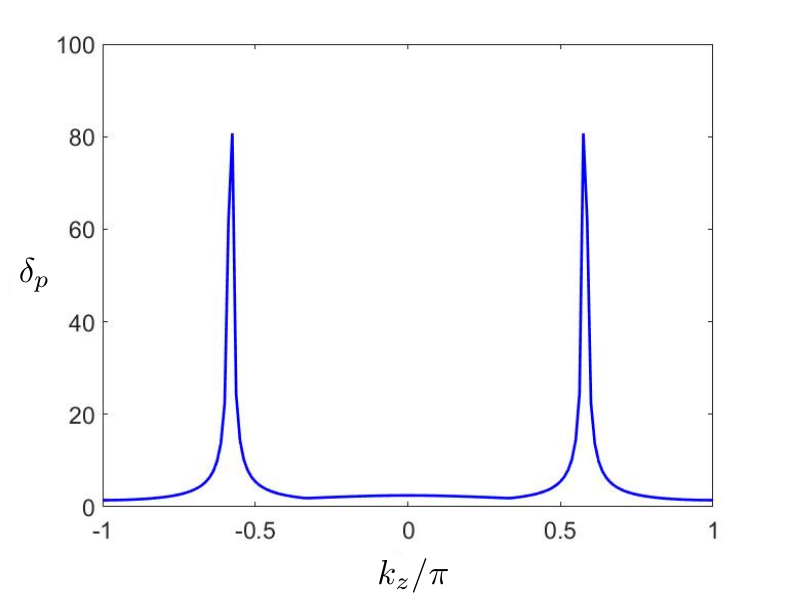}
\vspace*{-2mm}
\caption{(Color online) Penetration depth (in units of the lattice constant) of 
flat-band Majoranas as a function of $k_z$.
The parameters are  
$\mu=0$, $u_{cd}=t=\lambda=1$, $\Delta=4$.
\label{fig:pendep}}
\end{figure}

Whether and how far the Majorana modes in the flat band penetrate in
the bulk is important from the point of view of scattering. Our generalized Bloch theorem 
allows us to obtain a good estimate of the penetration depth without diagonalizing the system.
In the large-$N$ limit, the wavefunction corresponding to a Majorana mode for a single wire described by 
$H_{k_z,N_x}$ must include left emergent solutions and decaying extended solutions, so that
\[ |\epsilon=0\rangle 
= 
\sum_{s=1}^{s_0}\alpha_s^-|\psi_{k_z s}^{-}\rangle+
\sum_{|z_\ell|<1}\sum_{s=1}^{s_\ell}\alpha_{\ell s}|\psi_{k_z\ell s}\rangle,
\]
for complex amplitudes $\{\alpha_s^-, \alpha_{\ell s}\}$.
The emergent solutions are perfectly localized, and so the penetration depth 
is determined by the extended solutions only. The latter are labeled by the roots $\{z_\ell\}$, 
computed at $\epsilon=0$, of the polynomial equation 
$z^{dR}\det (H_{k_z}(z)-\epsilon\mathds{1}_8)=0$, 
which is the dispersion relation. Each extended solution $|\psi_{k_z\ell s}\rangle$ 
corresponding to the root $z_\ell,\ |z_\ell|<1$ has penetration depth $(-\ln |z_\ell|)^{-1}$.
A useful estimate of the penetration depth $\delta_p$ of a zero energy mode may then 
by obtained by taking the maximum of the individual penetration depths of the bulk solutions 
\cite{Lee81}, leading to the expression 
\[\delta_p \equiv (-\ln |z_p|)^{-1},\quad |z_p| \equiv \max\,\{|z_\ell|,\ |z_\ell|<1 \} . \]
Since the roots $\{z_\ell\}$
depend on the value of the transverse momentum $k_z$, so does 
the penetration depth $\delta_p$. As seen in Fig.\,\ref{fig:pendep}, 
the Majoranas penetrate more inside the bulk near the critical values of the transverse 
momentum, where the excitation gap closes. At these points, the penetration 
depth diverges, signifying that the corresponding Majorana excitations become 
part of the bulk bands.

\subsubsection{Impact of a Majorana flat band on Josephson current}

Beside resulting in an enhanced local DOS at the surface \cite{Deng14}, 
one expects that the MFB may impact the nature of the equilibrium (DC) 
Josephson current at zero temperature. We now show (numerically) that the 
Josephson current flowing through a strip of finite width is $4\pi$-periodic, 
irrespective of the width of the strip. This is at variance with the behavior 
expected for a gapped $D=2$ $s$-wave TSC, in which case the $4\pi$-periodic
contribution resulting from a fixed number of Majorana modes is washed 
away once the strip width becomes large.
   
We model a SNS
junction of the SC under investigation by letting 
the normal part be a weak link with the same type of hopping, spin-orbit coupling 
and hybridization as the SC, but weaker by a factor of $w=0.2$. 
The DC Josephson current can be calculated using the formula \cite{lesovik11}
\[ 
I(\phi) = \frac{2e}{\hbar}\frac{\partial E_0}{\partial \phi} = 
-\frac{2e}{\hbar}\sum_{\epsilon_n>0}\frac{\partial \epsilon_n}{\partial \phi}, 
\]
where $E_0$ is the energy of the many-body ground state, 
$\epsilon_n$ are single-particle energy levels, and $\phi$ is the SC phase 
difference (or flux). As $\phi$ is varied, 
at the level crossings 
of low-lying energy levels with the many-body ground state associated 
with the $4\pi$-periodic effect, the system continues in the state which respects 
fermionic parity and time-reversal symmetry in all the virtual wires. 

\begin{figure}
\centering
\includegraphics[width = 8cm]{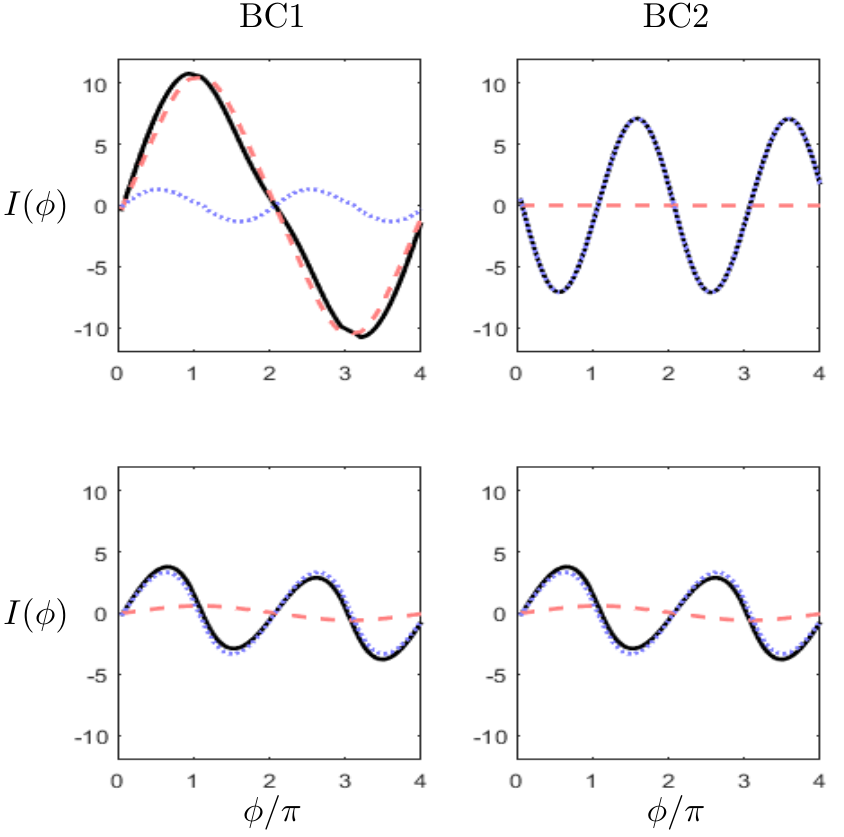}
\vspace*{-2mm}
\caption{(Color online) Total Josephson current $I(\phi)$ (black solid line),
$2\pi$-periodic component $I_{2\pi}(\phi)$ (blue dotted line) and 
$4\pi$-periodic component $I_{4\pi}(\phi)$ (red dashed line) in units of $2e/\hbar$, 
as a function of flux $\phi$. Top (bottom) panels correspond to the gapless (gapped) model of 
a $D=2$ $s$-wave TSC, whereas left (right) panels correspond to BC1 (BC2), respectively. 
The parameters used for both models 
are $\mu=0$, $u_{cd}=t=\lambda=1$, $\Delta=4$, $N_x=N_z=60$. 
\label{Josephson}}
\end{figure}

The upper panels of Fig.\,\ref{Josephson} show the Josephson response $I(\phi)$ 
of the gapless TSC under the two BCs. While in the 
BC1 configuration the behavior of the current $I(\phi)$ (solid black line) 
is $4\pi$-periodic, the BC2 configuration displays standard $2\pi$-periodicity, reflecting 
the presence of the MFB {\em only} under BC1. The lower panels of Fig.\,\ref{Josephson} 
show the Josephson response of the gapped $s$-wave TSC model
introduced and analyzed in Ref.\,[\onlinecite{swavePRL,swavePRB}]. It can be seen that 
the Josephson current is now identical under BC1 and BC2, as expected 
from the fact that a standard bulk-boundary correspondence is in place. 

Let us separate the total Josephson current $I(\phi)$ into 
$2\pi$- and $4\pi$-periodic components by letting $I(\phi) = I_{2\pi}(\phi) + I_{4\pi}(\phi)$, 
with 
\[I_{2\pi}(\phi) \equiv \left\{ 
\begin{array}{lcl}
\frac{1}{2}[I(\phi)+I(\phi+2\pi)] & \text{if} & 0\le \phi < 2\pi\\[2pt] 
\frac{1}{2}[I(\phi)+I(\phi-2\pi)] & \text{if} & 2\pi \le \phi < 4\pi
\end{array} \right.,\]
\vspace{-0.4cm}
\[I_{4\pi}(\phi) \equiv \left\{ 
\begin{array}{lcl}
\frac{1}{2}[I(\phi)-I(\phi+2\pi)] & \text{if} & 0\le \phi < 2\pi\\[2pt] 
\frac{1}{2}[I(\phi)-I(\phi-2\pi)] & \text{if} & 2\pi \le \phi < 4\pi
\end{array} \right.,\]
In the four panels of Fig.\,\ref{Josephson}, the $2\pi$- and $4\pi$-periodic 
components are individually shown by (blue) dotted and (red) dashed lines, 
respectively.
The nature of the supercurrent in the gapped TSC 
(lower panels) is predominantly $2\pi$-periodic, with only a small 
$4\pi$-periodic component due to the presence of a finite number of 
Majoranas (two per edge). Further numerical simulations (data not shown) 
reveal that the amplitude of the $2\pi$-periodic current relative to the 
$4\pi$-periodic current increases linearly with the width of the strip, 
so that for large strip width, the Josephson current is essentially 
$2\pi$-periodic. The origin of such a degradation of the $4\pi$-periodicity
lies in the fact that the number of Majorana modes is constant,  
irrespective of the width of the strip, as only one virtual wire hosts 
Majorana modes in this gapped model. Since only the Majorana modes can support
$4\pi$-periodic current, their contribution relative to the extensive $2\pi$-periodic
current arising from the bulk states diminishes as the strip width becomes large. 
In contrast, for the gapless TSC in the MFB phase (top panels), the number of virtual wires hosting
Majorana modes grows linearly with the width of the strip in the BC1 configuration.
This leads to an {\em extensive contribution from the $4\pi$-periodic component}, 
which may be easier to detect in experiments.

\section{Summary and Outlook}
\label{outlook}

As mentioned in the Introduction, this paper constitutes the sequel, Part II, to 
Ref.\,[\onlinecite{PRB1}], where we introduced a generalization of Bloch's theorem 
for arbitrary boundary conditions. In clean
systems translation symmetry is only broken by surface terminations and boundary constraints 
that encode physical or 
experimental conditions. The conventional Bloch theorem is not in force
because translational symmetry is explicitly broken. However, since such a symmetry is only
mildly broken, one wonders whether one can one continue to label single-particle electronic excitations  
in terms of some kind of ``generalized momenta". Our generalized Bloch theorem \cite{PRL,PRB1} 
provides a precise answer to that question. The mathematical framework makes the idea of 
approximate translation precise by relating the spectral properties of certain shift operators
to non-unitary representations of the group of translations \cite{JPA}.  According to the 
generalized Bloch theorem, the exact eigenstates of a clean system of independent 
fermions with terminations are linear combinations of eigenstates of non-unitarily represented translations. 
It is because of this lack of unitarity that complex momenta arise. The latter leads to the emergence of 
localized edge modes and more involved power-law corrections to the Bloch-like wavefunctions. 
The amplitudes that weigh the relative contribution of the generalized Bloch states to the 
exact energy eigenstates are determined by a boundary matrix. This piece of our formalism, 
the {\em boundary matrix}, optimally combines information about the translation-invariant bulk 
and the boundary conditions: it allows one to compactly parametrize the manifold of boundary 
conditions and may eventually suggest new ways of accessing effective edge theories. 

Part II focused on presenting two new theoretical developments and several non-trivial 
applications to higher-dimensional systems. New developments include the extension of the 
generalized Bloch theorem formalism to incorporate:
(1) Surface reconstruction and surface disorder; and (2) Interface physics involving multiple bulks. 
Within our framework, boundary conditions for \(D\)-dimensional systems must be imposed 
on two parallel hyperplanes, but are otherwise arbitrary. Thus, the generalized Bloch theorem
yields highly-effective tools for diagonalizing systems subject to anything from pristine 
terminations to surface relaxation, reconstruction and disorder. The extension to interfaces 
between multiple bulks allows us to study arbitrary junctions, including interface modes 
resulting from putting in contact two exotic topologically non-trivial bulks. 

It is interesting to digress on what happens when one tries to formulate a generalized Bloch
theorem for clean systems cut into hypercubes. The bulk-boundary separation goes through 
essentially unchanged: for example, the range of the boundary projector consists of a 
hypercubic surface layer of thickness determined by the bulk structure of the system. The 
challenge in higher dimensions is solving the bulk equation explicitly and in full generality. 
It is a worthy challenge, because it would yield insight into the plethora of corner states that 
can appear in such systems \cite{benalcazar17,hashimoto17,flore18}. While special 
cases may still be able to be handled on a case-by-case basis, in general we see little hope
of using the same mathematical techniques (crucially, the Smith decomposition \cite{JPA}) 
that work so well in our setup. In general, the analytic continuation of the Bloch Hamiltonian become 
then a matrix-valued analytic function of \(D\) complex variables. The passage from one complex 
variable to several makes a critical difference. 

We have illustrated our formalism with several applications to models of  
current interest in condensed matter physics. Table \ref{MainTable} 
summarizes all systems that we have solved so far by our techniques, 
where exact analytic solutions were unknown prior to our 
findings,  to the best of our knowledge. 
For example, we  showed that it is possible to {\em analytically} determine 
Andreev bound states for an idealized SNS junction.  
More importantly, the existence of power-law modes
would not have been unveiled without our mathematical formalism. 
Among the challenging applications presented in this paper, we investigated
in detail the Creutz ladder system, where thanks to a 
Gaussian duality \cite{equivalence}, we can map this topological insulator 
to a pair of coupled Kitaev Majorana chains.  The presence 
of power-law topological modes in the Creutz ladder insulator is noteworthy, 
see Sec.\,\ref{creutzladder}. We also find power-law modes on the surface of 
the \(p+ip\) chiral superconductor as part of our closed-form full calculation
of the surface states of this system, see Sec.\,\ref{pwavetoponductor}. It 
seems reasonable now to accept that power-law modes, topological or otherwise, 
are a general, if fine-tuned, feature of {\it short range} tight-binding models. 
We have also included applications 
to other $D=2$ systems, such as the full closed-form diagonalization of graphene 
ribbons for zigzag-bearded and armchair surface terminations. While the edge
modes for zigzag-bearded graphene have been computed before in closed form,
the closed-form band states appear to be new in the literature. It seems a distinctive
feature of the generalized Bloch theorem that {\em both} edge and bulk bands can 
be treated analytically on equal footing. Finally, we investigated in detail the Majorana 
flat bands of the gapless $s$-wave topological superconductor we previously introduced 
\cite{Deng14}. There, we find an extensive contribution of the surface Majorana flat band to the 
$4\pi$-periodic component of the Josephson current,  which would serve as a smoking gun
for experimental detection should a candidate material realization be identified.

In view of these results it seems fair to grant that
the generalized Bloch theorem bestows a higher level of control over surface and interface 
physics, and opens the door for a deeper investigation of the interplay between surface/interface 
and bulk critical phenomena \cite{book8,quelle15,kempkes16}. Let us conclude by recalling a main 
motivation behind the formulation of our generalized Bloch theorem. That motivation was to investigate
the bulk-boundary correspondence in {\it boundary space}, that is, the space of boundary conditions, 
as opposed to the usual parameter space, in order to quantitatively express stability and  
robustness in this new space that clearly affects boundary invariants most directly 
\cite{prodanBook}. Physically, boundary conditions are idealized representations of interfaces
between the system of interest and an ``environment" that we choose not to characterize, 
and so they capture matching conditions that can have a big impact on the 
energy spectrum of the system. This interpretation suggests that it might be very
illuminating to bring closer together precise mathematical ideas of stability and robustness from quantum 
information processing and control engineering, and more qualitative concepts in condensed matter physics. 
We have not carried out this systematic task in this paper which is, strictly speaking, still an exploration of the 
power of the generalized Bloch theorem. We will return to the study of the relation between boundary and bulk 
topological invariants in future publications.

\section*{Acknowledgements}

Work at Dartmouth was partially supported by the NSF through Grant No. 
PHY-1066293 and the Constance and Walter Burke Special Projects Fund in Quantum 
Information Science.

\appendix

\section{A criterion for the absence of localized eigenstates} 
\label{app:condition}

Symmetry conditions paired with suitable BCs can exclude completely edge modes, topological
or otherwise. We have identified one particularly useful sufficient 
condition that guarantees the absence of edge modes. 
It relies on the analytic diagonalization of the matrices 
\[ T_\theta+T_\theta^\dagger\equiv  e^{-i\theta}T+e^{i\theta}T^\dagger , \quad \theta 
\in [0, 2\pi).  \]
Physically, the phase $\theta$ may arise from \(\k_\parallel\),  
see for example Sec.\,\ref{armpitsec}, or one may think of \(\theta\)
as an applied electric field. 
The combination \(T+T^\dagger\) is singled out by a symmetry argument.
The \(\mathds{Z}_2\) mirror symmetry,  
\[ U_{\sf m} \equiv \sum_{j=1}^N|N-j+1\rangle\langle j|,\quad U_{\sf m}^\dagger=U_{\sf m}^{-1}=U_{\sf m}, \]
exchanges the two shift operators, \(U_{\sf m}TU_{\sf m}=T^\dagger\), so that
\[ U_{\sf m} (T+T^\dagger)U_{\sf m}=T^\dagger+T. \]
The eigenstates and eigenvalues of \(T+T^\dagger\) are known \cite{MouraSin}, 
and were recomputed by way of the generalized Bloch theorem in Part I (see Sec. V A therein):
\[
(T+T^\dagger)|k_q\rangle=2\cos\big(\frac{\pi q}{N+1}\big)|k_q\rangle ,\quad q=1,\dots,N , \]
with unnormalized eigenvectors
\[ |k_q\rangle=\sum_{j=1}^N\sin\big(\frac{\pi qj}{N+1}\big)|j\rangle  .
\]
Let \(X \equiv \sum_{j=1}^N j \, |j\rangle\langle j|\) denote the position operator. As 
explained in Part I (see Appendix B), \([X,T]=-T\), and thus 
\(e^{i\theta X}Te^{-i\theta X}=e^{-i\theta}T\). 
In particular,
\[
e^{i\theta X}(T+T^\dagger)e^{-i\theta X}=e^{-i\theta}T+e^{i\theta}T^\dagger.
\]
It follows that the eigenstates of \( T_\theta+T_\theta^\dagger\)
are given by 
\[
|k_q,\theta\rangle=\sum_{j=1}^N\sin\big(\frac{\pi qj}{N+1}\big)e^{i\theta j}|j\rangle,\quad 
q=1,\dots,N.
\]

Assume now that {\em all} the matrices ${h}_r$ entering the single-particle Hamiltonian of interest 
satisfy the relation 
\[h_r^\dagger=e^{i 2r\theta} h_r ,\]
for some choice of \(\theta\), that is, 
\[
H =\mathds{1}_N\otimes h_0+\sum_{r=1}^R(T_\theta^r+T_\theta^{\dagger\,r})\otimes e^{ i r\theta}h_r + W.  \] 
Then, it is easy to see that $H$ rewrites as
\[
H=
\mathds{1}_N\otimes h_0+
\sum_{r=1}^R (T_\theta+T_\theta^{\dagger})^r\otimes  \tilde{h}_r+W'+W, 
\]
in terms of new hopping matrices \(\tilde{h}_r\) and boundary contribution \(W'\) with the 
{\em same} range finite $R$ (for example, \((T+T^\dagger)^3=T^3+3T-|1\rangle\langle 2|-|N-1\rangle\langle N|+{\rm H.c}.\)). 
If the original BCs are such that $W=-W'$, then
$H$  can be expressed as a function of 
$T_\theta+T_\theta^{\dagger}$. It follows that no localized eigenstate can exist. This is exactly the situation
for armchair graphene, see Sec.\,\ref{armpitsec}.

\section{The BCS chain} 
\label{appBCS}

A tight-binding BCS chain with $N$ lattice sites can be modeled in terms of the 
Hamiltonian \cite{bena12}
\begin{eqnarray*}
\widehat{H}&=&-\sum_{j,\sigma}(tc^\dagger_{j\sigma}c_{j+1 \sigma}
+\frac{\mu}{2}c^\dagger_{j\sigma}c_{j\sigma}+\text{H.c.}) \\
&-& \sum_j(\Delta
c_{j\uparrow}^\dagger c_{j\downarrow}^\dagger+\text{H.c.}).
\end{eqnarray*}
The single-particle Hamiltonian associated to \(\widehat{H}_{\sf S} \) is
\[ H_N=\mathds{1}_N\otimes h_0+(T+T^\dagger)\otimes h_1 ,
\]
where we assume open BCs, with  
\[
h_0=-\mu\tau_z\otimes \mathds{1}_2-\Delta \tau_y\otimes \sigma_y,\quad
h_1=-t\tau_z\otimes \mathds{1}_2. 
\]
$H_N$ commutes with 
$\mathcal{S} = \mathds{1}_N\otimes \mathds{1}_2\otimes \sigma_y$
because total spin is conserved. Thus, following the discussion in Sec. \ref{sub:gbt}, 
we can block-diagonalize $H_N$ as
\[
H_N=\sum_{{\rm s}=\pm 1}H_{N,{\rm s}}\otimes|{\rm s}\rangle\langle {\rm s}|,
\] 
where \(|{\rm s}\rangle\) denotes the eigenstate of \(\sigma_y\) 
for the eigenvalue \({\rm s}=\pm 1\). The internal matrices for 
\(H_{N,{\rm s}}\) are 
\[
h_{{\rm s},0}=-\mu\tau_z-{\rm s}\, \Delta \tau_y,\quad 
h_{{\rm s},1}=-t \tau_z,  \] 
and the action of the particle-hole symmetry on the blocks is 
\begin{align}
&\mathcal{P}H_{N,{\rm s}}\otimes |{\rm s}\rangle\langle {\rm s}|
\mathcal{P}^{-1}= 
\label{beN} \\
&=\tau_x H_{N,s}^*\tau_x\otimes (|{\rm s}\rangle\langle {\rm s}|)^*
=-H_{N,-{\rm s}}|-{\rm s}\rangle\langle-{\rm s}|. \nonumber
\end{align}
Hence, the two blocks are exchanged by particle-hole symmetry, whereas the 
full Hamiltonian only changes sign. Note that, 
taken individually, these blocks do {\em not} respect the particle-hole symmetry 
because of Eq.\,\eqref{beN}. Therefore, the many-body Hamiltonian
does {\it not} decouple into two blocks.

The nontrivial spatial structure of each of the two blocks
is encoded in the matrix \(T+T^\dagger\).  According to Appendix \ref{app:condition},
this fact suffices to guarantee the absence of edge modes and 
goes a long way towards analytic solvability. For open
BCs the eigenstates are 
\begin{eqnarray*}
|\epsilon_{n,q},{\rm s}\rangle=
\sum_{j=1}^N\sin(\frac{\pi q j}{N+1})|j\rangle\begin{bmatrix}  i \rm{s}\Delta \\ \epsilon_{n,q}+\mu+2t\cos(\frac{\pi q}{N+1})
\end{bmatrix},
\end{eqnarray*}
with $q=1,2,\dots,N$ and \(n=1,2\) the band index for spin \(\rm{s}\) along the \(y\) direction.
The energy \(\epsilon_{n,q}\) satisfies the relation
\begin{eqnarray}
\label{BCSdispersion}
\epsilon_{n,q} =(-1)^n \sqrt{\Big(\mu+2t\cos (\frac{\pi q}{N+1})\Big)^2+|\Delta|^2}.
\end{eqnarray}

\section{An SNS junction}
\label{snsApp}

With reference to Sec. \ref{interfaces}, our aim is to find the exact Andreev bound 
states that form on the normal region. The block-diagonalization in spin space reduces to
solving the boundary value problem for the blocks with reduced internal space. 
Because of Eq.\,\eqref{beN}, note that each spin block does {\em not} individually 
describe an SNS junction Hamiltonian.
The SNS junction is modeled as the system formed by attaching a finite metallic {\sf N} chain to two 
semi-infinite SC chains, {\sf S1} and {\sf S2}, with the 
length of the metallic chain being $N=4{\sf L}-1$ for some positive integer ${\sf L}$. 
The projectors corresponding to the left and right semi-infinite {\sf S1} and {\sf S2} regions are 
\[
P_1 = \sum_{j=-\infty}^{-2{\sf L}}|j\rangle\langle j|,\quad P_2 = \sum_{j=2{\sf L}}^{\infty}|j\rangle\langle j|,  \]
whereas the region {\sf N} is finite with 
an associated projector
\[ P_3 = \sum_{j=-2{\sf L}+1}^{2{\sf L}-1}|j\rangle\langle j|. \] 
The links connecting the SC regions {\sf S1, S2} to the metal
region \text{\sf N} at $j=-2{\sf L}$ and $j=2{\sf L}-1$ have weaker hopping strength
$t'$, and we set the chemical potential $\mu\equiv 0$. 
The metallic chain is therefore modeled by only the NN hopping of strength $t$, 
and links to the two SC leads by way of a hopping amplitude $t' <t$. The 
Hamiltonian of the full system is $\widehat{H}_{\sf SNS}=\widehat{H}_{\sf S1}+
\widehat{H}_{\sf S2}+\widehat{H}_{\sf T}+ \widehat{H}_{\sf N},$ with 
the SC and tunneling Hamiltonians given in Eqs. (\ref{BCS})-(\ref{Htunnel}) in the 
main text. Note that the relevant matrices $h_0,h_1$ for the metal part can be obtained from the 
ones for the SC part (in Appendix \ref{appBCS}) by setting $\Delta=0$. 

The single-particle Hamiltonian $H_{\sf SNS}$
of the junction is block-diagonalized in the basis of the spin operator $\sigma_y$, and the 
two blocks are related to each other by the particle-hole symmetry in the 
same way as described by Eq.\,\eqref{beN}. Let us focus on the 
${\rm s}=+1$ block, and denote it by $H_+$. This system has three translation-invariant regions 
(bulks) connected by two internal boundaries. The energy eigenvector
ansatz in this case is obtained by extending the ansatz in Eq.\,\eqref{ansatzmultbulk} 
in the main text to a system of three bulks. 

Consider first the case with no phase difference between the two 
SC leads {\sf S1} and {\sf S2}, that is, $\Delta_1=\Delta_2=\Delta$ for a real value of $\Delta$. 
Note that $H_+$ obeys a mirror symmetry about $j=0$, 
\[
\mathcal{S}_1 = \sum_{j \in \mathds{Z}} |-j\rangle\langle j | \otimes \mathds{1}_2 , \]
and another local symmetry,
\[
\mathcal{S}_2 = \sum_{j \in \mathds{Z}} (-1)^j|j\rangle\langle j | \otimes \tau_y .
\]
Since we are only interested in the states bound on the metal {\sf N} region, we restrict the value of
energy to be in the band gap of the SCs, which is $(-\Delta,\Delta)$.
For these bound states to carry a superconducting current, they must be of
extended nature on the metallic region, which is allowed by energies such that 
$|\epsilon|<|t|$.The eigenstate ansatz for any such energy in each of the 
three bulks will be in terms of the roots of Eq. \eqref{BCSdispersion}, with $\mu=0$. 
Noting appropriate symmetries of the polynomial, we denote the four roots 
in the bulks of {\sf S1} and {\sf S2} by $\{z_1,z^{-1}_1,-z_1,-z^{-1}_1\}$. 
Without loss of generality, we can choose $|z_1|>1$ and $t(z_1+z^{-1}_1)=
i {\sf D}$, with ${\sf D} \equiv \sqrt{\Delta^2-\epsilon^2}$.
From Eq. \eqref{BCSdispersion} and the above constraints, we find that
\begin{equation}
\label{snsz}
z_1 =-\frac{{\sf D}+\sqrt{{\sf D}^2+4t^{2}}}{2it} .
\end{equation}
For an exponentially decaying mode in the {\sf S1} and {\sf S2} region, the ansatz is given by 
\begin{multline*}
P_1|\epsilon,{\rm s}_1,{\rm s}_2,\bm{\alpha}\rangle = \alpha_{1}
\Big(P_1|z_1,1\rangle\begin{bmatrix}i\Delta \\
\epsilon+i{\sf D}
\end{bmatrix} \\+ {\rm s}_2 P_1|-z_1,1\rangle\begin{bmatrix}{\sf D}-i\epsilon\\
-\Delta
\end{bmatrix}\Big), 
\end{multline*}
\begin{multline*}
P_2|\epsilon,{\rm s}_1,{\rm s}_2,\bm{\alpha}\rangle  =\alpha_{1}\Big({\rm s}_1P_2| \frac{1}{z_1},1\rangle\begin{bmatrix}i\Delta\\
\epsilon+i{\sf D}
\end{bmatrix}  \\ + {\rm s}_1{\rm s}_2 P_2|-\frac{1}{z_1},1\rangle\begin{bmatrix}{\sf D}-i\epsilon\\
-\Delta \end{bmatrix}\Big),
\end{multline*}
respectively, where ${\rm s}_1,{\rm s}_2$ denote eigenvalues of symmetries
$\mathcal{S}_1$ and $\mathcal{S}_2$ respectively. 
For the metallic region, since $\Delta=0$, all four roots lie
on the unit circle. We denote them by 
$\{w_1,w^{-1}_1,-w_1,-w^{-1}_1\}$, with the convention 
$t(w_1 + w_1^{-1}) = -\epsilon$.
Then the ansatz for the {\sf N} region can be written as
\begin{multline*}
P_3|\epsilon,{\rm s}_1,{\rm s}_2,\bm{\alpha}\rangle=
\alpha_{2}\big(P_3|w_1,1\rangle+{\rm s}_1P_3|1/w_1,1\rangle\begin{bmatrix}1\\0
\end{bmatrix}+\\
{\rm s}_2P_3|-w_1,1\rangle+{\rm s}_1{\rm s}_2P_3|-1/w_1,1\rangle\big)\begin{bmatrix}0\\i
\end{bmatrix}.
\end{multline*}

Therefore, we have obtained four eigenstate ans\"{a}tze corresponding to 
the four cases $\{{\rm s}_1=\pm1,\ {\rm s}_2=\pm1\}$,
which we denote by $|\epsilon,{\rm s}_1,{\rm s}_2,\bm{\alpha}\rangle$,
where $\bm{\alpha} = [\alpha_{1} \; \alpha_{2}]^{\rm T}$ are the free parameters.
The BCs are provided by the weak links,
that is, $j = \pm2{\sf L}, \pm(2{\sf L}-1)$.
We choose the basis $\{|2{\sf L},{\rm s}_1,{\rm s}_2\rangle,\ 
|2{\sf L}-1,{\rm s}_1,{\rm s}_2\rangle,\ {\rm s}_1,{\rm s}_2 = 1,-1\}$
of the boundary subspace, where
\begin{eqnarray*}
|2{\sf L},{\rm s}_1,{\rm s}_2\rangle \!&\equiv& \frac{1}{2}(|-2{\sf L}\rangle+{\rm s}_1|2{\sf L}\rangle)\begin{bmatrix}1 \\ 
i{\rm s}_2 \end{bmatrix},\\
|2{\sf L}-1,{\rm s}_1,{\rm s}_2\rangle\! &\equiv& \frac{1}{2}(|-2{\sf L}+1\rangle+
{\rm s}_1|2{\sf L}-1\rangle)\begin{bmatrix}1 \\ -i{\rm s}_2
\end{bmatrix}\!.
\end{eqnarray*}
There will be four boundary matrices $B(\epsilon,{\rm s}_1,{\rm s}_2)$, 
for ${\rm s}_1,{\rm s}_2=1,-1$, arising from the equations
\begin{eqnarray*}
\langle 2{\sf L},{\rm s}_1,{\rm s}_2| (H-\epsilon\mathds{1}_2)) 
|\epsilon,{\rm s}_1,{\rm s}_2,\bm{\alpha}\rangle = 0,\\
\langle 2{\sf L}-1,{\rm s}_1,{\rm s}_2|(H-\epsilon\mathds{1}_2)) 
|\epsilon,{\rm s}_1,{\rm s}_2,\bm{\alpha}\rangle = 0.
\end{eqnarray*}
The boundary matrix corresponding to $({\rm s}_1,{\rm s}_2)$ is
\begin{multline*}
B(\epsilon,{\rm s}_1,{\rm s}_2) = \\
\begin{bmatrix}
tz_1^{-2{\sf L}+1}(i\Delta -{\rm s}_2({\sf D} - i\epsilon)) & -t'(w_1^{-2{\sf L}+1}+s_1w_1^{2{\sf L}-1})\\
-t'z_1^{-2{\sf L}}(i\Delta +{\rm s}_2({\sf D} - i\epsilon)) & t(w_1^{-2{\sf L}}+s_1w_1^{2{\sf L}})
\end{bmatrix}.
\end{multline*}
In writing the above, we made use of the identity 
$$(h_0-\epsilon\mathds{1} +z_\ell h_1)|u_\ell\rangle = -z_\ell^{-1}h_1^\dagger|u_\ell\rangle,$$ 
which follows from the bulk equation.
The condition for non-trivial kernel of the boundary matrix in the four cases
leads, after simplification using Eq.\,\eqref{snsz}, to the following four boundary equations:
\begin{subequations}
\begin{eqnarray}
\label{snsboundaryeq1}
-\left(\frac{t'}{t}\right)^{2}\!\!
\left(\frac{2t}{\epsilon +{\rm s}_2 \Delta}\right)
\left(1 + \sqrt{1+\frac{4t^2}{\Delta^2-\epsilon^2}}\right)^{-1}= 
\nonumber \\
\frac{\cos k(2{\sf L}-1)}{\cos k(2{\sf L})},
\quad \text{if} \quad {\rm s}_1 = +1, {\rm s}_2 = \pm 1 \qquad \\
-\left(\frac{t'}{t}\right)^{2}\!\!
\left(\frac{2t}{\epsilon +{\rm s}_2 \Delta}\right)
\left(1 + \sqrt{1+\frac{4t^2}{\Delta^2-\epsilon^2}}\right)^{-1} =
\nonumber \\
\frac{\sin k(2{\sf L}-1)}{\sin k(2{\sf L})},
\quad \text{if} \quad {\rm s}_1 = -1, {\rm s}_2 =\pm 1, \qquad 
\label{snsboundaryeq2}
\end{eqnarray}
\end{subequations}
where $e^{ik} \equiv w_1$. 
Whenever any one of these conditions is satisfied,
$\epsilon$ is an eigenvalue. 
The coefficients $\alpha_{1},\alpha_{2}$, that completely determine the 
eigenstates in the four cases, in turn satisfy
\begin{eqnarray*}
\frac{\alpha_{2}}{\alpha_{1}} = \Big(\frac{t'}{t}\, \Big)\frac{z_1^{-2{\sf L}}(i\Delta +{\rm s}_2({\sf D} - i\epsilon))}{2\cos k(2{\sf L})},
\;\;\text{if} \; {\rm s}_1 = +1, {\rm s}_2 = \pm 1, \\
\frac{\alpha_{2}}{\alpha_{1}} = \Big(\frac{t'}{t}\, \Big)\frac{z_1^{-2{\sf L}}(i\Delta +{\rm s}_2({\sf D} - i\epsilon))}{2\sin k(2{\sf L})},
\;\; \text{if} \;{\rm s}_1 = -1, {\rm s}_2 = \pm 1. 
\end{eqnarray*}
 
The structure of the above boundary equations explains how the number of bound
modes increases as we increase $N$ or $\Delta$. Notice that the
function on the right hand side of Eq. \eqref{snsboundaryeq1}
assume all real values between any two adjacent poles, given by 
\[k=\frac{\pi q}{2{\sf L}-1}+\frac{\pi}{2(2{\sf L}-1)},\quad q=0,1,\dots,2{\sf L}-2.\]
When the metal strip is completely disconnected from the SC, 
that is, when $t'=0$, the bound states corresponding to 
${\rm s}_1=+1$ in the metal are given by 
\[k=\frac{\pi q}{2{\sf L}} +\frac{\pi}{4{\sf L}},\quad q=0,1,\dots,2{\sf L}-1,\]
each of which lie singularly 
between two adjacent poles. This can
be seen from the relation 
\[\frac{\pi q}{2{\sf L}-1}>\frac{\pi q}{2{\sf L}}>\frac{\pi(q-1)}{2{\sf L}-1}.\]
This analysis 
indicates that each metallic state at energy less 
than $\Delta$ gets converted into a bound state
with slightly different value of energy in the presence of weak 
tunneling. For a fixed value of $\Delta$, increasing $N$ implies more poles
for the functions on the right hand-side, therefore allowing more solutions 
of the boundary equations, as discussed in the main text.

\section{The Creutz ladder}
\label{creupendix}

In terms of the array \(\hat{\Psi}_{j}^{\dagger}=
\begin{bmatrix}a_{j}^{\dagger} & b_{j}^{\dagger}\end{bmatrix},\) 
the single-particle Hamiltonian for the Creutz ladder, given by 
Eq.\,\eqref{CreutzHam} in the main text, is specified by the matrices
\begin{equation*}
{h}_{0}=-\begin{bmatrix}0 & {\sf M}\\
{\sf M} & 0
\end{bmatrix},\quad{h}_{1}=-\begin{bmatrix}{\sf K}e^{i\theta} & {\sf K}{\sf r}\\
{\sf K}{\sf r}& {\sf K}e^{-i\theta}
\end{bmatrix}.
\end{equation*}
It is more convenient, however, to work with the equivalent ladder Hamiltonian 
\(\widetilde{H}_N=\mathds{1}_N\otimes \tilde{h}_0+(T\otimes \tilde{h}_1+\text{H.c.})\)
defined in Eq. (\ref{tildeCreutz}), where the new matrices
\begin{eqnarray*}
\tilde{h}_{0}=\!-\!\begin{bmatrix}{\sf M} \!& 0\\
0 \!& -{\sf M}
\end{bmatrix}\!, 
\tilde{h}_1=-\begin{bmatrix}{\sf K}({\sf r}+\cos\theta) & {\sf K}\sin\theta\\
-{\sf K}\sin\theta & {\sf K}(-{\sf r}+\cos\theta)
\end{bmatrix}. \nonumber
\end{eqnarray*}
The analytic continuation of the corresponding Bloch Hamiltonian is
\begin{widetext}
\begin{align*}
\widetilde{H}(z) = 
-\begin{bmatrix}
{\sf M}+{\sf K}({\sf r}+\cos\theta)(z+z^{-1}) & {\sf K}\sin\theta \, (z-z^{-1})\\
-{\sf K}\sin\theta\,(z-z^{-1}) & -{\sf M} + {\sf K}(-{\sf r}+\cos\theta)(z+z^{-1})
\end{bmatrix},
\end{align*}
and the condition \(\det(\widetilde{H}(z)-\epsilon\mathds{1}_2)=0\) yields the
polynomial equation
\begin{align}
\label{creutzdispersion}
P(\epsilon,z) = -z^2\big[({\sf r}^{2}-1){\sf K}^2(z+z^{-1})^{2}+2{\sf K}({\sf M}{\sf r}-\epsilon \cos\theta)(z+z^{-1})+
{\sf M}^{2}+4{\sf K}^{2}\sin^{2} \theta -\epsilon^{2}\big]=0.
\end{align}
\end{widetext}

For fixed but arbitrary values of the parameters,
the singular, that is, flat-band energies, can be determined as
the solutions in \(\epsilon\) of the system of equations
\begin{align*}
({\sf r}^{2}-1){\sf K}^2=0 ,&\\
{\sf K}({\sf M}{\sf r}-\epsilon \cos \theta)=0,&\\
{\sf M}^{2}+4{\sf K}^{2}\sin^{2}\theta-\epsilon^{2}=0,&
\end{align*} 
For any combination of parameter values that exclude flat bands, 
the generalized Bloch theorem can be used to determine {\it all} the 
(regular) energy eigenvalues 
and eigenstates. For this system, there are \(2Rd=4\) independent 
solutions of the bulk equation for each value of \(\epsilon\), 
and they are all extended. Excluding 
power-law modes, these extended bulk solutions are labeled by 
the distinct roots of Eq.\,\eqref{creutzdispersion}. 
The solution $|u(\epsilon,z_\ell)\rangle$ of the kernel equation 
\((\widetilde{H}(z_\ell)-\epsilon\mathds{1}_2)|u\rangle=0\) can be taken 
to be 
\begin{eqnarray}
\label{creautzu}
&|u(\epsilon,z_\ell)\rangle=\begin{bmatrix}a(z_\ell)\\
\epsilon+b(z_\ell)\end{bmatrix},\\
&a(z)=-{\sf K}\sin\theta\,(z-z^{-1}),\nonumber\\
&b(z)={\sf M}+{\sf K}({\sf r}+\cos\theta)\,(z+z^{-1}) .\nonumber
\end{eqnarray}
For ${\sf r}\ne \pm1$, $h_1$ is invertible, and we get total four roots 
which come in reciprocal pairs. We choose the convention   
\(
z_{1}=z_3^{-1}\),  \(z_2=z_{4}^{-1}, \quad |z_1|,|z_2|\le 1
\) to denote them.
Then the ansatz is
\begin{equation*}
|\epsilon,\bm{\alpha},\bm{\beta}\rangle=\sum_{\ell=1,2} \Big(
\alpha_{\ell}|z_{\ell},1\rangle|u(\epsilon,z_{\ell})\rangle+
\beta_{\ell}|z_{\ell}^{-1},1\rangle|u(\epsilon,z_{\ell}^{-1})\rangle\Big),
\end{equation*}
with amplitudes \((\bm{\alpha},\bm{\beta}) = (\{\alpha_\ell\}, \{\beta_\ell\})\) to be determined by the 
boundary matrix. 

It is useful at this point to cast the ansatz in a more 
symmetric form. The unitary operator  
\begin{equation}
\label{symC}
\mathcal{S}=-\sum_{j=1}^{N}|N+1-j\rangle\langle j|\otimes \begin{bmatrix}1 &0\\0 &-1\end{bmatrix},
\end{equation}
describes a \(\mathds{Z}_2\) symmetry of the Hamiltonian 
in Eq.\,\eqref{CreutzHam}. 
It commutes with the bulk projector $P_B$, so that
\[
[\mathcal{S},P_B(H_N-\epsilon)]=0.
\]
Following Sec. \ref{sub:gbt}, this allows us to 
partition the bulk solution space into ${\rm s}=+1$ 
and ${\rm s}=-1$
eigenspaces of $\mathcal{S}$. 
Notice that under this transformation,
\begin{multline*}
\mathcal{S}|z_{\ell},1\rangle|u(\epsilon,z_{\ell})\rangle 
=\sum_{j=1}^{N}|N+1-j\rangle\langle j|z_{\ell},1\rangle
\otimes 
\begin{bmatrix}-a(z_\ell)\\
\epsilon+b(z_\ell)\end{bmatrix}\\
=z_\ell^{N+1}|z_{\ell}^{-1},1\rangle\begin{bmatrix}a(z_\ell^{-1})\\
\epsilon+b(z_\ell^{-1})\end{bmatrix}
=z_\ell^{N+1}|z_{\ell}^{-1},1\rangle|u(\epsilon,z_{\ell}^{-1})\rangle.
\end{multline*}
This equation is a consequence of the 
symmetry 
\[
\sigma_z \widetilde{H}(z) \sigma_z = \widetilde{H}(z^{-1}),\quad \sigma_z = \begin{bmatrix}1 &0\\0 &-1\end{bmatrix}
\] 
of the reduced bulk Hamiltonian.
Therefore, the ansatz yields eigenstates of 
$\mathcal{S}$ provided $z_\ell^{N+1}\alpha_{\ell}=\pm\beta_{\ell}$. 
For each energy, we obtain two ans\"{a}tze,
\begin{align}\label{creutzansatz2}
&|\epsilon,{\rm s},\bm{\alpha}\rangle=\\
&\sum_{\ell=1,2}\alpha_{\ell}\{|z_{\ell},1\rangle|u(\epsilon,z_{\ell})\rangle+{\rm s} 
z_\ell^{N+1}|z_{\ell}^{-1},1\rangle|u(\epsilon,z_{\ell}^{-1})\rangle\},\nonumber
\end{align}
corresponding to the eigenvalues ${\rm s}=\pm 1$. 
Each of these ans\"{a}tze,
with only two free parameters, is representative of the 
$D=2$ bulk solution space compatible with the 
corresponding eigenvalue of the symmetry.

The next step is to construct the boundary matrices 
corresponding to ${\rm s}=\pm 1$. 
We need to find a basis of the boundary subspace in which 
${\rm s}$ is block-diagonal. 
One such basis is $\{|{\rm s},m\rangle,\,{\rm s}=1,-1, m=1,2\}$, 
where
\[
|{\rm s},1\rangle \equiv \frac{1}{\sqrt{2}}(|1\rangle - {\rm s}|N\rangle)
\begin{bmatrix}1 \\ 0\end{bmatrix},\;\;
|{\rm s},2\rangle \equiv \frac{1}{\sqrt{2}}(|1\rangle + {\rm s}|N\rangle)
\begin{bmatrix}0 \\ 1\end{bmatrix}\!.
\]
The two boundary matrices are then 
\begin{multline*}
\!\!\!B(\epsilon,{\rm s})=\!\\
-\sqrt{2}h_1^\dagger\begin{bmatrix}
a(z_{1})(1 -{\rm s} z_1^{N+1})
& a(z_{2})(1 -{\rm s} z_2^{N+1})\\
(\epsilon+b(z_{1}))(1 +{\rm s} z_1^{N+1}) 
& (\epsilon+b(z_{2}))(1 +{\rm s} z_2^{N+1}) 
\end{bmatrix}\!.
\end{multline*}
In simplifying the boundary matrix, we have used
\begin{multline*}
\langle N|\langle m|(\widetilde{H}_N-\epsilon)|\epsilon,s,\bm{\alpha}\rangle \\
= (-1)^m s \langle 1|\langle m|(\widetilde{H}_N-\epsilon)|\epsilon,s,\bm{\alpha}\rangle,
\end{multline*}
which follows from the symmetry $\mathcal{S}$ of $\widetilde{H}_N$,
and also 
\[
(\tilde{h}_0-\epsilon\mathds{1}_2+z_\ell \tilde{h}_1)|u(\epsilon,z_\ell)\rangle
=-z_\ell^{-1}\tilde{h}_1^\dagger|u(\epsilon,z_\ell)\rangle,\] 
which follows from the bulk equation.

\subsubsection{The parameter regime \({\sf M}=0, \theta=\pi/2, {\sf r} \ne 1\)} 

We now derive explicit solutions for energy eigenstates 
in the parameter regime ${\sf M}=0,\ \theta=\pi/2$, for the non-trivial case ${\sf K}\neq 0$ and 
odd values of $N$. The calculation for $\theta=-\pi/2$ can be carried out in a 
similar way. We will also assume ${\sf r}\neq \pm 1$ for this analysis, which 
makes $\tilde{h}_1$ invertible. In this parameter regime, the 
Creutz ladder is dual to two decoupled copies of Kitaev's Majorana 
chain. Notice from Eq.\,\eqref{creutzdispersion} that in this 
case, we get $z_1=-z_2$ for any value of $\epsilon$.  
This leads to the simplification $a(z_1)=-a(z_2)$ and $b(z_1)=-b(z_2)$
for the quantities appearing in the boundary matrices. 
Further, for odd $N$, we get $z_2^{N+1}=z_1^{N+1}$.
This allows us to determine the solutions of $\det B(\epsilon,{\rm s})=0$
analytically. Observe that for $\epsilon=0$, the two columns 
of the boundary matrices differ by a minus sign. Therefore, the kernel 
vector
of the boundary matrix is $\bm{\alpha}=[1 \quad 1]^{\rm T}$.
We get two eigenvectors corresponding to exact zero energy, 
which are given by the unified expression (up to normalization)
\begin{eqnarray}
\label{creutzev0}
& |\epsilon=0,{\rm s},\bm{\alpha}\rangle = 2\sum_{j \ \text{odd}} |j\rangle
\begin{bmatrix} a(z_1)(z_1^j -{\rm s}\, z_1^{N+1-j}) \\ 
b(z_1)(z_1^j +{\rm s}\, z_1^{N+1-j})\end{bmatrix},\nonumber\\
& z_1 = \left\{
\begin{array}{lcl}
i\sqrt{({\sf r}-1)/(1+{\sf r})} & \text{if} & {\sf r} > 1\\
\sqrt{(1-{\sf r})/(1+{\sf r})} & \text{if} & 0<{\sf r} <1 \\
\sqrt{(1+{\sf r})/(1-{\sf r})} & \text{if} & -1<{\sf r} < 0 \\
i\sqrt{(1+{\sf r})/({\sf r}-1)} & \text{if} & {\sf r} < -1
\end{array}\right. . \quad \quad \quad 
\end{eqnarray}
The symmetry $\mathcal{S}$ is spontaneously broken by these 
zero energy eigenvectors.
It is worth a remark that, following the exact same analysis, 
the energy of the edge mode is found to be exact zero
for any value of $\theta$ as long as ${\sf M}=0$ and $N$ is odd. 
A similar phenomenon was uncovered in Kitaev's Majorana chain 
in Ref.\,[\onlinecite{Katsura17}]. 

If the eigenvalue is non-zero, then $\det B(\epsilon,{\rm s})=0$ 
leads 
to the condition $z_1^{N+1}=\pm 1$, in which case either the 
upper or lower
row of the boundary matrix vanishes. This condition is satisfied 
if $z_1$ takes any 
value from the set $\{e^{i\pi q/(N+1)}, q=1,\dots,2N+2\}$.  
Out of these $2N+2$ values, 
the four values $z_1=\pm1,\pm i$ do not fit this analysis, since 
each of them are double
roots of the characteristic equation, and lead to power-law bulk 
solutions.
We will now find eigenvectors corresponding to the remaining $2N-2$ 
values of $z_1$. First, consider $q$ odd, so that $z_1^{N+1}=-1$. 
The corresponding energy values found from Eq.\,\eqref{creutzdispersion}
are $\epsilon = \pm \epsilon_q$, where
\[
\epsilon_{q}= 2{\sf K}\sin[\pi q/(N+1)]\sqrt{1+\gamma_q^2},\quad 
\gamma_q= {\sf r}\cot[\pi q/(N+1)].
\]
For either of these two energy values, the lower row of the 
boundary matrix 
$B(\epsilon,{\rm s}=+1)$ is identically zero, and its kernel is determined 
by the upper row, which is spanned by $[1 \quad 1]^{\rm T}$.
For simplicity of calculations, we choose 
\[
\bm{\alpha}=i\Big(8{\sf K}\sin (\pi q/(N+1))\Big)^{-1}\begin{bmatrix}1 \\ 1\end{bmatrix}.
\]
This leads us to the eigenvectors
\begin{eqnarray}
&&|\epsilon=\pm\epsilon_q,{\rm s}=+1,\bm{\alpha}\rangle = 
\!\!\!\!\label{creutzev1} \\ 
&&\!\!\sum_{j \text{ odd}} |j\rangle \!
\begin{bmatrix} \cos\big(\frac{\pi qj}{N+1}\big) \\ 
-\gamma_q\sin\big(\frac{\pi qj}{N+1}\big)\end{bmatrix}
\!-\!\!\!\sum_{j \text{ even}}|j\rangle \! 
\begin{bmatrix} 0 \\ 
\pm\sqrt{1+\gamma^2_q}\sin\big(\frac{\pi qj}{N+1}\big)
\end{bmatrix}\!. \nonumber 
\end{eqnarray}
One can repeat the same procedure to calculate the eigenvectors 
$|\epsilon=\pm\epsilon_q,{\rm s}=-1,\bm{\alpha}\rangle$ from the boundary 
matrix $B(\epsilon,{\rm s}=-1)$. However, notice that the operator
\[
\mathcal{C} = \mathds{1}_N\otimes \begin{bmatrix} 0&1\\1&0\end{bmatrix}
\]
satisfies the anti-commutation relation 
$\mathcal{C}\widetilde{H}_N\mathcal{C}^{-1}=-\widetilde{H}_N$, 
and therefore is a chiral symmetry of $\widetilde{H}_N$.
This allows us to 
write 
\begin{widetext}
\begin{equation}
\label{creutzev2}
|\epsilon=\mp\epsilon_q,{\rm s}=-1,\bm{\alpha}\rangle = \mathcal{C}|\epsilon=\pm\epsilon_q,{\rm s}=+1,\bm{\alpha}\rangle= 
\sum_{j \text{ odd}}|j\rangle
\begin{bmatrix}  -\gamma_q\sin\big(\frac{\pi qj}{N+1}\big) 
\\ \cos\big(\frac{\pi qj}{N+1}\big) \end{bmatrix}
-\sum_{j \text{ even}}|j\rangle\begin{bmatrix} 
\pm\sqrt{1+\gamma^2_q}\sin\big(\frac{\pi qj}{N+1}\big)\\ 0
\end{bmatrix}.
\end{equation}
\end{widetext}
Repeating this analysis for even values of $q$ reveals that 
Eqs.\,\eqref{creutzev1} and \eqref{creutzev2} still provide 
the expressions for the corresponding eigenvectors, but in contrast to the 
situation for odd $q$, the expression in Eq.\,\eqref{creutzev1}
is in the symmetry sector $\mathcal{S}=-1$ and the one in 
Eq.\,\eqref{creutzev2} lies in the sector $\mathcal{S}=+1$.

Finally, let us tally the total number of eigenvectors that 
we have found. 
Each value of $z_1$ (other than $\pm1$ and $\pm i$) provided 
us four eigenvectors 
($|\epsilon=\pm\epsilon_q,{\rm s}=+1,\bm{\alpha}\rangle,\ |\epsilon=\pm\epsilon_q,{\rm s}=-1,\bm{\alpha}\rangle$).
However, for each value of $z_1$, three other roots, 
namely $-z_1,\ z_1^{-1}$
and $-z_1^{-1}$ lead to the exact same set of four eigenvectors. 
Effectively, we get one eigenvector per $q$. These account for 
$2N-2$ energy eigenstates. Combined with
the two zero eigenstates in Eq.\,\eqref{creutzev0}, we have 
accounted for all $2N$ eigenstates.

\vspace*{1cm}

\subsubsection{The parameter regime ${\sf r}=\pm1$}

In the parameter regime \({\sf r}=\pm 1\) and arbitrary values 
of ${\sf M},\ {\sf K}$ and $\theta$, 
the matrix \(\tilde{h}_1\) is no longer invertible. In the 
cases with no flat energy bands, two out of four bulk solutions for any 
value of $\epsilon$ are then emergent solutions.
In this section, we will shed light on the bulk-boundary 
correspondence of the Hamiltonian $\widetilde{H}_N$ assuming open BCs, and by 
assuming that the emergent solutions
have no contribution in forming the eigenstates. The latter 
assumption is validated by numerical calculations.

Let us set ${\sf r}=1$ for concreteness. Then,
\begin{equation*}
\tilde{h}_{1}=-{\sf K}\begin{bmatrix}2\cos^{2}(\theta/2) & \sin\theta\\
-\sin\theta & -2\sin^{2}(\theta/2)
\end{bmatrix},
\end{equation*}
is a rank-$1$ matrix, and this is the reason for the quadratic 
dependence on $z$ of the dispersion relation, as opposed to quartic 
in general. The two roots \(z_\ell,\ \ell=1,2\) must 
be reciprocals of each other, that is 
$z_1= z_2^{-1}$, with $|z_1|\le1$. The vector $|u(\epsilon, z_\ell)\rangle$ of 
Eq.\,\eqref{creautzu} simplifies to
\begin{equation*}
|u(\epsilon, z_\ell)\rangle=\begin{bmatrix}-{\sf K}\sin\theta(z_\ell-z_\ell^{-1})\\
\epsilon+{\sf M}+2{\sf K}\cos^{2}(\theta/2)(z_\ell+z_\ell^{-1})
\end{bmatrix}.
\end{equation*}
This accounts for half of the solutions 
of the bulk equation. The other two bulk solutions 
are emergent solutions localized on the edges of the system.  
The appropriate submatrices in this case 
are ${K}^- = h_1^\dagger$ and ${K}^+=h_1$, and so the emergent solutions are 
\begin{align*}
|j=1\rangle|u^-\rangle &= 
|j=1\rangle
\begin{bmatrix}
\sin(\theta/2) \\ 
-\cos(\theta/2)
\end{bmatrix},\\
|j=N\rangle|u^+\rangle &= 
|j=N\rangle
\begin{bmatrix}
\sin(\theta/2) \\ 
\cos(\theta/2)
\end{bmatrix} ,
\end{align*}
independent of \(\epsilon\).
Having found all four bulk solutions, the boundary matrix can be constructed 
as usual. For analytical, as opposed to computer-assisted, work it is 
advantageous to focus on obtaining the energy  eigenstates that have no 
contribution from the emergent solutions. 

The ansatz for propagating states is
\[
|\epsilon\rangle = 
\alpha_1 |z_1,1\rangle|u(\epsilon,z_1)\rangle + \alpha_2 |z_1^{-1},1\rangle|u(\epsilon,z_1^{-1})\rangle .
\] 
We can once again use the symmetry of Eq.\,\eqref{symC}.
The boundary equation for the symmetric 
(${\rm s}=+1$) 
and antisymmetric (${\rm s}=-1$) ans\"{a}tze in this 
case leads to the polynomial equation
\[
4{\sf K}\cos^2(\theta/2)(z_1+{\rm s} z_1^N) + (\epsilon+{\sf M})(1+{\rm s} z_1^{N+1})=0.
\]
With some algebraic manipulation, the two conditions 
can be recast into the transcendental equations
\begin{eqnarray*}
\label{creutzboundary2}
&&-\frac{4{\sf K}\cos^2{\frac{\theta}{2}}}{\epsilon+{\sf M}}=\frac{\cos[k(N+1)/2]}{\cos[k(N-1)/2]}
\quad \text{if }{\rm s}=+1,\\
&&-\frac{4{\sf K}\cos^2{\frac{\theta}{2}}}{\epsilon+{\sf M}}=\frac{\sin[k(N+1)/2]}{\sin[k(N-1)/2]}
\quad \text{if }{\rm s}=-1,
\end{eqnarray*}
respectively, where we have substituted $z_1=e^{ik}$.
When any one of these conditions is satisfied, the 
corresponding eigenstate is found to be 
\begin{widetext}
\[
|\epsilon, {\rm s}\rangle = 
\sum_{j=1}^{N}|j\rangle\begin{bmatrix}-{\sf K}\sin\theta\sin k \sin [(\frac{N+1}{2}-j)k + \frac{(1-{\rm s})\pi}{4}]\\
(\epsilon+{\sf M}+2{\sf K}\cos^{2}(\theta/2)\cos k) \cos [(\frac{N+1}{2}-j)k - \frac{(1-{\rm s})\pi}{4}]
\end{bmatrix}.
\]
\end{widetext}

In the large-$N$ limit, the condition for the existence of edge 
state on the left edge can be derived by substituting $\lim_{N\mapsto\infty} z_1^{N}=0$, 
that leads to 
$-\frac{4{\sf K}}{ \epsilon+{\sf M}} \cos^2 (\theta/2) =1/{z_1}.$
Therefore, if there exists a solution ($\epsilon$, $z_1$) to this 
equation
that is compatible with the dispersion relation of Eq.\,\eqref{creutzdispersion}
and satisfies $|z_1|<1$, then $\widetilde{H}_N$ hosts a 
localized mode on the left edge. By substituting the value of $z_1$ 
in terms of energy and other parameters 
in Eq.\eqref{creutzdispersion}, we obtain a cubic polynomial 
equation in energy. Two of the roots of this equation are $\epsilon=-({\sf M}\pm 4{\sf K}\cos^2(\theta/2))$, 
that correspond to $z_1=\pm1$. We throw away these roots, because 
they do not correspond to bound states, since $z_1$ lies on the unit 
circle. The third root, that is the root of our interest, is 
$\epsilon={\sf M}\cos{\theta}$. The corresponding value of $z_1$ is $z_1=-{\sf M}/2{\sf K}$. 
Now we impose the final condition, which is $|z_1|<1$. 
This is satisfied for the values $|{\sf M}|<|2{\sf K}|$. Therefore, we conclude 
that if $|{\sf M}|<|2{\sf K}|$, then $\widetilde{H}_N$ hosts a localized state on the left edge,
with non-zero energy in general. This calculation is consistent with 
the original observation by Creutz that the system 
hosts edge states for the parameter regime $|{\sf M}|<|2{\sf K}{\sf r}|$,
which coincides with $|{\sf M}|<|2{\sf K}|$ for ${\sf r}=1$.

\section{Dimerized chains}
\label{basic_examples}

The $D=1$ model Hamiltonian of the form 
\begin{eqnarray*} 
\widehat{H}& =& \sum_{i=1}^{2N} \, [v-(-1)^i\delta_v ] c_i^\dagger c_i \\
& -& \sum_{i=1}^{2N-1} \![(t-(-1)^i\delta_t)
c_{i}^\dagger c_{i+1} +\text{H.c.}],
\end{eqnarray*}
where the parameters 
$$ t=\frac{t_1+t_2}{2},\; \delta_t=\frac{t_1-t_2}{2},\; v=\frac{v_1+v_2}{2},
\; \delta_v=\frac{v_1-v_2}{2},$$
subsumes several interesting spin-insensitive phenomena of 
$D=1$ electronic matter. At half-filling, the model is mostly insulating 
(the gap only closes only if \(\delta_t=0=\delta_v\)), and 
has been used for investigating solitons in polyenes (the Rice-Mele 
model at \(v=0\)), ferroelectricity, and charge fractionalization (the SSH  
model, or even Peierls chain sometimes, at \(v=0=\delta_v\)); see 
Ref.\,[\onlinecite{BerryPhaseRMP}] and references therein. If \(\delta_t=0\), 
our dimerized chain can also be regarded as a special instance of the Aubrey-Harper 
family of Hamiltonians. 

At present, the SSH model is regarded as the simplest particle-conserving
topological state of independent electrons (see again Ref.\,[\onlinecite{BerryPhaseRMP}]
for a discussion of the Berry phase if \(v=0\)). In this sense, 
it is the natural counterpart of the Kitaev's Majorana chain, and more 
is in fact true: {\it as a many-body Hamiltonian}, the SSH model is dual 
to the Majorana chain at vanishing chemical potential \cite{equivalence}. 
In contrast, if \(\delta_t=0\neq \delta_v\) (we will informally call 
this regime the Aubrey-Harper chain), the model is topologically trivial.
The Aubrey-Harper chain is exactly solvable for open BCs. For generic 
parameters, the full model is {\it not} analytically solvable for open BCs, 
but we will introduce distorted open BCs that yield analytic rather than just exact solvability. 
Fortunately, these unconventional open BCs map by duality to the standard ones 
for the Majorana chain.
In all cases, a very precise picture of intra-gap states can be obtained 
in a well-controlled large-size approximation that does not remove the 
geometric inversion operation \(j\leftrightarrow N+1-j\), as passing to a 
half-infinite system geometry does.      

The single-particle Hamiltonian for our dimerized chain subject
to open BCs is
\begin{align*}
H_N&=\mathds{1}_N\otimes h_0+(T\otimes h_1+\text{H.c.}),\\
h_0&=\begin{bmatrix}
v_1& -t_1\\
-t_1^{*}& v_2
\end{bmatrix},\quad
h_1=
\begin{bmatrix}
0& 0\\
-t_2& 0
\end{bmatrix},\quad v_1,v_2,t_2 \in \mathds{R}.
\end{align*}
For $v_1=v_2=0$, the Hamiltonian has a chiral symmetry 
\[\mathcal{C}_1=\mathds{1}_N \otimes 
\begin{bmatrix}1 & 0 \\ 0 & -1\end{bmatrix}.
\]
For real values of $t_1$ and $v_2=-v_1$, the system has another non-local chiral symmetry,
\[\mathcal{C}_2=\sum_{j=1}^{N}|N+1-j\rangle\langle j| \otimes 
\begin{bmatrix}0 & -i \\ i & 0\end{bmatrix},
\]
which is, however, absent in the limit $N \rightarrow \infty$.
The analytic continuation of the Bloch Hamiltonian is then 
\begin{eqnarray}
\label{precisely0}
H(z)=
\begin{bmatrix}
v_1 & -t_1-t_2z^{-1}\\
-t_1^*-t_2z& v_2
\end{bmatrix},
\end{eqnarray} 
so that the condition \(\det(H(z)-\epsilon\mathds{1}_2)=0\) is
equivalent to the ``dispersion relation" $P(\epsilon,z)=0$, where
\begin{equation}
\label{peierlschar}
P(\epsilon,z)=z^2[(\epsilon-v_1)(\epsilon-v_2)-
(t_1+t_2z^{-1})(t_1^*+t_2z)].
\end{equation}

Before we continue investigating this model with the aid
of the generalized Bloch theorem, it is convenient 
to isolate the occurrence of flat bands. For the dimerized 
chain, flat bands are only possible if $t_1=0$ or $t_2=0$.
The diagonalization of the system is then trivial. We will 
not pursue it further, assuming from now on that \(t_1,t_2\neq0\). 
In order to be able to diagonalize our dimerized chain in 
closed form, we will impose BCs
\[
W=|N\rangle\langle N|\otimes 
\begin{bmatrix}
0& t_1\\
t_1^*& 0
\end{bmatrix}.
\]

Since the range of 
hopping is \(R=1\) and the number of internal states is \(d=2\) 
(two atoms per unit cell), the number of boundary
degrees of freedom is \(2Rd=4\). This
number coincides with the number of solutions of the bulk equation
for each value of the ansatz parameter \(\epsilon\). There are two 
emergent bulk soluctions, and two extended ones labelled by the roots 
\(z_\ell,\ \ell=1,2\) of Eq.\,\eqref{peierlschar}. The two 
roots coincide, that is, \(z_1=z_2\), only if \(\epsilon\) 
takes one of the four values values 
$$\big\{v \pm \sqrt{\delta_v^2+(|t_1|+t_2)^2},\ v \pm \sqrt{\delta_v^2+(|t_1|-t_2)^2}\big\}.$$
For these special values of the energy, one of the extended 
solutions shows power-law behavior. 

Ignoring power-law solutions for the moment, the propagating 
solutions are \(|z_\ell,1\rangle|u_\ell\rangle,\ \ell=1,2\), with  
\begin{eqnarray}\label{presentation1}
|u_\ell\rangle=|u(\epsilon,z_\ell)\rangle\equiv
\begin{bmatrix}
t_1+t_2z_\ell^{-1}\\
v_1-\epsilon
\end{bmatrix}
\end{eqnarray}
such that \((H(z_\ell)-\epsilon\mathds{1}_2)|u_\ell\rangle=0\).
Notice that for $\epsilon = v_1$ and $z_2 = -t_2/t_1$, the vector 
$|u(v_1,-t_2/t_1)\rangle$ vanishes. Therefore, we will deal with 
the case $\epsilon=v_1$ separately.
For our dimerized chain, the matrices \(K^{\pm}\) that determine
the emergent solutions are simply $K^-=h_1$ and $K^+=h_1^\dagger$, 
so that the solutions themselves are $|\psi^-\rangle = |1\rangle|u^-\rangle$
and $|\psi^+\rangle = |N\rangle|u^+\rangle$, with
\[
|u^-\rangle = 
\begin{bmatrix}
 1 \\ 0 
\end{bmatrix},	\quad 
|u^+\rangle=
\begin{bmatrix} 
0 \\ 1 
\end{bmatrix} ,
\]
{\em independently of $\epsilon$}. We emphasize
that emergent solutions are {\em not} always independent of \(\epsilon\) \cite{JPA}. 
Then our ansatz for energy eigenstates of the dimerized chain is
\[
|\epsilon\rangle = 
\alpha_-|1\rangle|u^-\rangle+
\sum_{\ell=1}^{2}\alpha_{\ell}|z_\ell,1\rangle|u(\epsilon,z_\ell)\rangle 
+ \alpha_+|N\rangle|u^+\rangle.
\]

Our generalized Bloch theorem guarantees that the eigenstates of 
the model are necessarily contained in the ansatz, with amplitudes 
\(\bm{\alpha}=[\alpha_-\,\alpha_1\,\alpha_2\,\alpha_+]^{\rm T}\) 
determined by the boundary matrix  
\begin{multline}
B(\epsilon)=  \\
\begin{bmatrix}
v_1-\epsilon & t_2(v_1-\epsilon) & t_2(v_1-\epsilon) & 0\\
-t_1^* & 0 & 0 & 0\\
0 & z_1^Nt_1(v_1-\epsilon) & z_2^Nt_1(v_1-\epsilon) & 0\\
0 & z_1^N(v_1-\epsilon)(v_2-\epsilon) & z_2^N(v_1-\epsilon)(v_2-\epsilon) & v_2-\epsilon
\end{bmatrix} \hspace*{-4mm} \label{bmpeierls}
\end{multline}
as \(B(\epsilon)\bm{\alpha}=0\).  The first and the last columns of the
boundary matrix are contributed by the emergent modes, 
and the remaining two by the propagating 
modes. The kernel of the boundary matrix is nontrivial only if 
\begin{eqnarray}
\label{peierlscond1}
\epsilon=v_2\quad\mbox{or}\quad z_1^N=z_2^N \quad\mbox{or}\quad\epsilon=v_1. 
\end{eqnarray} 
Out of these three possibilities, $\epsilon=v_2$ yields the 
kernel vector 
$\bm{\alpha} = [ 0 \; 0 \; 0 \; 1]^{\rm T}$ of the boundary 
matrix,
which represents the decoupled fermion at site $j=N$,
\[
|\epsilon=v_2,\bm{\alpha}\rangle = |N\rangle\begin{bmatrix} 0 \\ 1 \end{bmatrix}.
\]
In order to solve the second equation ($z_1^N=z_2^N$) fully, 
it is necessary
to notice a ``symmetry" of Eq.\,\eqref{peierlschar}: The
roots of the dispersion relation must satisfy the constraint
\begin{eqnarray}
z_1z_2=t_1^*/t_1 \equiv e^{-2i\phi},
\end{eqnarray}
This leads to the allowed values
\begin{eqnarray}
&z_1=e^{2i\phi}z_2^{-1}= e^{i\frac{\pi}{N} q-i\phi},\quad q=-N-1,\dots,N. \quad 
\label{pu1}
\end{eqnarray}
Combining this equation with Eq.\,\eqref{peierlschar},
we find that the corresponding energy values are 
\begin{align*}
&\epsilon_n(q)=v + (-1)^n\sqrt{\delta_v^2 + |t(q)|^2},\quad n=1,2,\\
&|t(q)|^2 \equiv |t_1|^2 + t_2^2 + 2|t_1|t_2\cos (\pi q/N),
\end{align*}
independent of \(\phi\). 
The last step is putting together the stationary-wave states
associated to Eq.\,\eqref{pu1}. 
The kernel of the boundary matrix is spanned by 
$\bm{\alpha}=\begin{bmatrix}0 & 1 & -1 & 0\end{bmatrix}^{\rm T}$.
The actual eigenstates are
\begin{align}
\nonumber
&|\epsilon_n(q),\bm{\alpha}\rangle=|\chi_1(q)\rangle
\begin{bmatrix} t_1\\ v_1-\epsilon_n(q)\end{bmatrix}
+|\chi_2(q)\rangle\begin{bmatrix}t_2\\ 0\end{bmatrix},
\nonumber
\end{align}
with \(|\chi_i(q)\rangle, i=1,2,\) as in Eqs.\,\eqref{chiwf1}, \eqref{chiwf2}.
Notice that the eigenvectors $|\epsilon_n(q),\bm{\alpha}\rangle$
and $|\epsilon_n(-q),\bm{\alpha}\rangle$ obtained in this way 
are identical. Further, the energies corresponding to $q=\{0,N\}$ 
are precisely the ones that have associated power-law bulk solutions, 
and
the boundary matrix in Eq.\,\eqref{bmpeierls} is not valid for these 
energies. Therefore,
the above analysis has revealed only $2(N-1)$ bulk eigenstates
(along with the one localized eigenstate at $\epsilon=v_2$ found 
earlier).
We are still missing one eigenstate, because we have not yet 
analyzed the
case $\epsilon = v_1$, and also not considered the situation 
corresponding to power-law modes.

Let us now focus on the case $\epsilon=v_1$. 
$B(\epsilon)$ in Eq.\,\eqref{bmpeierls} is not the correct boundary 
matrix for 
$\epsilon=v_1$, since $|u(\epsilon=v_1,z_2=-t_2/t_1)\rangle$ vanishes 
as mentioned before. 
For this energy and $z_2$ (which satisfy $P(\epsilon,z_2)=0$), we have
\[
H(-t_2/t_1)-v_1\mathds{1}_2 = \begin{bmatrix}
0 & 0 \\
-t_1^* + t_2^2/t_1 & v_2-v_1,
\end{bmatrix} ,
\]
whose kernel is spanned by 
\[
|u_2\rangle = \begin{bmatrix}
t_1(v_2-v_1)\\
|t_1|^2-t_2^2
\end{bmatrix}.
\]
We can still use $|u_1\rangle = |u(\epsilon=v_1,z_1 = -t_1^*/t2)\rangle$,
and the two emergent solutions ($|\psi^-\rangle$ and $|\psi^+\rangle$) 
are as before. Then the boundary matrix for $\epsilon=v_1$ is
\begin{multline*}
B(\epsilon=v_1)=\\
\begin{bmatrix}
0 & 0 & t_2(|t_1|^2-t_2^2) & 0\\
-t_1^* & 0 & 0 & 0\\
0 & 0 & (-t_2/t_1)^Nt_1(|t_1|^2-t_2^2) & 0\\
0 & 0 & (-t_2/t_1)^N(v_2-v_1)(|t_1|^2-t_2^2) & v_2-v_1
\end{bmatrix}.
\end{multline*}
The kernel of $B(\epsilon=v_1)$ is $D=1$, and is spanned 
by $\bm{\alpha} = [0 \; 1 \; 0 \; 0]^{\rm T}$. Thus, there is only 
one eigenstate at $\epsilon=v_1$,
\[
|\epsilon=v_1,\bm{\alpha}\rangle=
|z_1=-t_1^*/t_2,1\rangle
\begin{bmatrix}
(|t_1|^2-t_2^2)/t_1\\
0
\end{bmatrix}.
\]
For $t_2<t_1$, this energy eigenstate is exponentially localized on 
the right edge, whereas for $t_2>t_1$, it is localized on the left edge. 
This 
behavior is characteristic of the topological phase transition that 
occurs at $t_2=t_1$. It is not possible to continue this eigenvector
into the parameter regime \(t_2=t_1\). With this localized eigenstate,
we have found all $2N$ eigenstates of $H_N+W$.
According to these results, the emergent solution on the left does not
enter the physical spectrum for open BCs. The one on the right does, at energy 
\(\epsilon=v_2\). 

Since we have already found the eigenbasis of \(H_N+W\) in terms
of the ansatz pertaining to those $\epsilon$ for which \(z_1\neq z_2\),
the boundary matrix calculated at those values of $\epsilon$ which bear 
coinciding roots should produce no more eigenvectors. It is instructive
to check explicitly that this is the case. By looking at the discriminant 
of Eq.\,\eqref{peierlschar},
we find that such double roots appear for the energy values in the set 
$\big\{v \pm \sqrt{\delta_v^2+|t(0)|^2},\ v \pm \sqrt{\delta_v^2+|t(N)|^2}\big\}$.
Let us consider $\epsilon_{\pm} = v \pm \sqrt{\delta_v^2+|t(0)|^2}$, 
for which $z_1=z_2=e^{-i\phi}$ is a double root. In addition to the 
generic solution 
$|\psi_1\rangle = |z_1\rangle|u(\epsilon,z_1)\rangle$, the bulk equation 
also has a power-law solution in this case, which is
\begin{eqnarray*}
|\psi_2\rangle = \partial_{z_1}|\psi_1\rangle
=\big(|z_1\rangle\partial_{z_1} + |z_1,2\rangle\big)|u(\epsilon,z_1)\rangle.
\end{eqnarray*} 
This effectively replaces each entry in the second column
of the boundary matrix $B(\epsilon)$ in Eq.\,\eqref{bmpeierls}
by its derivative with respect to $z_2$. Therefore, the resulting 
boundary matrix is 
\begin{widetext}
\begin{eqnarray*}
B(\epsilon_{\pm})=
\begin{bmatrix}
v_1-\epsilon_{\pm} & t_2(v_1-\epsilon_{\pm})  & 0 &  0 \\
t_1e^{i\phi} & 0 & 0 &  0 \\
0 &  t_1 e^{-iN\phi}(v_1-\epsilon_{\pm}) & N(v_1-\epsilon_{\pm}) t_1 e^{i(N-1)\phi} &  0 \\
0 &  e^{-iN\phi}(v_1-\epsilon_{\pm})(v_2-\epsilon_{\pm}) & 
Ne^{-i(N-1)\phi}(v_1-\epsilon_{\pm})(v_2-\epsilon_{\pm}) &  v_2-\epsilon_{\pm}
\end{bmatrix}.
\end{eqnarray*}
\end{widetext}
This boundary matrix is found to have a non-trivial kernel 
if and only if $\epsilon_\pm=v_1$. But since this analysis pertains to the points away from 
the phase transition ($|t_0|\ne|t_1|$), neither of the conditions $\epsilon_\pm=v_1$ can be satisfied. 
A similar analysis for the energy values 
$v \pm \sqrt{\delta_v^2 + |t(N)|^2}$, 
for which $z_1=z_2 = -e^{-i\phi}$ is the double root, 
leads to the conclusion that, away from the critical 
points, no eigenvector of $H$ takes contributions 
from power-law solutions. 
This is a particular feature of this Hamiltonian.


\begin{thebibliography}{}
\bibitem{footpip}
There are two other solutions of the system in Eq.\,\eqref{rightedge}, 
\(z_{1,\pm}=-\frac{1}{2}(\xi\pm\sqrt{4+\xi^2})\). These solutions
are excluded because the internal
state \(|u_\ell\rangle\) vanishes identically if evaluated 
at \(z_\ell=z_{1,\pm}\), see Eq.\,\eqref{upipeasy}. Similar remarks apply to the system in Eq.\,\eqref{leftedge}.
\bibitem{Notenotation}
Reference [\onlinecite{Deng14}] used the notation $P_{B,k_{\|}}$ ($\equiv Q_{k_{\|}}$),  
which however would be confusing in the present content.
\end{thebibliography}
\end{document}